\documentclass[a4paper,10pt]{elsarticle}
\usepackage[utf8]{inputenc}
\usepackage{import}

\usepackage{fullpage}

\usepackage[english]{babel}
\usepackage{amsmath,amssymb,amsthm}
\usepackage{cases}
\usepackage{mathtools}
\usepackage{mathrsfs}
\usepackage{setspace}
\usepackage{tabularx}
\usepackage{booktabs}

\usepackage{multicol}
\usepackage{multirow}
\usepackage{xparse}
\usepackage{xspace}
\usepackage{relsize}
\usepackage{lscape}
\usepackage[normalem]{ulem}
\usepackage[usenames,dvipsnames,table]{xcolor}
\usepackage{graphicx}
\usepackage{placeins}
\usepackage{svg}
\usepackage{float}
\usepackage{overpic}
\svgpath{{./images/}}
\definecolor{ultramarine}{RGB}{0,32,96}
\usepackage[colorlinks=true]{hyperref}
\usepackage{cleveref}
\usepackage{caption}          
\usepackage{subcaption}
\usepackage[notref,notcite,final]{showkeys}
\usepackage[linesnumbered,ruled,vlined]{algorithm2e}
\SetKwInput{KwInput}{Input}                
\SetKwInput{KwOutput}{Output}     
\usepackage[final,color=gray, backgroundcolor=gray!10, textwidth=2cm, textsize=footnotesize]{todonotes}
\usepackage{framed}
\usepackage{enumitem}

\newtheorem{remark}{Remark}

\newcommand{\Dm}{\mathcal{D}}
\newcommand{\Em}{\mathcal{E}}
\newcommand{\Fm}{\mathcal{F}}

\newcommand{\Lm}{\mathcal{L}}

\newcommand{\Pm}{\mathcal{P}}

\newcommand{\Rm}{\mathcal{R}}
\newcommand{\Sm}{\mathcal{S}}

\newcommand{\Wm}{\mathcal{W}}

\newcommand{\Rbb}{\mathbb{R}}

\newcommand{\Sec}[1]{Sec.~\ref{#1}}
\newcommand{\Fig}[1]{Fig.~\ref{#1}}
\newcommand{\Tab}[1]{Tab.~\ref{#1}}

\newcommand{\Alg}[1]{Alg.~\ref{#1}}

\newcommand{\FED}{\texttt{FED}\xspace}

\newcommand{\bra}[1]{\left( #1 \right)}

\newcommand{\cbra}[1]{\left\lbrace #1 \right\rbrace}

\renewcommand{\l}[1]{\text{\scalebox{0.9}{\upshape #1}}}

\newcommand{\coh}{\frac{1}{2}}
\newcommand{\ar}[1]{\!\bra{#1}}
\newcommand{\abs}[1]{\lvert #1 \rvert}
\newcommand{\diff}{\mathop{}\!\mathrm{d}}
\newcommand{\Diff}{\mathop{}\!\mathrm{D}}

\newcommand{\pd}[2]{\frac{\partial #1}{\partial #2}}
\NewDocumentCommand\argmin{o}{\IfNoValueTF{#1}{\arg\min\,}{\underset{#1}{\arg\min}\,}}

\AtBeginDocument{
  \let\div\relax
  \DeclareMathOperator{\div}{div}
}

\newcommand{\vect}[1]{{\boldsymbol{#1}}}
\newcommand{\vnorm}[1]{\| #1 \|}

\newcommand{\grad}{\nabla}
\newcommand{\vinner}{\cdot\!} 

\DeclareMathOperator{\tr}{tr}
\DeclareMathOperator{\Sym}{Sym}

\newcommand{\domain}{\Omega}
\newcommand{\bdomain}{\partial \domain}
\newcommand{\bdomainD}{\partial_{\l{D}} \domain}
\newcommand{\bdomainN}{\partial_{\l{N}} \domain}

\NewDocumentCommand\intargV{O{\domain}m}{\int_{#1} #2 \,\diff \vpos}
\NewDocumentCommand\intargt{O{0}O{t}m}{\int_{#1}^{#2} #3 \,\diff \tau}

\newcommand{\cw}{c_{\fDam}} 
\newcommand{\Gf}{G_\l{f}}
\newcommand{\Gc}{G_\l{c}}
\newcommand{\Gceff}{\bar{G}_\l{c}}

\newcommand{\sdisp}{u}

\newcommand{\seps}{\varepsilon}
\newcommand{\sepsv}{\seps_\l{v}}
\newcommand{\sepss}{\seps_\l{s}}
\newcommand{\sepsd}{\seps_\l{d}}
\newcommand{\sepsvbar}{{\bar\seps_\l{v}}}
\newcommand{\sepsdbar}{{\bar\seps_\l{d}}}

\newcommand{\sepsvcheck}{{\check\seps_\l{v}}}
\newcommand{\sepsdcheck}{{\check\seps_\l{d}}}

\newcommand{\sepsvtilde}{{\tilde\seps_\l{v}}}
\newcommand{\sepsdtilde}{{\tilde\seps_\l{d}}}

\newcommand{\sstr}{\sigma}

\newcommand{\sstrh}{\sstr_\l{h}}
\newcommand{\sstrd}{\sstr_\l{d}}

\newcommand{\sstrhbar}{{\bar\sstr_\l{h}}}
\newcommand{\sstrdbar}{{\bar\sstr_\l{d}}}

\newcommand{\sstrhhat}{{\hat\sstr_\l{h}}}
\newcommand{\sstrdhat}{{\hat\sstr_\l{d}}}

\newcommand{\sstrhtilde}{{\tilde\sstr_\l{h}}}
\newcommand{\sstrdtilde}{{\tilde\sstr_\l{d}}}

\newcommand{\sla}{\vartheta}

\newcommand{\sdam}{\alpha}
\newcommand{\vdamd}{\grad{\sdam}}
\newcommand{\sdamdt}{\dot\sdam}
\newcommand{\vdamddt}{\grad{\sdamdt}}

\newcommand{\sdamvar}{\tilde\sdam}
\newcommand{\vdamdvar}{\grad{\sdamvar}}

\newcommand{\vpos}{\vect{x}}
\newcommand{\vdisp}{\vect{\sdisp}}
\newcommand{\vdispvar}{\tilde\vdisp}
\newcommand{\vdispdt}{\dot{\vect{\sdisp}}}

\newcommand{\teps}{\vect{\seps}}

\newcommand{\tstr}{\vect{\sigma}}

\newcommand{\Uspace}{\mathscr{U}}
\newcommand{\Uvarspace}{\tilde{\Uspace}}
\newcommand{\Dspace}{\mathscr{A}}
\newcommand{\Dvarspace}{\tilde{\Dspace}}
\newcommand{\Allspace}{\mathscr{Q}}
\newcommand{\Allvarspace}{\tilde{\Allspace}}

\newcommand{\fDam}{w}

\newcommand{\Enfree}{\Em}
\newcommand{\Enpot}{\Pm}

\newcommand{\Dissdist}{\mathrm{D}_\mathcal{D}}
\NewDocumentCommand\dissdist{o}{\IfNoValueTF{#1}{D}{D\ar{#1}}}
\newcommand{\Disswork}{\Dm}
\newcommand{\disswork}{\delta}

\newcommand{\enfree}{\psi}

\newcommand{\enfreeD}{\enfree_\l{D}}
\newcommand{\enfreeR}{\enfree_\l{R}}
\newcommand{\disspot}{\phi}
\newcommand{\Disspot}{\Phi}

\newcommand{\setR}{\mbox{$\mbox{$\Rbb $}$}}
\NewDocumentCommand\endiss{o}{\IfNoValueTF{#1}{\disswork}{\disswork\ar{#1}}}

\definecolor{DarkerGreen}{RGB}{0,170,0}
\definecolor{DarkerRed}{RGB}{170,0,0}

\definecolor{myRed}{rgb}{0.450385, 0.157961, 0.217975}
\definecolor{myBlue}{rgb}{0.139681, 0.311666, 0.550652}
\definecolor{myLightRed}{rgb}{0.717647, 0.113725, 0.207843}
\definecolor{myLightBlue}{rgb}{0.247059, 0.603922, 0.694118}

\graphicspath{
  {images/}
  {images/mathematica/}
  {images/numerical_results/}
  }

\journal{}

\begin{document}

\begin{frontmatter}

\title{Endowing phase-field fracture models \\ with custom strength criteria}

\author[1]{Roberto Alessi\corref{cor1}}
\cortext[cor1]{Corresponding author}
\ead{roberto.alessi@unipi.it}

\author[2]{Matteo Brunetti}
\ead{matteo.brunetti@uniud.it}

\author[3]{Roshan Udaram Patil}
\ead{roshan.patil@iitjammu.ac.in}

\author[4]{Jacinto Ulloa}
\ead{julloa@umich.edu}

\affiliation[1]{organization={Department of Civil and Industrial Engineering, University of Pisa},
addressline={Largo Lazzarino 1},
city={Pisa},
postcode={56122},
country={Italy}}

\affiliation[2]{organization={Polytechnic Department of Engineering and Architecture, University of Udine},
addressline={Via delle Scienze 206},
city={Udine},
postcode={33100},
country={Italy}}

\affiliation[3]{organization={Department of Mechanical Engineering, Indian Institute of Technology Jammu},
addressline={Jagti, NH-44, PO Nagrota},
city={Jammu},
postcode={181221},
country={India}}

\affiliation[4]{organization={Department of Mechanical Engineering, University of Michigan},
addressline={1221 Beal Ave.},
city={Ann Arbor},
postcode={MI 48109},
country={USA}}

\begin{abstract}
By now, several approaches have been proposed to endow phase-field fracture models with the ability to describe crack nucleation under multiaxial stress states. These include techniques for splitting the free energy, direct modifications of the phase-field driving or resisting forces, and the introduction of additional internal variables, such as plastic strains or other nonlinear strains. In this paper, we propose a fundamentally different strategy for incorporating arbitrary elastic domains into phase-field fracture models, formulated within the energetic framework of generalized standard materials. The proposed approach relies on letting the dissipation potential depend on the current state of the material. In this way, the elastic degradation and the strength criterion remain two distinct and independently controllable aspects of the material response, while the governing equations directly descend from a local stability condition and energy balance. Simple yet representative models are presented and thoroughly discussed to demonstrate the effectiveness of the proposed methodology. 
The resulting evolution of the elastic domain is investigated in both strain and stress spaces under proportional and non-proportional loading paths, with the latter highlighting the path dependence of the dissipated work.
Moreover, numerical simulations demonstrate a range of crack nucleation processes under multiaxial loading conditions for various analytical strength surfaces. 
The present work paves the way for future developments and applications in several directions, including crack propagation studies and dedicated numerical solution strategies.
\end{abstract}

\end{frontmatter}

\clearpage

\tableofcontents

\clearpage

\section{Introduction}
\label{sec_intro} 

The phase-field modeling of brittle fracture~\cite{Bourdin2000b} is nowadays a consolidated theory within the
realm of fracture mechanics. The defining features of the phase-field approach that
have contributed to its efficacy and popularity are, among others, its underlying variational structure, its ability to capture geometrically complex and a priori undefined crack patterns, and its ease of implementation. Nowadays, numerous models based on the phase-field approach have been developed and explored, often coupled with other physical behaviors. Examples include extensions to plasticity~\cite{Alessi2014,Ambati2015,Miehe2016}, viscoelasticity~\cite{Dammass2021,Kamarei2025,Ciambella2025}, fatigue in brittle~\cite{Alessi2018,Alessi2019c,Mesgarnejad2019a} and ductile~\cite{Seiler2020,Ulloa2021a} materials, dynamic crack propagation~\cite{Borden2012,Li2016}, and various multiphysics problems~\cite{Mikelic2015,Martinez-Paneda2018,Dittmann2020}, to cite a few.

Despite the widespread adoption and remarkable success of the phase-field approach to fracture, several important issues remain open and deserve further investigation. Among the most relevant is the possibility of arbitrarily prescribing a strength criterion and consistently describing the evolution of the corresponding elastic domain as a function of the phase-field variable. This aspect is crucial for an accurate modeling of crack nucleation under mixed-mode loading conditions, which play a fundamental role in virtually all relevant fracture problems.

Several approaches have been proposed in the literature to address this issue. These can be broadly classified according to the adopted modeling strategy as follows:

\begin{enumerate}[label={(\alph*)}]
 \item \label{item_a} Modification of the free energy through energy splits;
 \item \label{item_b} Direct modification of the PDE system;
 \item \label{item_c} Coupling with plastic strains or other nonlinear strains.
\end{enumerate}

Approach~\ref{item_a} is undoubtedly the most widely adopted one
\cite{Amor2009a,Miehe2010a,Freddi2010,Freddi2011,DeLorenzis2021,Navidtehrani2022,Vicentini2024}. Its main advantage is that the problem retains a variational structure. However, this approach generally lacks flexibility in describing arbitrary elastic domains and mixed-mode strength criteria. Moreover, as highlighted in \citet[Tab. 2]{Vicentini2024}, the majority of the most examined energy decompositions suffer from spurious non-zero tangential stiffness in fully developed cracks.

An example of approach~\ref{item_b} that directly modifies the \emph{resisting force} is found in \citet{Bian2024}, where a mixed-mode cohesive-zone model is obtained through a suitable fracture toughness expression depending on different history variables. On the other hand, examples of approach~\ref{item_b} where the \emph{driving force} is modified can be found in several works~\cite{Miehe2015b,Zhang2017d,Wang2019,Kumar2020,Feng2022,AbrariVajari2023}. This approach allows for significant modeling versatility~\cite {Kumar2020,Kamarei2026}, particularly for introducing strength criteria. Nevertheless, in all approaches belonging to class~\ref{item_b}, the variational structure of the problem is lost, together with its theoretical and computational advantages. Moreover, a feature shared by approaches~\ref{item_a} and~\ref{item_b} is that the elastic degradation behavior and the strength criterion are strongly coupled and cannot be easily tuned independently.

Finally, the flexibility required to describe arbitrary strength criteria is achieved in approach~\ref{item_c} by coupling the phase-field variable with plastic strains or other nonlinear strain measures. These formulations remain variationally consistent, introducing yield criteria in the ductile case or strength criteria in the quasi-brittle case via the support function of the convex set defining the elastic domain. The idea was originally developed for ductile fracture~\cite{Alessi2014}, where degradation is coupled to plastic dissipation within a fully energetic formulation. Micromechanically motivated models have also been proposed within this framework, yielding natural energy splits for quasi-brittle frictional materials~\cite{Ulloa2021d}. More recently, plastic strains have been replaced by reversible nonlinear strain measures to describe quasi-brittle fracture with tunable strength criteria~\cite{Vicentini2026,Bourdin2025}. In general, the class of models~\ref{item_c} is associated with cohesive fracture in the sharp-crack limit~\cite{DalMaso2016,Maggiorelli2026}. Although flexible, these models require additional internal variables that increase the complexity of the formulation and the resulting mechanical response.

The objective of the present work is to propose a new modeling paradigm for incorporating arbitrary strength criteria into standard phase-field models. As a consequence, the resulting framework is also able to describe mixed-mode fracture behavior. 
This is achieved by allowing the dissipation potential to depend on the loading direction in the \emph{invariant stress--strain space} through its explicit dependence on the current material state, characterized by both the strain and the phase-field variable. Interestingly, a similar approach has also been explored to prescribe elastic domains in phase-field models by \citet[Sec.~5]{Maggiorelli2025}, letting the dissipation depend on the strain. Indeed,
considering a state-dependent dissipation potential has been the key modeling strategy to endow, for instance, Griffith's theory and phase-field models with fatigue effects~\cite{Alessi2018,Carrara2020,Ulloa2021a,Alessi2023}, or to recover a variational structure in non-associative plasticity~\cite{Francfort2018,Ulloa2021}. The proposed model in the present work is primarily formulated within the energetic formulation for generalized standard materials and rate-independent processes~\cite{Mielke2006}. But, differently from the \emph{classical energetic formulation}, global stability is replaced by a new notion of local stability. Moreover, elastic degradation behavior is kept well distinct from the underlying strength criterion, a feature we deem desirable as these mechanisms characterize different material properties, namely stiffness and strength. Indeed, the free energy, which defines the phase-field driving force, will encode which part of the elastic stiffness is degraded, while the dissipation potential, which defines the phase-field resisting force, will primarily encode the strength criterion and the corresponding elastic domain.  The resulting model turns out to be very flexible, easy to calibrate, and of clear mechanical interpretation. 

The paper is structured as follows. \Sec{sec_gen_mod} discusses the approach for incorporating an arbitrary strength criterion into standard phase-field models within an energetic framework. \Sec{sec_specificMatMod} presents specific material models to assess the flexibility of the modeling approach. These models, each owning different features, are investigated and compared through analytical and numerical analyses in \Sec{sec_examples}. The main conclusions and perspectives are drawn in \Sec{sec_conc}.

\paragraph{Notation}
 
Let us summarize the notation and some useful relations. Vectors and second-order tensors are denoted by boldface fonts, e.g., $\vdisp$ for the displacement vector, $\teps \coloneqq \rm{sym}\, \grad \vdisp$ for the infinitesimal strain tensor, and $\tstr$ for the stress tensor. For the standard orthogonal decomposition of a second-order tensor $\vect{a}$ into volumetric/hydrostatic and deviatoric parts, we use the following notation:
\begin{equation}
  a_\l{v} = \tr\vect{a}, \qquad 
  a_\l{h} = \dfrac{a_\l{v}}{n}, \qquad 
  \vect{a}_\l{d} = \vect{a} - a_\l{h}\vect{I}, \qquad
  a_\l{d} = \vnorm{\vect{a}_\l{d}}
  , \qquad
  a_\l{s} = \sqrt{\dfrac{n}{n-1}}\vnorm{\vect{a}_\l{d}},
\end{equation} 
where $n=2$ or $n=3$ for two- or three-dimensional problems, and $\vect{I}$ is the $n$-dimensional second-order identity tensor. For an isotropic elastic undamaged material with Young's modulus $E$ and Poisson's ratio $\nu$, we denote by $(\lambda, \mu)$ and $\kappa$ the Lamé parameters and the bulk modulus,
with
\begin{equation}
  \lambda = \dfrac{E \,\nu}{(1+\nu)(1-(n-1)\nu)},
  \qquad
  \mu = \dfrac{E}{2(1+\nu)},
  \qquad
  \kappa = \lambda + \dfrac{2}{n}\mu = \dfrac{E}{n(1-(n-1)\nu)}.
\end{equation} 
Depending on the context, we denote by $a^{\pm}$ either the positive part and negative part of a scalar-valued function, namely 
\begin{equation}
  a^{\mathsmaller{+}} = \dfrac{a + \abs{a}}{2}, \qquad a^{\mathsmaller{-}} = \dfrac{a - \abs{a}}{2}, 
\end{equation} 
or two independent quantities $a^\mathsmaller{+}=a_1$ and $a^\mathsmaller{-}=a_2$. 

The time evolution parameter is denoted by $t$, with $t=0$ being the initial instant of the evolution process. A dot ($\dot{\Box}$) denotes differentiation with respect to either the time-evolution parameter or the arc-length parameter ($s$ or $\varsigma$) along an evolution path.

Finally, we refer to the $\sepsv$--\thinspace$\sepsd$ and $\sstrh$--\thinspace$\sstrd$ half-planes as volumetric--deviatoric \emph{strain space} and hydrostatic--deviatoric \emph{stress space}, or as (invariant) \emph{strain space} and \emph{stress space}, respectively.

\section{The general model}
\label{sec_gen_mod} 

In this section, the general model is presented, and the evolution equations are derived within the energetic formulation framework. 
The modeling setting is kept as simple as possible without loss of generality. Possible modeling extensions are addressed throughout the text.

\subsection{Fracture energy depending on the load direction in stress/strain space}
\label{sec_mod_strat} 

\begin{figure}[H]
  \centering
  \small
    \includeinkscape[scale=1.4]{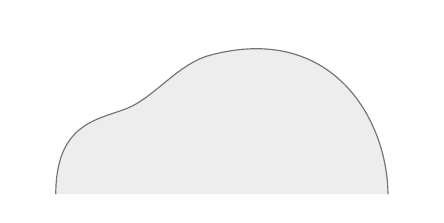}
  \caption{Qualitative representation of the elastic domain $\Rm(\sdam)$, evolving as a function of the phase-field variable $\alpha$, and associated with the fracture function with respect to the loading angle~$\vartheta$ in the $\sepsv$--\thinspace$\sepsd$ half-plane.}
  \label{fig_Gctheta_example} 
\end{figure}

In order to incorporate a sufficiently general strength criterion into a phase-field fracture model, we let the fracture energy depend on a material function, called hereafter \emph{fracture function}, which depends on suitable variables characterizing the local loading state of each material point. In this work, we limit our analysis to strength criteria that depend only on the first and second strain/stress invariants, that is, fully described in the volumetric--deviatoric strain and hydrostatic--deviatoric stress half-planes, $\sepsv$--\thinspace$\sepsd$ and $\sstrh$--\thinspace$\sstrd$, respectively. Roughly speaking, the strength criterion is identified by the corresponding elastic domain that defines the states for which the mechanical response is purely elastic and no dissipative phenomena take place. A rigorous definition of elastic domain will be provided later.
In such a framework, the implicit loading direction can be completely characterized by a single scalar parameter, such as the loading angle $\sla$ representing the loading direction in the volumetric--deviatoric strain space or hydrostatic--deviatoric stress space. We then let the fracture function depend on this parameter,
\begin{equation}
  \Gf = \Gf(\sla),
\end{equation}
as depicted in \Fig{fig_Gctheta_example}.
In this schematic illustration, the elastic domain in the invariant strain space is linked to the polar plot of~$\Gf(\sla)$ and evolves as a function of the phase-field variable $\alpha$. For proportional monotonic loading, this quantity can be linked to the fracture toughness of the material $\Gc$ for the given loading direction.

The $\sla$-dependence of the fracture function cannot be totally arbitrary as it must comply with fundamental thermodynamic requirements \cite{Pham2010a}. In particular, we assume that the null state always belongs to the elastic domain and that the elastic domain is at least star-convex with respect to $\vect{0}$~\cite{Marigo1989,Vicentini2024}.

\subsection{The modeling framework}
\label{sec_var_form}
\FloatBarrier

The local material behavior is first characterized within the framework of generalized standard materials (GSM)~\cite{Nguyen1975,NguyenB2000,Nguyen2010}. Global energetic quantities are then defined and the evolution equations are derived consistently with the energetic formulation~\cite{Mielke2006,Alessi2016b}, but with a new notion of local stability. 

We consider a body occupying the domain $\domain \subset \setR^n$, where $n$ is the spatial dimension. The state at a material point $\vpos\in\domain$ is characterized by the displacement field $\vdisp(\vpos)$ and the irreversible phase-field variable $\sdam(\vpos) \in [0,1]$, with $\sdam= 0$ and $\sdam=1$ corresponding to sound and fully damaged states, respectively.
Throughout the manuscript, $\sdam$ will be referred to interchangeably as \emph{damage} or \emph{phase-field variable}.
Imposed displacements $\bar{\vdisp}_t$ are applied on the Dirichlet part of the boundary $\bdomainD$, while imposed forces $\bar{\vect{f}}_t$ are applied on the Neumann part $\bdomainN$, such that $\bdomain = \bdomainD \cup \bdomainN$ and $\bdomainD \cap \bdomainN = \varnothing$. The body may be subjected to volume forces $\vect{b}_t$. The space of kinematically admissible displacement fields is denoted as $\Uspace_t\coloneqq\{\vdisp \in H^1(\domain;\setR^n) \;:\; \vdisp = \bar{\vdisp}_t \text{ on }\bdomainD\}$, while the corresponding space of admissible displacement variations reads $\Uvarspace\coloneqq\{\vdispvar \in H^1(\domain;\setR^n) \;:\; \vdispvar = \vect{0} \text{ on }\bdomainD\}$.  Similarly, the space of admissible damage fields is given by $\Dspace\coloneqq\{\sdam \in H^1(\domain;\setR) \;:\; 0 \leq \sdam \leq 1\}$, while the corresponding space of admissible damage variations reads $\Dvarspace(\sdam)\coloneqq\{\sdamvar \in H^1(\domain;\setR) \;:\; \sdamvar \geq 0 \text{ if $\sdam <1$, $\sdamvar = 0$ \text{ otherwise}}\}$.
Let us also define the space of state fields as $\Allspace \coloneqq \Uspace \times \Dspace$ and the space of admissible state field variations as $\Allvarspace(\sdam) \coloneqq \Uvarspace \times \Dvarspace(\sdam)$.

\subsubsection{The local energy quantities}
The material behavior is fully characterized once the way the material stores and dissipates energy is defined. In this regard, within the GSM framework, this is accomplished by defining two local energy quantities, the \emph{free energy density} and the \emph{dissipation potential}.

The free energy density, here also called \emph{elastic energy density}, is assumed to be convex with respect to strains and decomposable into a fully \emph{degradable} part and a \emph{residual} part as follows:
\begin{equation}
  \boxed{
    \enfree(\teps,\sdam) \coloneqq g(\sdam)\,\enfreeD(\teps) + \enfreeR(\teps)
  ,}
  \label{def_enfree} 
\end{equation} 
where 
$\sdam \mapsto g(\sdam)$ is the elastic degradation function fulfilling \emph{Hypothesis 1} in \citet{Pham2011}, namely being continuously differentiable in $\sdam \in [0,1)$ and satisfying 
\begin{equation}
  g(0) = 1, \qquad g(1) = 0, \qquad  g'(\sdam) < 0.
  \label{def_assg} 
\end{equation} 
Energy expressions more complex than \eqref{def_enfree} can be considered. For instance, instead of a single degradation function, several functions degrading different parts of the elastic energy density can be used~\cite{Bian2024,Zolesi2024}.

The conjugate variables dual to the state variables $\teps$ and $\sdam$ are respectively given by
\begin{equation}
  \tstr \coloneqq \pd{}{\teps}\enfree(\teps,\sdam) = g(\sdam)\pd{}{\teps}\enfreeD(\teps) + \pd{}{\teps}\enfreeR(\teps), 
  \qquad 
  Y \coloneqq -\pd{}{\sdam}\enfree(\teps,\sdam) = -g'(\sdam)\enfreeD(\teps)
  \label{def_stress}
\end{equation} 
where $\tstr$ is the \emph{mechanical stress} and $Y$ is the \emph{phase-field/damage driving force}. In the spirit of \cite{Pham2010a}, a potential dual to the free energy density, that is, the \emph{complementary energy density}, can be defined as $$\enfree^*(\tstr,\sdam) \coloneqq \sup_{\teps\in\Sym} \bra{\tstr:\teps - \enfree(\teps,\sdam)},$$
from which
\begin{equation}
 \teps = \teps(\tstr,\sdam) = \pd{}{\tstr}\enfree^*(\tstr,\sdam).
\label{def_epsstr} 
\end{equation} 

The dissipative aspects of the model are encoded in the following state-dependent dissipation potential:
\begin{equation}
  \boxed{
  \disspot(\sdamdt,\vdamddt;\teps,\sdam,\vdamd) \coloneqq \Gf(\teps,\sdam)\bra{ \dfrac{w'(\sdam)}{\ell}\sdamdt + \ell \,\vdamd\vinner\vdamddt },
  }
  \label{def_disspot} 
\end{equation} 
where $\ell$ is the internal material length, related to the phase-field localization band width, and $\sdam \mapsto w(\sdam)$ is the phase-field dissipation function fulfilling \emph{Hypothesis 1} in \citet{Pham2011}, namely being continuously differentiable $\sdam \in [0,1)$ and satisfying 
\begin{equation}
  0 \leq w(\sdam) \leq 1, \quad w(0) = 0, \quad w(1) = 1, \quad w'(\sdam) \geq 0. 
  \label{def_assw} 
\end{equation} 
For rate-independent processes, such as the quasi-static fracture problem considered in this work, the dissipation potential \eqref{def_disspot} coincides with the dissipated power. It is worth noting that the dissipation potential \eqref{def_disspot} depends on both the strain and phase-field states, and in general cannot be expressed as a time derivative of any primitive function. Therefore, differently from standard phase-field models, the dissipated work considered here is not necessarily a state function. State-dependent dissipation potentials have by now proven effective in the modeling of a wide variety of complex phenomena, as discussed in the~introduction.
The dissipation potential is supplemented by the irreversibility condition $\sdamdt\geq0$ throughout the entire evolution. This constraint can be incorporated directly into the dissipation potential by adding a characteristic function%
\footnote{
The characteristic function $\chi_{\raisebox{-0.5ex}{S}}$ of a set $S$ is defined as
\begin{equation*}
  \chi_{\raisebox{-0.5ex}{S}}(x) =
  \begin{cases}
    0 & \text{if $x\in S$,} \\
    +\infty & \text{if $x\notin S$.}
  \end{cases}
\end{equation*}
}
of $\sdamdt$ associated with $\setR_{+}$ for $\sdam<1$ and with $\cbra{0}$ for $\sdam=1$, while adapting the functional space of admissible phase-field variations accordingly.

From a mechanical viewpoint, the \emph{fracture function} $(\teps,\sdam) \mapsto \Gf(\teps,\sdam)$ encodes the load-direction dependence of the fracture energy, as discussed in \Sec{sec_mod_strat}, and must be positive to comply with the second law of thermodynamics~\cite{Marigo2016}. In standard phase-field models, the fracture function is constant and directly linked to the material fracture toughness; specifically, $\Gf = \Gc/\cw$ with $\cw=8/3$ for the \texttt{AT1} model and $\cw=2$ for the \texttt{AT2} model~\cite{Tanne2018}. For the more general cases considered here, specifically when the dissipated work is not a state function, such a link is not readily available and requires an assessment of the crack propagation behavior. We set this worthy task aside for future work.

As will be clear in \Sec{sec_R_hom}, the free energy form~\eqref{def_enfree} primarily encodes which part of the elastic energy is degraded, whereas the state-dependent dissipation potential~\eqref{def_disspot} tunes the shape of the elastic domain by means of $\Gf(\teps,\sdam)$. 
In the present model, these two aspects are therefore kept clearly separated.

\subsubsection{The global energy quantities}
\label{sec_Geq} 

To set up an energetic formulation, we define global energy quantities that represent global energy storage, dissipation, and exchange with the surrounding environment. The \emph{total stored energy} functional reads
\begin{equation}
  \Enfree(\vect{q}) \coloneqq \intargV{\enfree(\teps,\sdam)},
  \label{def_Enfree}
\end{equation} 
with $\vect{q}\coloneqq(\vdisp,\sdam)$. 
The \emph{total dissipated energy} or \emph{total dissipated work} functional is a path-dependent quantity given in a time interval $[0,t]$ by 
\begin{equation}
\Disswork\bra{\vect{q};t} \coloneqq \intargt{\Disspot(\dot{\vect{q}};\vect{q})},
\label{eq_Disswork} 
\end{equation} 
where 
\begin{equation}
  \Disspot(\dot{\vect{q}};\vect{q}) \coloneqq \intargV{\disspot(\sdamdt,\vdamddt;\teps,\sdam,\vdamd)}
\end{equation}
is the  \emph{total dissipation potential}. Let us also introduce, according to the classical energetic formulation~\cite{Mielke2015}, the \emph{dissipation distance} functional characterizing the energy dissipation between two states $\vect{q}_0$ and $\vect{q}_1$, 
\begin{multline}
  \Dissdist(\vect{q}_0,\vect{q}_1) := \int_{\domain}\inf\Big\{
    \int_0^1 \disspot(\dot\beta,\grad\dot\beta;\vect{\gamma},\beta,\grad\beta)(\varsigma) \diff \varsigma \: : \: (\vect{\gamma},\beta,\grad\beta) \in \mathrm{C}^1([0,1],\mathrm{Q}),
    \\
    (\vect{\gamma},\beta,\grad\beta)(0)=\bra{\teps,\sdam,\vdamd}_0(\vpos), \ (\vect{\gamma},\beta,\grad\beta)(1)=\bra{\teps,\sdam,\vdamd}_1(\vpos)
  \Big\}
  \diff \vpos,
\label{eq_Dissdist} 
\end{multline}
where $\mathrm{Q}=\mathbb{R}_{\rm sym}^{n\times n}\times\mathbb{R}\times\mathbb{R}^n$ is the corresponding local state variable space at a given material point, and $\varsigma$ is a normalized arc-length. Minimizing paths associated with the infimum in \eqref{eq_Dissdist} will be referred to as \emph{geodesic (energetic) paths}.

The \emph{potential of the given external forces} takes the following linear
form:
\begin{equation}
  \Wm_t(\vdisp) \coloneqq \intargV{\vect{b}_t\cdot\vdisp} + \int_{\bdomainN} \vect{f}_t\cdot\vdisp \diff \vpos,
\end{equation} 
which is defined over all the admissible displacement fields. Moreover, the \emph{work of external actions} is defined as the time integral of the external power,
\begin{equation}
  \Lm(\vdisp;t) \coloneqq \intargt{\bra{
    \Wm_t(\dot\vdisp) 
    + \int_{\bdomainD}{\vect{f}_{\rm r} \cdot \dot{\bar{\vdisp}}_t}\diff\vpos
    }
  },
\end{equation} 
where $\vect{f}_{\rm r}$ is the reaction force on $\bdomainD$. The \emph{total potential energy} then reads 
\begin{equation}
  \Pm_t(\vect{q}) \coloneqq \Enfree(\vect{q}) - \Wm_t(\vdisp).
  \label{def_Totpot}
\end{equation}

\subsubsection{Local energetic formulation}
\label{sec_EF} 

The global stability condition of the classical energetic formulation presents challenges when applied to many material models~\cite{Mielke2015,Alessi2016b}, including phase-field models. Thus, we adopt here an energetic formulation based on a notion of local stability, together with the standard energy-balance condition. Hereafter, we refer to this formulation as the \emph{local energetic formulation}.

For phase-field/gradient-damage models in which the dissipated energy is a state function, which is non-decreasing as a consequence of the irreversibility constraint imposed on the phase-field evolution, the notion of local stability commonly adopted in the literature is, for instance, that introduced in~\citet[Eq.~$(12)$]{Marigo2016}. In some contexts, this local stability notion is also referred to as \emph{directional stability}, see, e.g.,~\cite[Def.~3.3]{Pham2010a}. Let us first recall this standard notion of local stability, hereafter referred to as \emph{local state-stability}, within Mielke's framework~\cite{Mielke2006}, i.e., by employing the dissipation distance~\eqref{eq_Dissdist}.

\smallskip
The state $\vect{q}$ is said to be \emph{locally state-stable} if, for every $\tilde{\vect{q}} \in \Allvarspace(\sdam)$, there exists $\bar{h}>0$ such that, for every $h\in[0,\bar{h}]$,
\begin{equation}
  \Pm_t(\vect{q}) \leq \: \Pm_t(\vect{q} + h \tilde{\vect{q}}) + \Dissdist(\vect{q},\vect{q} + h \tilde{\vect{q}}).
  \label{con_LSS}
  \tag{\texttt{LSS}}
\end{equation}
In the present framework, the dissipated work is not a state function but depends on the path followed during the evolution. In this setting, the local state-stability condition~\eqref{con_LSS} may no longer provide an appropriate notion of local stability. Indeed, when evaluating $\Dissdist$ between two states, no matter how close these states are, the minimizing geodesic path entering the definition of the dissipation distance may extend arbitrarily far from the initial state and may encounter intermediate states whose total energy exceeds that of the initial configuration. Consequently, a comparison involving only the initial and final states through~\eqref{con_LSS} does not account for possible energy barriers encountered along the geodesic path.

Let us thus introduce a different notion of local stability, hereafter referred to as \emph{local path-stability}. The central idea is to assess stability by comparing the energetic state of the system not only at the initial and final configurations, but also at all intermediate states along every admissible perturbation path emanating from the current state.

Accordingly, we say that the state $\vect{q}$ is \emph{locally path-stable} if for every admissible path $(\vect{\gamma},\beta)\eqqcolon\vect{w}\in\mathrm{C}^1([0,s^*],\Allspace)$ satisfying $\vect{w}(0)=\vect{q}$, $\vect{w}(s>0)\neq\vect{q}$, and $\dot{\beta}(s)\geq0$, where $s$ is the (non-normalized) arc-length parameter of the curve~$\vect{w}$, there exists $\bar{s}: 0 < \bar{s} \leq s^*$ such that, for every $s\in[0,\bar{s}]$, the following condition holds:
\begin{equation}
  \Pm_t(\vect{q}) \leq \: \Pm_t(\vect{w}(s)) + \Disswork\bra{\vect{w};s}.
  \label{con_LPS}
  \tag{\texttt{LPS}}
\end{equation}

\noindent
Here, $s$ can also be interpreted as a virtual pseudo-time evolution parameter. Moreover, $\vect{w}$ is understood as a pseudo-temporal--spatial field, whereas $\vect{w}(s)$ denotes the corresponding spatial state attained at the path parameter~$s$. The quantity $\Disswork\bra{\vect{w};s}$ represents the (virtual) work dissipated along the portion of the path~$\vect{w}$ between $0$ and $s$, according to~\eqref{eq_Disswork}. The standard \emph{energy-balance} condition, instead, requires that a process $\vect{q}:[0,T]\to\Allspace$ satisfies, for every $t\in[0,T]$,
\begin{equation}
  \Enfree(\vect{q}(t)) + \Disswork\bra{\vect{q};t}
  =
  \Enfree(\vect{q}(0)) + \Lm\bra{\vdisp;t}.
  \tag{\texttt{EB}}
  \label{con_EB}
\end{equation}

Then, in the present framework, a process $\vect{q}:[0,T]\to\Allspace$ is said to be a \emph{local energetic solution} if, for every $t\in[0,T]$, it satisfies both the local path-stability condition~\eqref{con_LPS} and the energy-balance condition~\eqref{con_EB}. 
Energy balance \eqref{con_EB} is an expression of the first law of thermodynamics, whereas the local stability condition \eqref{con_LPS} incorporates a state-dependent selection criterion for equilibrium states at a given time.

In case of smooth evolutions, necessary conditions are given by \emph{first-order local stability} and \emph{first-order energy balance}~\cite{Marigo2016}. Let us thus derive the first-order local path-stability condition associated with~\eqref{con_LPS}. To this end, we consider the expansions of $\Pm(\vect{w}(s))$ and $\Disswork(\vect{w};s)$ up to first order with respect to the arc-length parameter~$s$ in a neighborhood of $s=0$ at fixed path~$\vect{w}$. For readability, the subscript~$t$ is omitted from the total potential energy. The expansions read
\begin{subequations}
  \begin{align}
    \Pm(\vect{w}(s)) &= \Pm(\vect{q}) + s\,\Diff\Pm(\vect{q})(\dot{\vect{w}}(0)) + o(s),
    \label{def_TE} 
    \\[1ex]
    \Disswork(\vect{w};s) &= \underbrace{\Disswork(\vect{w};0)}_{=0} + s\,\Disspot(\dot{\vect{w}}(0);\vect{q}) + o(s),
    \label{def_TD}
  \end{align}
  \label{eq_T} 
\end{subequations}
where 
\begin{equation}
  \Diff\Pm(\vect{q})(\vect{p}) := \left.\dfrac{\diff}{\diff h}\Pm(\vect{q}+ h\vect{p})\right|_{h=0} 
\end{equation} 
is the first directional derivative of the functional $\Pm(\vect{q})$ in the direction
$\vect{p}$. Let us provide a formal proof for the second term in \eqref{def_TE}, that is,
\begin{equation}
  \dfrac{\diff}{\diff s} \Pm(\vect{w}(s)) = \Diff\Pm(\vect{w}(s))(\dot{\vect{w}}(s)).
  \label{eq_diffE} 
\end{equation} 
Since $\Pm$ is (Fréchet) differentiable,
\begin{equation*}
  \Pm(\vect{w}(s) + \vect{p}) = \Pm(\vect{w}(s)) +  \Diff\Pm(\vect{w}(s))(\vect{p}) + o(\|\vect{p}\|).
\end{equation*} 
Taking 
\begin{equation*}
  \vect{p} = \vect{w}(s + h) - \vect{w}(s) = h\, \dot{\vect{w}}(s)+ o(h),
\end{equation*} 
we have 
\begin{equation}
  \begin{aligned}
    \Pm(\vect{w}(s + h)) - \Pm(\vect{w}(s)) 
      &= \Diff\Pm(\vect{w}(s))(h\, \dot{\vect{w}}(s)+ o(h)) + o(h) \\
      &= h\,\Diff\Pm(\vect{w}(s))(\dot{\vect{w}}(s)) + o(h).
  \end{aligned}
  \label{eq_Texp} 
\end{equation} 
Dividing \eqref{eq_Texp} by $h$ and passing to the limit for $h \to 0$ gives \eqref{eq_diffE}.
\noindent

Replacing \eqref{eq_T} in \eqref{con_LPS}, dividing by $s$, and passing to the limit $s \to 0$ yields the \emph{first-order local path-stability} (\texttt{FO-LPS}) condition. Specifically, the state $\vect{q}$ is locally path-stable at first order if 
\begin{equation}
  \Diff\Pm(\vect{q})(\tilde{\vect{q}}) + \Disspot(\tilde{\vect{q}};\vect{q}) \geq 0, \qquad \forall \tilde{\vect{q}} \in \Allvarspace(\sdam),
  \label{con_FO-LPS} 
  \tag{\texttt{FO-LPS}}
\end{equation} 
with $\tilde{\vect{q}}$ having the physical meaning of a (virtual) velocity vector ($\dot{\vect{w}}$) at $s=0$ for a generic path.
Furthermore, the state $\vect{q}$ is locally path-stable, satisfying~\eqref{con_LPS}, if \eqref{con_FO-LPS} is satisfied with a \emph{strict} inequality. 
On the other hand, the \emph{first-order} energy balance reads
\begin{equation}
  \dfrac{\diff}{\diff t} \Enfree(\vect{q}(t)) 
  + \Disspot(\dot{\vect{q}};\vect{q}) 
  - \Wm_t(\dot\vdisp) 
  - \int_{\bdomainD}{\vect{f}_{\rm r} \cdot \dot{\bar{\vdisp}}_t}\diff\vpos
  = 0.
 \label{con_FO-EB}  
 \tag{\texttt{FO-EB}}
\end{equation} 

\begin{remark}[Second-order local path-stability]\label{rem:1}
One may also assess \emph{second-order local path-stability}, along the lines of \citet{Zolesi2024}. If \eqref{con_FO-LPS} does not hold as a strict inequality, a sufficient condition for~\eqref{con_LPS} can be obtained by considering the second-order terms of the expansions \eqref{eq_T},
\begin{subequations}
\begin{align}
    \Enpot(\vect{w}(s)) &= \underbrace{o(1/s) + o(1)}_{\text{given in \eqref{def_TE}}}
      \;+\; \dfrac{s^2}{2} \bra{\vphantom{\dfrac{1}{1}} \Diff^2\Enpot(\vect{q})(\dot{\vect{w}}(0)) + \Diff\Enpot(\vect{q})(\ddot{\vect{w}}(0))} + o(s^2),
    \\[1ex]
    \Disswork(\vect{w};s) &= \underbrace{o(1/s) + o(1)}_{\text{given in \eqref{def_TD}}}
      \;+\; \dfrac{s^2}{2} \bra{\Diff_{1}\Disspot(\dot{\vect{w}}(0);\vect{q})(\ddot{\vect{w}}(0)) + \vphantom{\dfrac{1}{1}} \Diff_{2}\Disspot(\dot{\vect{w}}(0);\vect{q})(\dot{\vect{w}}(0))}
      + o(s^2),
\end{align}
\label{eq_T2}
\end{subequations}
along paths whose initial rate $\tilde{\vect{q}}$ renders \eqref{con_FO-LPS} an equality. We define such rates as
\begin{equation}
  \mathscr{N}_0(\sdam) := \cbra{\tilde{\vect{q}} \in \Allvarspace(\sdam)\;:\;\Diff\Enpot(\vect{q})(\tilde{\vect{q}}) + \Disspot(\tilde{\vect{q}};\vect{q}) = 0},
\end{equation}
where $\Diff_{1}$ and $\Diff_{2}$ denote functional derivatives with respect to the first and second argument, respectively.
Then, a state $\vect{q}$ satisfying \eqref{con_FO-LPS} with a non-strict inequality is locally path-stable if
\begin{equation}
  \Diff^2\Enpot(\vect{q})(\tilde{\vect{q}}) 
    + \Diff_{2}\Disspot(\tilde{\vect{q}};\vect{q})(\tilde{\vect{q}})
    + \Diff\Enpot(\vect{q})(\breve{{\vect{q}}})
    + \Diff_{1}\Disspot(\tilde{\vect{q}};\vect{q})(\breve{{\vect{q}}}) > 0,
  \quad
  \forall \tilde{\vect{q}} \in \mathscr{N}_0(\sdam), \: \tilde{\vect{q}} \ne \vect{0}, \: \breve{{\vect{q}}} \in \breve{\Allspace}
  ,
  \label{con_SO-LPS} 
  \tag{\texttt{SO-LPS}}
\end{equation}
\noindent
where $\breve{{\vect{q}}}$ represents the (virtual) curvature $(\ddot{\vect{w}})$ at $s=0$ of a generic path and $\breve{\Allspace}$ the state curvature space, to be specified. By the arbitrariness of $\breve{{\vect{q}}}$ and noting that $\Diff_{1}\Disspot(\tilde{\vect{q}};\vect{q})(\breve{{\vect{q}}}) = \Disspot(\breve{{\vect{q}}};\vect{q})$, the last two terms in~\eqref{con_SO-LPS} coincide with the first-order condition~\eqref{con_FO-LPS} and thus vanish if $\breve{\vect{q}} \in \Allvarspace(\sdam)$. Consequently, only the first two terms need to be considered for assessing second-order local path-stability. The analysis could in principle be extended recursively to higher-order conditions, but this lies beyond the scope of the present work. Finally, we note that derivatives of the dissipated energy with respect to the state do arise in the proposed formulation, but only at the level of second-order conditions. This analysis remains the subject of ongoing research.
\end{remark}

Using \eqref{def_Enfree}--\eqref{def_Totpot}, and in view of the stress–strain relation \eqref{def_stress}, the first-order local path-stability condition \eqref{con_FO-LPS} yields
\begin{multline}
  \intargV{\bra{
    \tstr \vinner \teps(\vdispvar) - \vect{b}_t \vinner \vdispvar
  }}
  + \int_{\bdomainN}\bra{\tstr\cdot \vect{n} - \vect{f}_t}\cdot \vdispvar \diff \vpos \\
  + \intargV{
        \bra{\pd{}{\sdam}\enfree(\teps,\sdam)\,\sdamvar +
         \Gf(\teps,\sdam)\bra{\dfrac{w'(\sdam)}{\ell}\sdamvar + \ell\, \vdamd \cdot \vdamdvar }        
    }}
  \geq 0
  \label{eq_ExpFO-LPS}
\end{multline} 
for all $\tilde{\vect{q}}\in\Allvarspace(\sdam)$. Likewise, the first-order energy balance condition \eqref{con_FO-EB} yields
\begin{multline}
  \intargV{\bra{
    \tstr \vinner \teps(\vdispdt) - \vect{b}_t \vinner \vdispdt}}
  + \int_{\bdomainN}\bra{\tstr\cdot \vect{n} - \vect{f}_t}\cdot \vdispdt \diff \vpos 
  + \int_{\bdomainD}\bra{\tstr\cdot \vect{n} - \vect{f}_{\rm r}}\cdot \vdispdt \diff \vpos \\
  + \intargV{
        \bra{\pd{}{\sdam}\enfree(\teps,\sdam)\,\sdamdt +
         \Gf(\teps,\sdam)\bra{\dfrac{w'(\sdam)}{\ell}\sdamdt + \ell\, \vdamd \cdot \vdamddt }        
    }}
  = 0.
  \label{eq_ExpFO-EB}
\end{multline} 

\noindent
Then, by standard arguments of calculus of variations, and proceeding as in~\citet{Ulloa2021a}, we find that these first-order conditions are equivalent to the equilibrium equation
\begin{equation}
  \div \tstr + \vect{b}_t = \vect{0} \qquad \text{in $\domain$},
  \label{pb_eqi} 
\end{equation}
with boundary conditions
\begin{equation}
  \tstr \cdot \vect{n} = \vect{f}_t \quad \text{on $\bdomainN$},
  \qquad
  \tstr \cdot \vect{n} = \vect{f}_{\rm r} \quad \text{on $\bdomainD$},
  \label{pb_eqib}
\end{equation} 
and the Karush--Kuhn--Tucker (KKT) conditions
\begin{subnumcases}{\label{def_dam_con}}
  \pd{}{\sdam}\enfree(\teps,\sdam) + \Gf(\teps,\sdam)\bra{\dfrac{w'(\sdam)}{\ell} - \ell \Delta\sdam } \geq 0 \quad \text{ in $\domain$},
  \label{con_stsdam} 
  \\
  \sdamdt \geq 0 \quad \text{ in $\domain$},
  \label{def_dam_irr}
  \\
  \bra{\pd{}{\sdam}\enfree(\teps,\sdam) + \Gf(\teps,\sdam)\bra{\dfrac{w'(\sdam)}{\ell} - \ell \Delta\sdam }}\sdamdt = 0 \quad \text{ in $\domain$},
  \label{def_dam_evo_cond} 
\end{subnumcases}
with boundary conditions
\begin{equation}
  \vdamd \cdot \vect{n} \geq 0,
  \quad
  \sdamdt \geq 0,
  \quad
  \sdamdt \bra{\vdamd \cdot \vect{n}} = 0,
  \qquad \text{on $\bdomain$}.
  \label{def_dam_evo_condb} 
\end{equation} 
Here, $\vect{n}$ is the outer unit normal to the boundary.  Conditions \eqref{def_dam_con} are referred to as \emph{damage} or \emph{phase-field} \emph{criterion}, \emph{irreversibility condition}, and \emph{loading--unloading condition}, respectively. We note that a full stability analysis of the model up to higher-order conditions, as performed in references~\cite{Pham2011b,Alessi2015a,LeonBaldelli2021} for other models, is possible but out of scope in the present work (Remark~\ref{rem:1}).

\begin{remark}[Local state-stability along radial paths]
    It is worth noting that the first-order condition~\eqref{con_FO-LPS} coincides with that obtained from local state-stability based on Mielke's dissipation distance~\eqref{eq_Dissdist} when the admissible perturbations are restricted to radial paths, as done in prior works~\cite{Alessi2018,Carrara2020,Ulloa2021,Ulloa2021a}. To establish this equivalence, let us observe that any radial path connecting $\vect{q}$ to $\vect{q}+h\tilde{\vect{q}}$ can be parametrized in terms of its normalized arc-length~$\varsigma$ as
\begin{equation}
  (1-\varsigma)\vect{q} + \varsigma(\vect{q} + h\tilde{\vect{q}}) = \vect{q} + \varsigma\, h\, \tilde{\vect{q}} = \vect{w}(\varsigma).
\end{equation} 
Then, according to this generic radial path, the dissipation distance~\eqref{eq_Dissdist} is given by 
\begin{equation}
  \begin{aligned}
    \Dissdist(\vect{q},\vect{q} + h\,\tilde{\vect{q}}) &= \int_0^1 \Disspot(\dot{\vect{w}}(\varsigma);\vect{w}(\varsigma)) \diff \varsigma \\
      &= \int_0^1 \Disspot(h  \tilde{\vect{q}}; \vect{q} + \varsigma h\tilde{\vect{q}}) \diff \varsigma \\
      &= \dfrac{1}{h}\int_0^h \Disspot(h \tilde{\vect{q}}; \vect{q} + r \tilde{\vect{q}}) \diff r \\
      &= \int_0^h \Disspot(\tilde{\vect{q}}; \vect{q} + r \tilde{\vect{q}}) \diff r \\
      &= h\, \Disspot(\tilde{\vect{q}};\vect{q} + \xi_h \tilde{\vect{q}}),
  \end{aligned}
  \label{con_Dm} 
\end{equation}
where the change of variable $r = \varsigma h$ and the integral mean-value theorem have been used with $\xi_h \in [0,h]$ and $\lim_{h\to 0}\xi_h = 0$, assuming a continuous dissipation potential with respect to the state.\footnote{The case of a dissipation potential discontinuous at state $\vect{q}$ must be examined separately.} 
Let us now consider condition~\eqref{con_LSS} with perturbations taken along radial paths. First, note that
\begin{equation}
  \lim_{h \to 0} \bra{\dfrac{\Pm(\vect{q} + h\tilde{\vect{q}}) - \Pm(\vect{q})}{h}} = \Diff\Pm(\vect{q})(\tilde{\vect{q}})
\end{equation} 
and that, in view of \eqref{con_Dm}, 
\begin{equation}
  \lim_{h \to 0} \dfrac{1}{h}\Dissdist(\vect{q},\vect{q} + h \tilde{\vect{q}}) 
    = \Disspot(\tilde{\vect{q}};\vect{q})
  \label{lim_DD}.
\end{equation} 
Then, bringing all terms to the right-hand side of~\eqref{con_LSS}, dividing by $h$, and passing to the limit $h \to 0$ exactly yields the first-order local path-stability condition~\eqref{con_FO-LPS}. 
\end{remark}

\begin{remark}[On the derivation of the governing equations]
The first-order conditions, in either weak form \eqref{eq_ExpFO-LPS}--\eqref{eq_ExpFO-EB} or strong form \eqref{pb_eqi}--\eqref{def_dam_evo_condb}, can also be derived directly within the theory of generalized standard materials~\cite[Sec.~15.1.3]{NguyenB2000}.
\end{remark}

\subsubsection{The elastic domain}
\label{sec_R_hom} 

The damage conditions \eqref{con_stsdam} and \eqref{def_dam_evo_cond} allow us to introduce the notion of elastic domain in both strain space and stress space~\cite{Pham2010a,Marigo2016}. In the present context, in agreement with \citet{Vicentini2024}, the elastic domain refers to the elastic limit of a material with a homogeneous damage distribution. Specifically, the elastic domain in strain space $\Rm(\sdam)$ and the elastic domain in stress space~$\Rm^*(\sdam)$ read
\begin{align}
  \Rm(\sdam) &= 
    \cbra{\teps \in \Sym : -\pd{}{\sdam}\enfree(\teps,\sdam) 
    \leq 
    \Gf(\teps,\sdam)\dfrac{w'(\sdam)}{\ell} },
  \label{def_Rm} 
\\
  \Rm^*(\sdam) &= 
    \cbra{\tstr \in \Sym : \pd{}{\sdam}\enfree^*(\tstr,\sdam) 
    \leq 
    \Gf^*(\tstr,\sdam)\dfrac{w'(\sdam)}{\ell} }.
  \label{def_Rms}
\end{align}
In the last expression, we have used $\Gf^*(\tstr,\sdam) = \Gf(\teps(\tstr,\sdam),\sdam)$ and the constitutive relation \eqref{def_epsstr}. The strength surface, representing the boundary of the elastic domain, is then denoted in each space by
\begin{equation}
  \Sm(\sdam) = \partial\Rm(\sdam), \qquad \Sm^*(\sdam) = \partial\Rm^*(\sdam).
\end{equation} 

In the following, we consider material models obeying \emph{strain-hardening} and \emph{stress-softening} properties for the elastic domain~\cite{Pham2010a}, namely
\begin{subnumcases}{}
  \Rm(\sdam) \subset \Rm(\beta) \quad \forall\sdam<\beta \qquad &\text{\small(strain-hardening)}, \\[1ex]
  \Rm^*(\sdam) \supset \Rm^*(\beta) \quad \forall\sdam<\beta \qquad &\text{\small(stress-softening)}.
\end{subnumcases}
The strain-hardening property ensures uniqueness of the homogeneous response with respect to any strain level, whereas the stress-softening property ensures that in a homogeneous response the stress decreases as the damage evolves, allowing for the occurrence of damage localization.

Moreover, we assume that $\Rm(\sdam)$ and $\Rm^*(\sdam)$ are $\vect{0}$-star-convex, that is, $\forall \teps \in \Rm(\sdam)$ and $\forall \eta\in[0,1]$, $\eta \teps \in \Rm(\sdam)$, and similarly for $\Rm^*(\sdam)$~\cite{Hansen2020}. 
The star-convex property ensures that the unloaded material state $\teps=\vect{0}$ and $\tstr=\vect{0}$ always belongs to the elastic domain, a fundamental thermodynamic requirement.

\subsubsection{On the numerical solution}
\label{sec_num} 

In this section, possible numerical solution strategies are first briefly discussed, and a specific scheme is then devised for the evolution problem introduced in \Sec{sec_EF}. 

Classical energetic formulations commonly approximate rate-independent evolution by minimizing a single incremental functional combining potential energy and dissipation distance~\cite[Sec.~3.2]{Mielke2006}. 
In the present setting, this approach faces two difficulties: the state dependence of the dissipation potential makes the dissipation distance difficult to compute, and the connection between classical incremental minimization and local path-stability remains unclear.
Moreover, a direct implicit treatment of the state-dependent dissipation in an incremental functional introduces additional terms, such as contributions proportional to $\partial_{\teps}\Gf$ that are absent from the continuous first-order equations, so constructing a consistent functional requires further investigation~\cite{Yang2006,Stainier2011}. Alternatively, the first-order equations~\eqref{pb_eqi}--\eqref{def_dam_evo_condb} could be discretized directly using standard numerical schemes~\cite{Msekh2015,Zhou2019,Jin2024}.

Another possibility is to solve the evolution problem through sequential minimization, preserving the variational structure of each subproblem without requiring a single incremental energy functional. Recent developments in phase-field fracture provide relevant examples of this approach. In~\citet{Larsen2024a}, a Griffith phase-field theory incorporating material strength, originally formulated as a coupled system of PDEs, is recast in variational form by showing that the displacement and phase-field variables separately minimize two distinct functionals. Related ideas are found in~\citet{Larsen2024}, where local variational principles are introduced to overcome limitations associated with global energy minimization.

\smallskip
Following this philosophy, a sequential minimization strategy is also devised for the present model. 
To this aim, let us consider $n_t + 1$ discrete time instants $0 = t_0 < \dots < t_n < t_{n+1} < \dots < t_{n_t} = \bar{t}$, where all state variables are known up to $t_n$, and the goal is to find the state at the current time step $t_{n+1}$. For the sake of convenience, we define $\Box_n$ as a generic quantity~$\Box$ evaluated at time step $t_n$. The same quantity evaluated at $t_{n+1}$ will be denoted without subscript, namely $\Box \coloneqq \Box_{n+1}$. 

The proposed scheme is reported in \Alg{algo_sol}, where $k$ denotes a staggered iteration within the load step $n+1$ and 
\begin{equation}
  \Fm_{\sdam}(\sdam, \Gf) = \intargV{\Gf\bra{ \dfrac{w(\sdam)}{\ell} + \ell \,\vdamd\vinner\vdamd}}.
  \label{def_Fa} 
\end{equation}
Here, $\Gf$ is a constant but spatially varying field, rendering the algorithmic phase-field subproblem analogous to that of functionally graded materials~\cite[Eq.~$(2)$]{Hirshikesh2019}. 

Two approximations for $\Gf$ are considered, denoted by (\texttt{Approx-I}) and (\texttt{Approx-E}) in~\Alg{algo_sol}. In both schemes, $\Gf$ is held fixed during the phase-field minimization. In (\texttt{Approx-I}), it is updated within the staggered iterations using the newly computed state, namely $\Gf = \Gf(\teps(\vdisp^{(k)}),\sdam^{(k)})$. The optimality conditions of the two subproblems in \Alg{algo_sol} then recover the first-order condition~\eqref{eq_ExpFO-LPS} at convergence, up to spatial discretization error and solver tolerance. This algorithm can thus be interpreted as a staggered fixed-point scheme with a (partially) lagged evaluation of the state-dependent dissipation. By contrast, in (\texttt{Approx-E}), $\Gf$ is updated only after convergence at the end of each load step, namely $\Gf = \Gf(\teps(\vdisp_n),\sdam_n)$, and held fixed throughout the next load step, thus treating its state dependence~explicitly.

Although (\texttt{Approx-I}) more closely approximates the governing equations at each load step, it exhibits numerical instabilities that prevent the algorithm from converging when the evolution involves a discontinuous jump, as observed in the post-nucleation regime of the numerical examples investigated in \Sec{sec_NT}. These instabilities are not observed when using (\texttt{Approx-E}) for the same examples, for which the algorithm converges without issues.

Regardless of the adopted approximation for $\Gf$, the equilibrium problem~\eqref{pb_eqi}--\eqref{pb_eqib} is solved for $\vdisp$ at fixed $\sdam$, and the damage problem~\eqref{def_dam_con}--\eqref{def_dam_evo_condb} is solved for $\sdam$ at fixed $\vdisp$ and $\Gf$. These steps are iterated until convergence, exploiting the convexity of each individual subproblem. This staggered solution strategy is attractive since each subproblem retains the structure of a standard phase-field model and can be handled with conventional solvers, requiring only minor modifications to existing codes.

\begin{algorithm}[!ht]
\setstretch{1.5}
\DontPrintSemicolon
  \vspace{0.5em}
  \KwInput{$(\vdisp_n,\sdam_n)$, primary fields at previous converged time-step~$t_n$}
  \KwOutput{$(\vdisp_{n+1},\sdam_{n+1})$, primary fields at current converged time-step~$t_{n+1}$}
  
  Initialize iterations $k = 0$, 
    set $(\vdisp^{(0)},\sdam^{(0)}) = (\vdisp_n,\sdam_n)$ 
    and $\Gf^{(0)} = \Gf(\teps(\vdisp_n),\sdam_n)$ 
  
  \Repeat{$\abs{\sdam^{(k)}-\sdam^{(k-1)}} \leq \text{\upshape\texttt{TOL}}_{\sdam}$}{
    
    Set $k=k+1$
    
    $\vdisp^{(k)} = \displaystyle\argmin[\vdisp] \Enpot(\vdisp,\sdam^{(k-1)})$
    
    $\sdam^{(k)} = \argmin[\sdam\geq\sdam_{n-1}] \bra{
      \Enpot(\vdisp^{(k)},\sdam) + \Fm_{\sdam}(\sdam, \Gf^{(k-1)})}$
    
    \vspace{0.5em}
    $\displaystyle 
    \Gf^{(k)} = 
    \begin{cases}
      \Gf(\teps(\vdisp^{(k)}),\sdam^{(k)}) & \text{(\texttt{Approx-I})} \\
      \Gf(\teps(\vdisp_n),\sdam_n)         & \text{(\texttt{Approx-E})}
    \end{cases}
    $
    \vspace{0.5em}
  }
  
  Set $(\vdisp_{n+1},\sdam_{n+1}) = (\vdisp^{(k)},\sdam^{(k)})$
  
\caption{Staggered solution strategy at a given time-step $t_{n+1}$.}
\label{algo_sol} 
\end{algorithm}

More broadly, staggered schemes such as \Alg{algo_sol} are primarily suited to time-continuous fracture evolutions and do not necessarily ensure accuracy or satisfaction of the energy balance during abrupt fracture events~\cite{Alessi2016b,Correas2024}. Indeed, even in standard phase-field models with path-independent dissipated work, unstable fracture events may not be adequately captured within a purely quasi-static setting, as inertial effects can significantly influence both the evolution and the associated energy balance~\cite{Larsen2010,Correas2024}. 

Generally, quasi-static descriptions of rate-independent processes face challenges in resolving discontinuous transitions and accounting for the associated energy transfers~\cite{Alessi2016b}. During a discontinuous crack event, the quasi-static loading parameter alone does not resolve the transition between two locally stable states. This issue is particularly relevant for state-dependent dissipation potentials, since evaluating $\Gf$ only at the pre- or post-jump state may fail to capture the dissipation accumulated along the path.\footnote{Even when the dissipated work is a state function, resolving the intermediate evolution may be relevant to enforcing irreversibility if the support of the localized phase-field profile shrinks during damage evolution; see, e.g., \citet[Fig.~6]{Pham2011c}.}

Possible strategies for addressing such jumps include: (i) \emph{path-following or fracture-control methods}, which retain a quasi-static description but require a suitable control quantity~\cite{Wambacq2021,Loiseau2026}; (ii) \emph{dynamic regularization}, which resolves the transition by incorporating inertial effects but departs from the quasi-static framework~\cite{Li2016,Li2016a}; and (iii) \emph{vanishing-viscosity regularization}, which introduces a small viscous contribution and connects to solution concepts of bounded variation, particularly Balanced-Viscosity solutions~\cite[Sec.~3.8]{Mielke2015}. The latter appears particularly promising here: an auxiliary parametrization could resolve the otherwise instantaneous crack jump as a continuous transition, allowing $\Gf$ and the associated dissipation to be evaluated along the transition path. Establishing the corresponding vanishing-viscosity limit for the present state-dependent dissipation potential, however, remains an open problem.

These considerations identify the robust numerical treatment of discontinuous evolutions as a central challenge for further development of the proposed framework. In particular, when the dissipation potential is state-dependent, accounting for the transition path appears necessary to evaluate the dissipated fracture work consistently and undertake a comprehensive investigation of unstable crack propagation. We leave these developments for future work and focus here on crack nucleation.

\section{Material models}
\label{sec_specificMatMod} 

In this section, we introduce and discuss a variety of material models that differ in elastic degradation and strength criterion. Modeling assumptions common to all discussed models are first presented. Most of these assumptions can be relaxed but are enforced for the sake of simplicity. In particular, they turn out to be sufficiently flexible to describe all key aspects of the present modeling approach.

\subsection{Assumptions and features common to all considered models}

\paragraph{Elastic behavior}

For all considered models, a linear isotropic elastic behavior is assumed at fixed damage. For an undamaged material, we then have
\begin{equation}
    \enfree(\teps,0) = \coh\bra{\kappa\,\sepsv^2 + 2\mu\,\sepsd^2}
    \label{def_linel} 
\end{equation} 
and
\begin{equation}
  \enfree^*(\tstr,0) = \coh \bra{\dfrac{\sstrh^2}{\kappa} + \dfrac{\sstrd^2}{2\mu}},
\end{equation} 
where $\kappa$ and $\mu>0$ are the bulk and shear moduli of the undamaged material, respectively.

\paragraph{Phase-field model}
  
The underlying phase-field fracture model we consider is \texttt{AT1}~\cite{Li2023}, characterized by the following material functions complying with assumptions \eqref{def_assg} and \eqref{def_assw}:  
\begin{equation}
  g(\sdam) = (1-\sdam)^2, \qquad w(\sdam) = \sdam.
  \label{def_PF} 
\end{equation}
Let us also define the damage-dependent compliance modulation function and its derivative as 
\begin{equation}
  s(\sdam)= g^{-1}(\sdam) 
    = (1-\sdam)^{-2}, \qquad s'(\sdam) 
    = - \dfrac{g'(\sdam)}{(g(\sdam))^2} 
    = 2 (1-\sdam)^{-3}.
\end{equation} 

It is worth noting that the proposed modeling strategy can also be extended to phase-field formulations with a vanishing initial elastic domain, such as the \texttt{AT2} model~\cite{Pham2011}. Although a conventional elastic domain cannot be defined in this case, each proportional loading path still identifies a peak stress, whose locus can be shaped through the present approach to control the multiaxial strength response.

\paragraph{Energy split and elastic degradation}

Recalling the elastic energy decomposition \eqref{def_enfree}, and taking into account the linear elastic material behavior described in \eqref{def_linel}, we consider two decompositions:
\begin{equation}
  \begin{dcases}
    \enfreeD(\teps) = \coh\kappa\,\sepsv^2 + \mu\,\sepsd^2,
    \\[1ex]
    \enfreeR(\teps) = 0,
    \label{def_FED} 
    \tag{\texttt{FED}}
  \end{dcases}
\end{equation}
and
\begin{equation}
  \begin{dcases}
    \enfreeD(\teps) = \coh\kappa^\mathsmaller{+}\,(\sepsv^\mathsmaller{+})^2 + \mu\,\sepsd^2, 
    \\[1ex]
    \enfreeR(\teps) = \coh\kappa^\mathsmaller{-}\,(\sepsv^\mathsmaller{-})^2,
    \label{def_PED} 
    \tag{\texttt{PED}}
  \end{dcases}  
\end{equation}
where $\sepsv^\mathsmaller{+}$ and $\sepsv^\mathsmaller{-}$ are the positive and negative parts of the volumetric strain, respectively,
and $\kappa^\mathsmaller{+}$ and $\kappa^\mathsmaller{-}$ are the tensile and compressive bulk moduli, respectively.
The first case,~\eqref{def_FED}, corresponds to a model for which the entire elastic energy density is degraded as damage evolves (\texttt{F}\emph{ull} \texttt{E}\emph{lastic} \texttt{D}\emph{egradation} model). The second case,~\eqref{def_PED}, corresponds to a model for which the elastic energy density is degraded only partially as damage evolves (\texttt{P}\emph{artial} \texttt{E}\emph{lastic} \texttt{D}\emph{egradation} model). More specifically, only the volumetric compressive elastic energy is not degraded, a condition ensuring non-interpenetration~\cite{Chambolle2017}. Accordingly, the effect of $g(a)$ can also be interpreted as a degradation of the elastic moduli:
\begin{equation}
  \begin{cases}
    \kappa^\mathsmaller{\pm}(\sdam) = \kappa(\sdam) = g(\sdam)\kappa,
    \\
    \mu(\sdam) = g(\sdam)\mu,
  \end{cases}    
  \label{def_FEDmod} 
\end{equation} 
for the \eqref{def_FED} model, and
\begin{equation}
  \begin{cases}
    \kappa^\mathsmaller{+}(\sdam) = g(\sdam)\kappa,
    \\
    \kappa^\mathsmaller{-}(\sdam) = \kappa,
    \\
    \mu(\sdam) = g(\sdam)\mu,
  \end{cases}
  \label{def_PEDmod}
\end{equation}
for the \eqref{def_PED} model.

\begin{figure}[H]
\centering
  \begin{subfigure}{0.32\linewidth}
    \centering\small
      \includeinkscape[width=\linewidth]{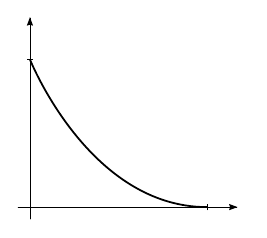}
    \caption{}
    \label{fig_FED}
  \end{subfigure}
  \hfill
  \begin{subfigure}{0.32\linewidth}
    \small\centering
        \includeinkscape[width=\linewidth]{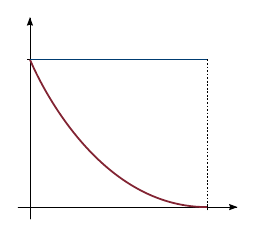}
    \caption{}
    \label{fig_PED}
  \end{subfigure}
  \hfill
  \begin{subfigure}{0.32\linewidth}
    \small\centering
        \includeinkscape[width=\linewidth]{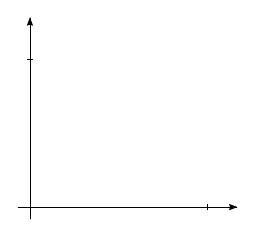}
    \caption{}
    \label{fig_SPED}
  \end{subfigure}
  \caption{Elastic moduli degradation: (\subref{fig_FED}) \texttt{FED} model, (\subref{fig_PED}) \texttt{PED} model, and (\subref{fig_SPED}) a model with three different degradation laws.}
  \label{}
\end{figure}

By means of the constitutive relation \eqref{def_epsstr}, we obtain
\begin{equation}
  \sstrh^{\mathsmaller{\pm}} = \sstrh^{\mathsmaller{\pm}}(\teps,\sdam) = \kappa^{\mathsmaller{\pm}}(\sdam)\,\sepsv^{\mathsmaller{\pm}}, 
  \qquad 
  \sstrd = \sstrd(\teps,\sdam) = 2\,\mu(\sdam)\,\sepsd,
\end{equation} 
and the inverse relations
\begin{equation}
  \sepsv^{\mathsmaller{\pm}} = \sepsv^{\mathsmaller{\pm}}(\tstr,\sdam) = \dfrac{\sstrh}{\kappa^{\mathsmaller{\pm}}(\sdam)} , 
  \qquad 
  \sepsd = \sepsd(\tstr,\sdam) = \dfrac{\sstrd}{2\,\mu(\sdam)},
\end{equation} 
with $\cbra{\kappa^{\mathsmaller{\pm}}(\sdam),\mu(\sdam)}$ given by \eqref{def_FEDmod} for \eqref{def_FED} and \eqref{def_PEDmod} for \eqref{def_PED}. The behavior of the elastic moduli for both models is represented in \Fig{fig_FED} and \Fig{fig_PED}. A model with different degradation laws for all three moduli, e.g., partially degrading the compressive bulk modulus and fully degrading the other two (\Fig{fig_SPED}), is also possible but not taken into account in the present work.

\paragraph{Fracture function}

Concerning the fracture function, we assume 
\begin{equation}
  \boxed{\Gf(\teps,\sdam) = \Gceff\bra{1 + f(\teps,\sdam)},}
  \label{def_fracfun} 
\end{equation} 
with $(\teps,\sdam)\mapsto f(\teps,\sdam)$ hereafter referred to as the \emph{fracture perturbation function}. Here, the constant $\Gceff$ denotes the effective fracture energy of the material, e.g., $\Gf = \Gceff=\Gc/\cw$ for standard phase-field models. Thus, the function~\eqref{def_fracfun} describes the deviation of the fracture criterion from that of standard phase-field models, accounting for the loading direction in stress/strain space and therefore representing the key ingredient for describing arbitrary elastic domains. In stress space, we have 
\begin{equation}
\Gf^*(\tstr,\sdam) = \Gceff \bra{1 + f^*(\tstr,\sdam)},	\qquad f^*(\tstr,\sdam) = f(\tstr(\teps,\sdam),\sdam),
\label{def_fracfun_stress} 
\end{equation}
where the constitutive relation \eqref{def_epsstr} has been used.
According to the modeling assumptions introduced in \Sec{sec_mod_strat}, the fracture perturbation function is assumed to depend on the first and second strain/stress invariants.

\paragraph{Elastic domain and its limit values}

For the adopted constitutive functions, and in particular owing to the dependence of both the elastic energy and the fracture perturbation function on the first two strain/stress invariants, the elastic domain and its evolution can be fully characterized in the $\sepsv$--\thinspace$\sepsd$ invariant strain space or, equivalently, in the $\sstrh$--\thinspace$\sstrd$ invariant stress space, as schematically illustrated in \Fig{fig_Gctheta_example}. To better highlight the elastic domain evolution with respect to damage, let us denote the intercept values of the elastic domain with the axes of the half-plane as $\bar\Box$. Formally, these limit values, if they exist for a given elastic domain, are defined as
\begin{equation}
  \begin{aligned}
    & \sepsvbar^{\mathsmaller{\pm}}(\sdam) = \max \cbra{\beta\geq 0 \;:\; \pm\beta \vect{I} \in\Rm(\sdam) },
    \\
    & \sepsdbar(\sdam) = \max \cbra{\beta \;:\; \beta \vect{a}_\l{d} \in \Rm(\sdam), \:\text{$\vect{a}\in\Sym$ and $a_{\l{d}}=1$}},
  \end{aligned}
  \label{def_limiteps} 
\end{equation} 
and
\begin{equation}
  \begin{aligned}
    & \sstrhbar^{\mathsmaller{\pm}}(\sdam) = \max \cbra{\beta\geq 0 \;:\; \pm\beta \vect{I} \in\Rm^*(\sdam) },
    \\
    & \sstrdbar(\sdam) = \max \cbra{\beta \;:\; \beta \vect{a}_\l{d} \in \Rm^*(\sdam), \:\text{$\vect{a}\in\Sym$ and $a_{\l{d}}=1$}}.
  \end{aligned}
  \label{def_limitstr}
\end{equation} 
For the initial elastic domain, which is the smallest one in strain space (strain hardening) and the largest one in stress space (stress softening), we denote the limit values by
\begin{equation}
  \sepsvcheck^{\mathsmaller{\pm}} = \sepsvbar^{\mathsmaller{\pm}}(0),
  \quad
  \sepsdcheck = \sepsdbar(0),
  \quad
  \sstrhhat^{\mathsmaller{\pm}} = \sstrhbar^{\mathsmaller{\pm}}(0),
  \quad
  \sstrdhat = \sstrdbar(0).
\label{def_initlim} 
\end{equation}

\subsection{Specific material models}

We consider the following material models, which differ in the type of elastic degradation (full elastic degradation \eqref{def_FED} or partial elastic degradation \eqref{def_PED}) and in the prescribed strength criterion:
\begin{description}[labelindent=5mm]
 \item[\texttt{M1}]-- Standard \texttt{AT1} phase-field model with~\eqref{def_FED};
 \item[\texttt{M2}]-- Double-ellipse strength criterion with~\eqref{def_FED} and arbitrary volumetric and deviatoric limits;
 \item[\texttt{M3}]-- Drucker--Prager strength criterion with~\eqref{def_FED};
 \item[\texttt{M4}]-- Drucker--Prager strength criterion with~\eqref{def_PED};
 \item[\texttt{M5}]-- Huber strength criterion with~\eqref{def_PED} and arbitrary tensile-volumetric and deviatoric limits.
\end{description}
In the following sections, the key features of each model are discussed, with emphasis on the \emph{degradation behavior} and the \emph{evolution of the elastic domain} as functions of damage. These features are summarized in \Tab{tab_mod}. Moreover, the standard \texttt{AT1} phase-field model \texttt{M1} is taken as a reference, with its limit values used to define non-dimensional stress and strain variables for comparing the different models.

\subsubsection{Standard \texttt{AT1} phase-field model with full elastic degradation (model \texttt{M1})}
\label{sec_stdPF} 

In standard phase-field models~\cite{Bourdin2000b}, all terms in the free energy are degraded by the same function, as in~\eqref{def_FED}. Concerning the fracture function, for standard phase-field models, the fracture perturbation vanishes, $f(\teps,\sdam)=0$, and therefore $\Gf$ is constant. In particular, for the \texttt{AT1} phase-field model with a non-vanishing elastic domain, we have $\Gf = 3\Gc/8$, with $\Gc$ the mode-independent fracture toughness.

As a result, the elastic domains \eqref{def_Rm} and \eqref{def_Rms} are given by
\begin{equation}
  \Rm(\sdam) = \cbra{\teps \in \Sym \;:\; \dfrac{\kappa}{2}\,\sepsv^2 + \mu\,\sepsd^2 \leq 
  \dfrac{\Gf}{2\ell (1-\sdam)}}
  \label{def_RstdPF} 
\end{equation}
in strain space and
\begin{equation}
  \Rm^*(\sdam) = \cbra{\tstr \in \Sym \;:\; \dfrac{\sstrh^2}{2\kappa} + \dfrac{\sstrd^2}{4\mu} 
    \leq 
    \dfrac{\Gf}{2\ell}(1-\sdam)^3}
  \label{def_RsstdPF}
\end{equation}
in stress space. It is clear from \eqref{def_RstdPF} and \eqref{def_RsstdPF} that for the standard phase-field model, once $\Gf/\ell$ is fixed, the elastic domains are uniquely identified. For any damage state, these correspond in the strain and stress spaces to symmetric half ellipses centered at the origin. The vertices coincide with the intercepts and are given by $\cbra{(\sepsvbar^{\mathsmaller{-}}(\sdam),0),(0,\sepsdbar(\sdam)),(\sepsvbar^{\mathsmaller{+}}(\sdam),0)}$ and $\cbra{(\sstrhbar^{\mathsmaller{-}}(\sdam),0),(0,\sstrdbar(\sdam)),(\sstrhbar^{\mathsmaller{+}}(\sdam),0)}$ 
with $\sepsvbar^{\mathsmaller{-}} = -\sepsvbar^{\mathsmaller{+}}$ and $\sstrhbar^{\mathsmaller{-}} = -\sstrhbar^{\mathsmaller{+}}$. Accordingly, the elastic domain limits \eqref{def_limiteps} and \eqref{def_limitstr} follow as

\begin{equation}
    \sepsvbar^{\mathsmaller{\pm}}(\sdam) 
    =  \pm\dfrac{\sepsvcheck}{\sqrt{1-\sdam}},
  \qquad 
  \sepsdbar(\sdam) 
    = \dfrac{\sepsdcheck}{\sqrt{1-\sdam}} 
  \label{def_eps_lim} 
\end{equation}
in strain space and
\begin{equation} 
  \sstrhbar^{\mathsmaller{\pm}}(\sdam) 
    = \pm \sstrhhat\sqrt{(1-\sdam)^{3}},
  \qquad 
  \sstrdbar(\sdam) 
    = \sstrdhat \sqrt{(1-\sdam)^{3}} 
  \label{def_str_lim}
\end{equation} 
in stress space, with initial values given by \eqref{def_limitstr},
\begin{equation}
  \sepsvcheck = \sqrt{\dfrac{\Gf}{\kappa\ell}}, 
  \quad
  \sepsdcheck = \sqrt{\dfrac{\Gf}{2\mu\ell}},
  \quad
  \sstrhhat = \sqrt{\dfrac{\kappa\,\Gf}{\ell}},
  \quad
  \sstrdhat = \sqrt{\dfrac{2\mu\,\Gf}{\ell }}.
\label{def_limval} 
\end{equation}

\begin{figure}[t!]
\footnotesize
\centering
	\begin{subfigure}{0.4\linewidth}
    \footnotesize
		\centering
      \includeinkscape{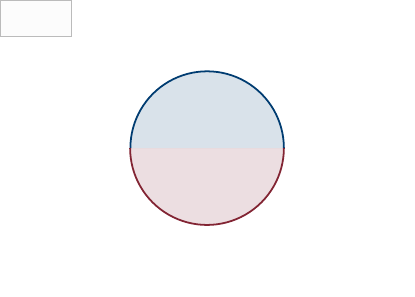}
		\caption{}
		\label{fig_M1_el0}
	\end{subfigure}
	\begin{subfigure}{0.58\linewidth}
    \footnotesize
		\centering
    \includeinkscape{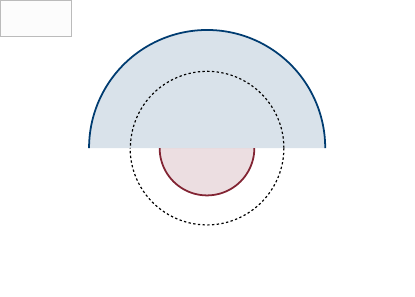}
		\caption{}
		\label{fig_M1_el}
	\end{subfigure}
	\caption{Elastic domain of \texttt{M1} for (\subref{fig_M1_el0}) $\sdam=0$ and (\subref{fig_M1_el}) $\sdam > 0$ with the strength surface of the sound material in dashed black.}
	\label{fig_M1} 
\end{figure}

The undamaged limit values of the standard model~\eqref{def_limval} allow us to define the following non-dimensional variables:
\begin{equation}
  \sepsvtilde = \dfrac{\sepsv}{\sepsvcheck},
  \quad
  \sepsdtilde = \dfrac{\sepsd}{\sepsdcheck},
  \quad
  \sstrhtilde = \dfrac{\sstrh}{\sstrhhat},
  \quad
  \sstrdtilde = \dfrac{\sstrd}{\sstrdhat}.
\label{def_nondimvar} 
\end{equation} 
With respect to these non-dimensional variables, \eqref{def_RstdPF} and \eqref{def_RsstdPF} can be expressed as
\begin{equation}
  \Rm(\sdam) = \cbra{\teps \in \Sym \;:\; \sepsvtilde^2 + \sepsdtilde^2  \leq \dfrac{1}{(1-\sdam)}},
  \label{def_RepsPFnorm} 
\end{equation}
\begin{equation}\
  \Rm^*(\sdam) = \cbra{\tstr \in \Sym \;:\; \sstrhtilde^2 + \sstrdtilde^2 \leq (1-\sdam)^3}.
  \label{def_RstrPFnorm} 
\end{equation}
With this reparameterization, the elastic domains~\eqref{def_RepsPFnorm} and~\eqref{def_RstrPFnorm} correspond, in both strain and stress spaces, to half-disks whose radius depends on the damage variable and is unitary in the undamaged state. The limit functions~\eqref{def_eps_lim}--\eqref{def_str_lim} with respect to the non-dimensional variables \eqref{def_nondimvar} follow as
\begin{equation}
  \epsilon(\sdam) 
    \coloneqq  \dfrac{1}{\sqrt{1-\sdam}}
    \qquad
    \text{and}
    \qquad
  \varsigma(\sdam) 
    \coloneqq  \sqrt{(1-\sdam)^3},
\end{equation} 
since $\epsilon(\sdam) 
  = \pm\sepsvbar^{\mathsmaller{\pm}}(\sdam)/\sepsvcheck
  = \sepsdbar(\sdam)/\sepsdcheck$
and $\varsigma(\sdam) 
  = \pm\sstrhbar^{\mathsmaller{\pm}}(\sdam)/\sstrhhat
  = \sstrdbar(\sdam)/\sstrdhat
$.
The normalized elastic domains \eqref{def_RepsPFnorm} and \eqref{def_RstrPFnorm} are shown in \Fig{fig_M1_el0} and \Fig{fig_M1_el} for $\sdam=0$ and $\sdam > 0$, respectively.

\FloatBarrier
\subsubsection{Double-ellipse strength criterion with full elastic degradation and arbitrary volumetric/deviatoric limits (model~\texttt{M2})}
\label{sec_M2} 

\begin{figure}[t!]
\footnotesize
\centering
	\begin{subfigure}{0.4\linewidth}
    \footnotesize
		\centering
      \includeinkscape[scale=1.2]{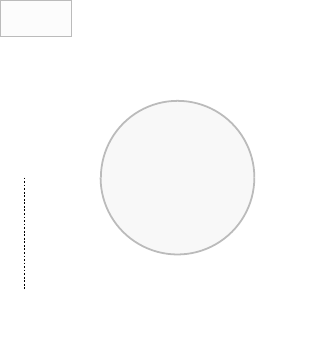}
		\caption{}
		\label{fig_M2_el0}
	\end{subfigure}
	\hfill
	\begin{subfigure}{0.58\linewidth}
    \footnotesize
		\centering
    \includeinkscape[scale=1.2]{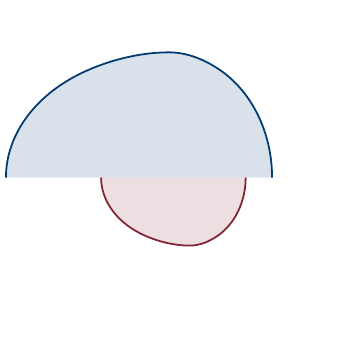}
		\caption{}
		\label{fig_M2_el}
	\end{subfigure}
	\caption{Elastic domain of \texttt{M2} for (\subref{fig_M2_el0})  $\sdam=0$ and (\subref{fig_M2_el}) $\sdam > 0$  with $c^*(\sdam)=c(\sdam)(1-\sdam)^2$. The gray region in (\subref{fig_M2_el0}) corresponds to the initial elastic domain of~\texttt{M1}. The black dashed curve in (\subref{fig_M2_el}) represents the strength surface of the sound~material.}
	\label{fig_M2} 
\end{figure}

For \texttt{M2}, as in standard phase-field models, all terms in the free energy are degraded by the same degradation function, corresponding to~\eqref{def_FED}. To highlight the flexibility of the proposed modeling approach in describing arbitrary strength criteria, a more general shape for the elastic domain is considered, namely going from a half-disk domain to a domain composed of two elliptical arcs of different eccentricities, with semi-axes depending on three positive parameters $\cbra{a^{\mathsmaller{-}},a^{\mathsmaller{+}},b}$ and one non-negative function $c(\sdam)$ with $c(0) = c$ and $c(1) = 0$.  Moreover, $c(\sdam)$ must be such that, for any damage level, the origin is always contained in the elastic domain. In strain space, the semi-axis lengths of the first and second elliptical arcs are $\cbra{a^{\mathsmaller{-}}\,\epsilon(\sdam),b\,\epsilon(\sdam)}$ and $\cbra{a^{\mathsmaller{+}}\,\epsilon(\sdam), b\,\epsilon(\sdam)}$, respectively, while in stress space these are given by $\cbra{a^{\mathsmaller{-}}\,\varsigma(\sdam),b\,\varsigma(\sdam)}$ and $\cbra{a^{\mathsmaller{+}}\,\varsigma(\sdam), b\,\varsigma(\sdam)}$, respectively. \Fig{fig_M2} illustrates the geometry of the elastic domain and the meaning of these parameters.

\begin{figure}[!htb]
  \centering
  \small
  \begin{overpic}[width=0.8\linewidth]{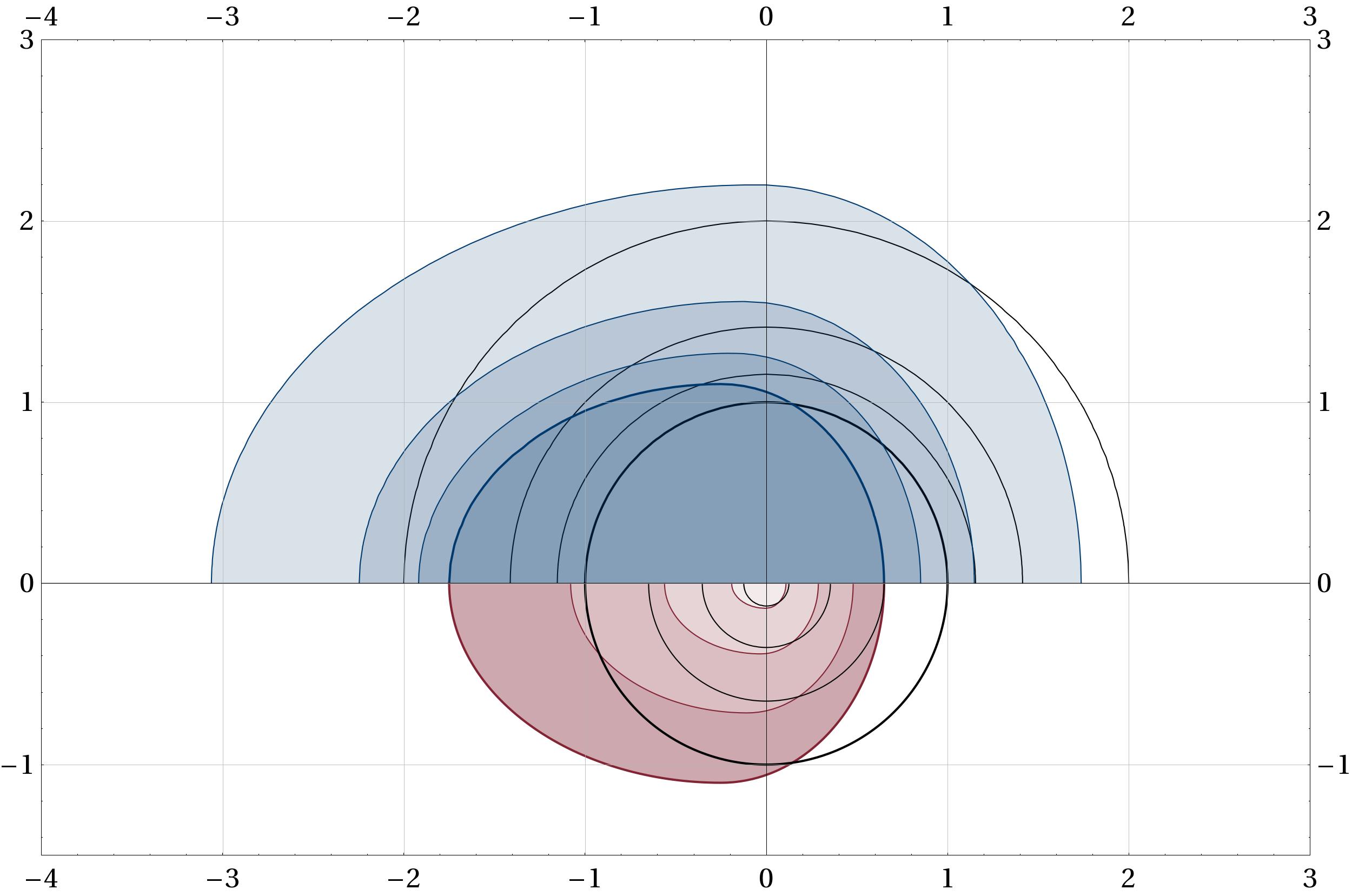}
    \put(-3,43){\color{myBlue}$\sepsdtilde$}
    \put(-3,15){\color{myRed}$\sstrdtilde$}
    \put(50,68){\color{myBlue}$\sepsvtilde$}
    \put(50,-3){\color{myRed}$\sstrhtilde$}
  \end{overpic}
  \vspace{3mm}
  \caption{Elastic domain evolution as a function of damage for \texttt{M2} with $a^\mathsmaller{+}=0.9$, $a^\mathsmaller{-}=1.5$, $b=1.1$, and $c(\sdam) = 0.1$. Contour plots are shown for $\sdam = \cbra{0,0.25,0.5,0.75}$ (from dark to light). The strength surfaces of the standard \texttt{M1} model are reported in black for the same damage levels.}
  \label{fig_M2_EDevo} 
\end{figure}

The elastic domain of \texttt{M2} is characterized by the following normalized fracture perturbation function:
\begin{equation}
   \tilde{f}(\tilde\seps_\l{v},\sdam) 
    = (1 - \sdam) \bra{\dfrac{
        (a^{\mathsmaller{\pm}})^2 \, \sepsvtilde^2 
        -b^2
        \,\bra{\sepsvtilde - c(\sdam)}^2
      }{(a^{\mathsmaller{\pm}})^2}}  
      + b^2 - 1,
  \label{def_f1} 
\end{equation} 
where $a^{\mathsmaller{+}}$ is used if $\sepsvtilde \geq c(\sdam)$, while $a^{\mathsmaller{-}}$ is used if $\sepsvtilde < c(\sdam)$. The corresponding function in stress space follows as $\tilde{f}^*(\sstrhtilde,\sdam) = \tilde{f}(s(\sdam)\sstrhtilde,\sdam)$. The fracture perturbation functions appearing in~\eqref{def_fracfun} and~\eqref{def_fracfun_stress} are then given by
\begin{equation}
    f(\teps,\sdam)  = \tilde{f}(\tilde\seps_\l{v}(\teps),\sdam), \qquad f^*(\tstr,\sdam) = \tilde{f}^*(\sstrhtilde(\tstr),\sdam).    
\end{equation}
The corresponding elastic domain reads
\begin{equation}
\begin{aligned}
  \Rm(\sdam) 
    &= \cbra{\teps \in \Sym \;:\; 
      \sepsvtilde^2 
      + 
      \sepsdtilde^2
      \leq 
      \dfrac{1+ \tilde{f}(\sepsvtilde,\sdam)}{(1-\sdam)}}
    \\[1ex]
    &= \cbra{\teps \in \Sym \;:\;
      \dfrac{(\sepsvtilde-c(\sdam))^2}{(a^{\mathsmaller{\pm}})^2} 
      + \dfrac{\sepsdtilde^2}{b^2} 
      \leq 
      \dfrac{1}{(1-\sdam)}
    }
\label{def_RstdPFnorm} 
\end{aligned}
\end{equation}
in strain space and
\begin{equation}
\begin{aligned}
  \Rm^*(\sdam) 
    &= \cbra{\tstr \in \Sym \;:\; 
        \sstrhtilde^2 
        + 
        \sstrdtilde^2
        \leq 
        \bra{1+ \tilde{f}^*(\sstrhtilde,\sdam)}(1-\sdam)^3}
    \\[1ex]
    &= \cbra{\tstr \in \Sym \;:\;
        \dfrac{\bra{\sstrhtilde-c(\sdam)(1-\sdam)^2}^2}{(a^{\mathsmaller{\pm}})^2} 
        + \dfrac{\sstrdtilde^2}{b^2} 
        \leq 
        (1-\sdam)^3
      }
\label{def_RstdPFnorm_stress} 
\end{aligned}
\end{equation}
in stress space, where the elliptical arcs and the role of the parameters and functions are evident. \Fig{fig_M2_EDevo} shows an example of the evolution of the elastic domain compared with that of the standard phase-field model \texttt{M1}. In particular, the \texttt{M2} model represents a straightforward extension of \texttt{M1}, with fully tunable intercepts.

\FloatBarrier
\subsubsection{Drucker--Prager strength criterion with full elastic degradation (model \texttt{M3})}

The Drucker--Prager (DP) criterion is a classical model in geomechanics, widely used to describe pressure-sensitive failure in cohesive-frictional materials such as rocks and concrete, where compressive-shear failure plays a crucial role. Its use has been explored in recent phase-field models~\cite{Ulloa2021d,Vicentini2024,Kumar2020}. Owing to its dependence on the hydrostatic stress, the DP criterion is suitable for distinguishing tensile and compressive-shear strengths. In \texttt{M3}, this criterion is embedded in the proposed framework, assuming full elastic degradation according to~\eqref{def_FED}.

\begin{figure}[b!]
\small
\centering
	\begin{subfigure}{0.4\linewidth}
    \footnotesize
		\centering
      \includeinkscape[scale=1.2]{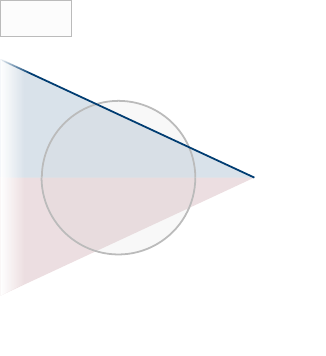}
		\caption{}
		\label{fig_M3_eD}
	\end{subfigure}
	\hfill
	\begin{subfigure}{0.58\linewidth}
    \footnotesize
		\centering
    \includeinkscape[scale=1.2]{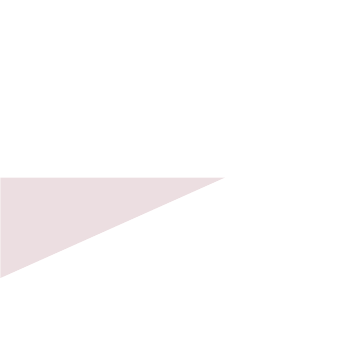}
		\caption{}
		\label{fig_M3_sD}
	\end{subfigure}
	\caption{Elastic domain of \texttt{M3} and \texttt{M4} for  (\subref{fig_M3_eD}) $\sdam=0$ and (\subref{fig_M3_sD}) $\sdam > 0$. The gray region in (\subref{fig_M3_eD}) corresponds to the initial elastic domain of \texttt{M1}. The black dashed curve in (\subref{fig_M3_sD}) represents the strength surface of the sound material.}
	\label{fig_M3} 
\end{figure}

The DP elastic domain is a cone in stress/strain space, identified by two constants $\cbra{a,b}$ defining the intercepts of the initial strength surface with the axes. Accordingly, during material degradation, these intercepts evolve as $\cbra{a \,\epsilon(\sdam),b\,\epsilon(\sdam)}$ in strain space and $\cbra{a \,\varsigma(\sdam),b\,\varsigma(\sdam)}$ in stress space. \Fig{fig_M3} illustrates the geometry of the elastic domain and the meaning of the parameters.

\begin{figure}[h!]
  \centering
  \vspace{1em}
  \small
  \begin{overpic}[width=0.6\linewidth]{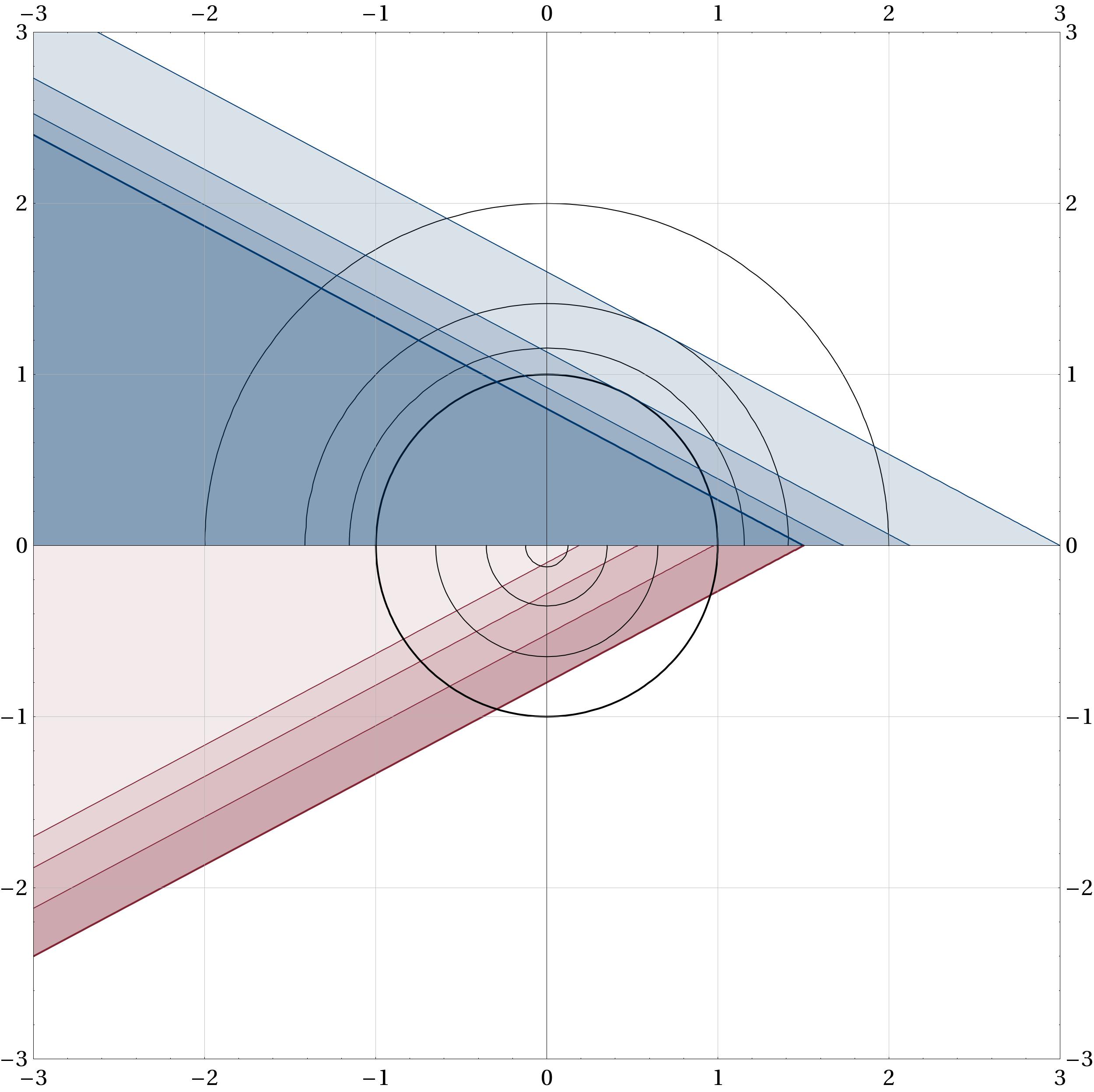}
    \put(-5,73){\color{myBlue}$\sepsdtilde$}
    \put(-5,30){\color{myRed}$\sstrdtilde$}
    \put(50,103){\color{myBlue}$\sepsvtilde$}
    \put(50,-4){\color{myRed}$\sstrhtilde$}
  \end{overpic}
  \vspace{3mm}
  \caption{Elastic domain evolution as a function of damage for \texttt{M3} and \texttt{M4} with  $a=1.5$ and $b=0.8$. Contour plots are shown for $\sdam = \cbra{0,0.25,0.5,0.75}$ (from dark to light). The strength surfaces of the standard \texttt{M1} model are reported in black for the same damage levels.}
  \label{fig_M3_EDevo} 
\end{figure}

The elastic domain of \texttt{M3} is characterized by the following normalized fracture perturbation function:
\begin{equation}
  \tilde{f}(\tilde\seps_\l{v},\sdam) = (1-\sdam)\bra{\dfrac{a^2+b^2}{a^2}}
  (\sepsvtilde)^2
  -
  \dfrac{2b^2}{a}(\sqrt{1-\sdam})\,\sepsvtilde
  + b^2-1.
  \label{def_fM3} 
\end{equation} 
The elastic domain then reads
\begin{equation}
\begin{aligned}
  \Rm(\sdam) 
    &= \cbra{\teps \in \Sym \;:\; 
      \sepsvtilde^2 
      + 
      \sepsdtilde^2
      \leq 
      \dfrac{1+ \tilde{f}(\sepsvtilde,\sdam)}{(1-\sdam)}}
    \\
    &= \cbra{\teps \in \Sym \;:\; 
        \dfrac{\sepsvtilde}{a} + \dfrac{\sepsdtilde}{b} \leq \dfrac{1}{\sqrt{1-\sdam}}
      }
\label{def_Reps3} 
\end{aligned}
\end{equation}
in strain space and
\begin{equation}
\begin{aligned}
  \Rm^*(\sdam) 
    &= \cbra{\tstr \in \Sym \;:\; 
      \sstrhtilde^2 
      + 
      \sstrdtilde^2
      \leq 
      \bra{1+ \tilde{f}^*(\sstrhtilde,\sdam)}(1-\sdam)^3}
    \\
    &= \cbra{\tstr \in \Sym \;:\; 
        \dfrac{\sstrhtilde}{a} + \dfrac{\sstrdtilde}{b} \leq \sqrt{(1-\sdam)^3}
      }
\label{def_Rstr3} 
\end{aligned}
\end{equation}
in stress space, where the physical meaning of the parameters $(a,b)$ is evident. \Fig{fig_M3_EDevo} shows an example of the evolution of the elastic domain compared with that of the standard phase-field model~\texttt{M1}.

\subsubsection{Drucker--Prager strength criterion with partial elastic degradation (model \texttt{M4})}

Differently from the previous models, we assume for \texttt{M4} that no degradation of the compressive bulk modulus can occur, corresponding to \eqref{def_PED} to account for non-interpenetration. Nevertheless, we wish to recover the same exact strength criterion of model \texttt{M3}, that is, the Drucker--Prager elastic domain~\eqref{def_Reps3}/\eqref{def_Rstr3}. To this aim, it is sufficient to slightly modify the fracture perturbation function~\eqref{def_fM3} as follows:
\begin{equation}
  \tilde{f}(\tilde\seps_\l{v},\sdam) 
  = (1-\sdam)\bra{\dfrac{b^2}{a^2}\sepsvtilde^2 + (\sepsvtilde^\mathsmaller{+})^2} - \dfrac{2b^2}{a}(\sqrt{1-\sdam})\,\sepsvtilde + b^2-1.
\end{equation}
By straightforward calculations, the resulting elastic domain turns out to be the same as~\eqref{def_Reps3} and \eqref{def_Rstr3}.

\FloatBarrier
\subsubsection{Huber strength criterion with partial elastic degradation and arbitrary tensile-volumetric/deviatoric limits (model~\texttt{M5})}

Unilateral frictionless contact under compression was first addressed by~\citet{Amor2009a} within the classical phase-field approach to brittle fracture, exploiting the volumetric--deviatoric decomposition of the elastic energy (corresponding here to~\eqref{def_PED}). This model has been rigorously proven, in the sense of $\Gamma$-convergence, to satisfy the non-interpenetration material condition, that is, to allow only non-negative normal jumps without affecting the tensile and shear behavior in the presence of a crack~\cite{Chambolle2017}. As in the case of full elastic degradation (\Sec{sec_stdPF}), when combined with a standard phase-field formulation, this energy split does not allow independent control of the shear and tensile strengths.

\begin{figure}[H]
\small
\centering
	\begin{subfigure}{0.4\linewidth}
    \footnotesize
		\centering
      \includeinkscape[scale=1.2]{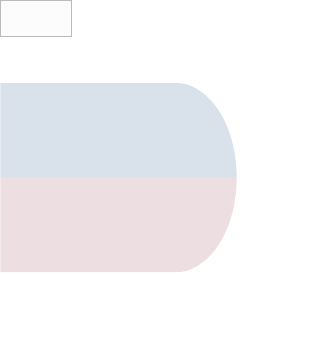}
		\caption{}
		\label{fig_M5_eD}
	\end{subfigure}
	\hfill
\begin{subfigure}{0.58\linewidth}
    \footnotesize
		\centering
    \includeinkscape[scale=1.2]{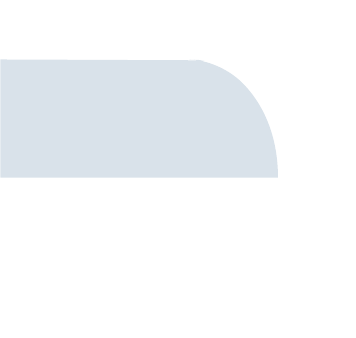}
		\caption{}
		\label{fig_M5_sD}
	\end{subfigure}
	\caption{Elastic domain of \texttt{M5} for (\subref{fig_M5_eD}) $\sdam=0$ and (\subref{fig_M5_sD}) $\sdam > 0$. The gray region in (\subref{fig_M5_eD}) corresponds to the initial elastic domain of  \texttt{M1}. The black dashed curve in (\subref{fig_M5_sD}) represents the elastic strength of the sound material.}
	\label{fig_M5} 
\end{figure}

In the \texttt{M5} model, we show one possibility to overcome these limitations in the present framework by combining the flexibility of \texttt{M2}, which allows for arbitrary volumetric--deviatoric limits, with the non-interpenetration condition. Specifically, for $\sepsv \geq 0$, we consider the same elliptical domain with prescribed vertices used in \texttt{M2}, taking $c(\sdam)=0$ without loss of generality. For $\sepsv \leq 0$, instead, the deviatoric limit is kept constant and is therefore unaffected by compression, as in a von Mises yield criterion. Accordingly, we employ two positive parameters $\cbra{a,b}$ representing the intercepts of the strength surface with the axes in the undamaged material. During material degradation, these intercepts evolve as $\cbra{a \,\epsilon(\sdam),b\,\epsilon(\sdam)}$ in strain space and $\cbra{a \,\varsigma(\sdam),b\,\varsigma(\sdam)}$ in stress space. \Fig{fig_M5} illustrates the geometry of the elastic domain and the meaning of these two parameters. The resulting model turns out to have the sought flexibility to prescribe the shear and tensile-volumetric strengths independently, while retaining non-interpenetration.

\begin{figure}[t!]
  \centering
  \small
  \begin{overpic}[width=0.6\linewidth]{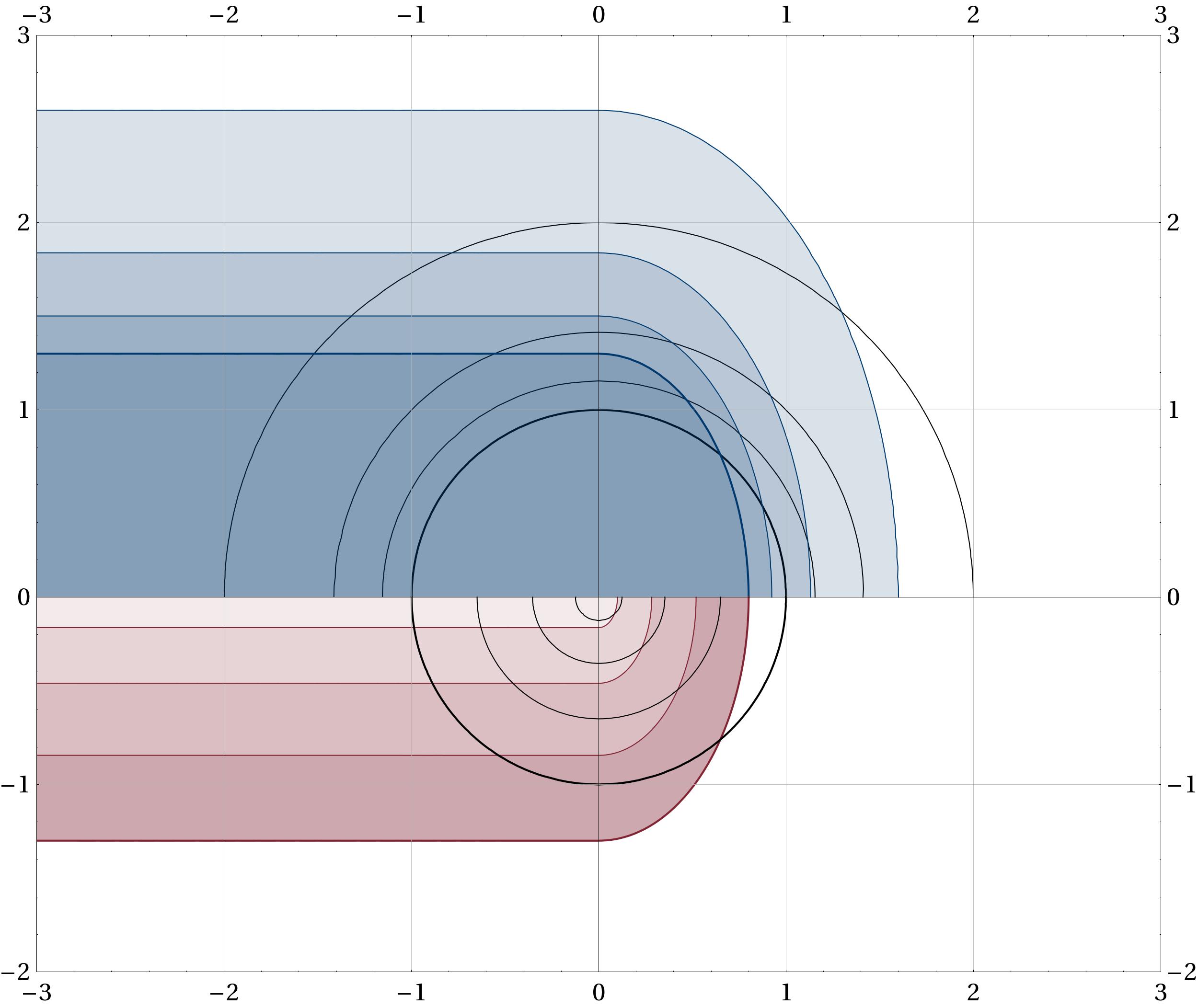}
    \put(-4,59){\color{myBlue}$\sepsdtilde$}
    \put(-4,22){\color{myRed}$\sstrdtilde$}
    \put(50,85){\color{myBlue}$\sepsvtilde$}
    \put(50,-4){\color{myRed}$\sstrhtilde$}
  \end{overpic}
  \vspace{3mm}
  \caption{Elastic domain evolution as a function of damage for \texttt{M5} with $a=0.8$ and $b=1.3$. Contour plots are shown for $\sdam = \cbra{0,0.25,0.5,0.75}$ (from dark to light). The strength surfaces of the standard \texttt{M1} model are reported in black for the same damage levels.}
  \label{fig_M5_EDevo} 
\end{figure}

In view of \eqref{def_f1}, the corresponding fracture perturbation function reads
\begin{equation}
  \tilde{f}(\tilde\seps_\l{v},\sdam) = (1-\sdam)\bra{\dfrac{a^2-b^2}{a^2}}
  (\sepsvtilde^{\mathsmaller{+}})^2
  + b^2-1.
\end{equation} 
The elastic domain is then given by
\begin{equation}
\begin{aligned}
  \Rm(\sdam) 
    &= \cbra{\teps \in \Sym : 
      (\sepsvtilde^{\mathsmaller{+}})^2 
      + 
      \sepsdtilde^2
      \leq 
      \dfrac{1+ \tilde{f}(\sepsvtilde,\sdam)}{(1-\sdam)}}
  \\
    &= \cbra{\teps \in \Sym :
        \dfrac{(\sepsvtilde^{\mathsmaller{+}})^2}{a^2} 
        + \dfrac{\sepsdtilde^2}{b^2} 
        \leq 
        \dfrac{1}{(1-\sdam)}
    }
\end{aligned}
\end{equation}
in strain space and
\begin{equation}
\begin{aligned}
  \Rm^*(\sdam) 
    &= \cbra{\tstr \in \Sym : 
      (\sstrhtilde^{\mathsmaller{+}})^2 
      + 
      \sstrdtilde^2
      \leq 
      \bra{1+ \tilde{f}^*(\sstrhtilde,\sdam)}(1-\sdam)^3}
  \\
    &= \cbra{\tstr \in \Sym :
        \dfrac{\bra{\sstrhtilde^{\mathsmaller{+}}}^2}{a^2} 
        + \dfrac{\sstrdtilde^2}{b^2} 
        \leq 
        (1-\sdam)^3
      }
\end{aligned}
\end{equation}
in strain space, where the physical meaning of the parameters $(a,b)$ is evident. \Fig{fig_M5_EDevo} shows an example of the evolution of these elastic domains compared with that of the standard phase-field elastic domains.

\subsection{Discussion and comparison of different models}

Let us briefly compare the considered models. \texttt{M1} and \texttt{M2} provide examples of phase-field models with the same degradation behavior but different elastic domains, whereas \texttt{M3} and \texttt{M4} exhibit different degradation behaviors but the same elastic domain. Models \texttt{M1} and \texttt{M5}, instead, differ in both degradation behavior and elastic domain. These comparisons emphasize that elastic degradation and the shape of the elastic domain can be tuned independently, and show how the general framework of \Sec{sec_gen_mod} allows phase-field models with custom strength criteria to be constructed by prescribing these properties. An overview of the considered models is reported in \Tab{tab_mod}. We note that models with more general shapes for the elastic domain and the corresponding strength surface can be obtained by fitting procedures.

\begin{landscape}
\begin{table}[!h]
\renewcommand{\arraystretch}{1.5}
\small
\centering
  \begin{tabularx}{\linewidth}{c>{\centering\arraybackslash}p{20mm}ccc>{\raggedright\arraybackslash\footnotesize}X}
  \toprule
    \textbf{Label} 
      & \textbf{Degradation model} 
      & \textbf{Strength criterion} 
      & \multicolumn{2}{c}{\textbf{Elastic domain}} 
      & \multicolumn{1}{c}{\textbf{Description}} 
  \\[-3ex]
      & & & $\Rm(\sdam)$ & $\Rm^*(\sdam)$ &  \\
  \midrule
    \texttt{M1} 
      & \multirow{3}{*}{\eqref{def_FED}} 
      & Standard \texttt{AT1}
      & $\cbra{\sepsvtilde^2 + \sepsdtilde^2  \leq \dfrac{1}{(1-\sdam)}}$
      & $\cbra{\sstrhtilde^2 + \sstrdtilde^2 \leq (1-\sdam)^3 \vphantom{\dfrac{1}{(1-\sdam)}}}$
      & Reference model \\
      \cmidrule(lr){1-1} \cmidrule(lr){3-6}
    \texttt{M2} 
      & 
      & Double ellipse
      & $\cbra{\dfrac{(\sepsvtilde-c(\sdam))^2}{(a^{\mathsmaller{\pm}})^2} 
          + \dfrac{\sepsdtilde^2}{b^2} 
          \leq 
          \dfrac{1}{(1-\sdam)}
          }$
      & $\cbra{\dfrac{\bra{\sstrhtilde-c(\sdam)(1-\sdam)^2}^2}{(a^{\mathsmaller{\pm}})^2} 
        + \dfrac{\sstrdtilde^2}{b^2} 
        \leq 
        (1-\sdam)^3
        }$
      & Tunable volumetric and deviatoric limits \\
      \cmidrule(lr){1-1} \cmidrule(lr){3-6}
    \texttt{M3} 
      & 
      & \multirow{2}{*}{Drucker--Prager} 
      & \multirow{2}{*}{$\cbra{
        \dfrac{\sepsvtilde}{a} + \dfrac{\sepsdtilde}{b} \leq \dfrac{1}{\sqrt{1-\sdam}}
        }$}
      & \multirow{2}{*}{$\cbra{
        \dfrac{\sstrhtilde}{a} + \dfrac{\sstrdtilde}{b} \leq \sqrt{(1-\sdam)^3}
        }$}
      & DP model with full elastic degradation  \\
  \cmidrule(lr){1-2} \cmidrule(lr){6-6}
    \texttt{M4} 
      & \multirow{2}{*}{\eqref{def_PED}} 
      &  
      &  
      &  
      & DP model with non-interpenetration \\
  \cmidrule(lr){1-1} \cmidrule(lr){3-6}
    \texttt{M5} 
      & 
      & Huber 
      & $\cbra{\dfrac{(\sepsvtilde^{\mathsmaller{+}})^2}{a^2} 
          + \dfrac{\sepsdtilde^2}{b^2} 
          \leq 
          \dfrac{1}{(1-\sdam)}
          }$
      & $\cbra{\dfrac{\bra{\sstrhtilde^{\mathsmaller{+}}}^2}{a^2} 
          + \dfrac{\sstrdtilde^2}{b^2} 
          \leq 
          (1-\sdam)^3
        }$
      & Tunable tensile-volumetric and deviatoric limits with non-interpenetration \\
  \bottomrule
 \end{tabularx}
\caption{Specific material models considered in this work.}
\label{tab_mod} 
\end{table}
\end{landscape}

\clearpage

\section{Analytical and numerical examples}
\label{sec_examples} 

\begin{figure}[H]
    \small
		\centering
		\includeinkscape[]{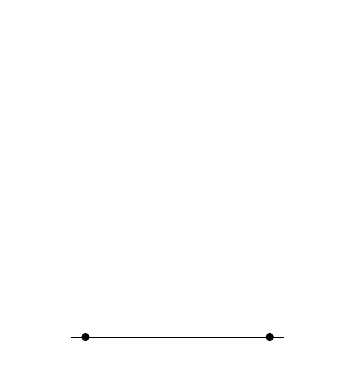}
		\caption{Boundary value problem for the biaxial disk.}
    \label{fig_HR_domain}
\end{figure}

In this section, we investigate analytical homogeneous responses under proportional and non-proportional loading conditions, together with numerical nucleation tests for models~\texttt{M1}--\texttt{M5}.
For this purpose, we employ the biaxial problem considered in \citet[Sec.~5.1]{Vicentini2024} (\Fig{fig_HR_domain}): a disk $\domain$ of diameter $D$ with its center at the origin of the Cartesian reference system and a Dirichlet boundary condition
\begin{equation}
  \bar{\vdisp}_t = \vect{\epsilon}(t)\,\vpos \quad \text{on $\bdomainD$},
  \label{eq_bc}
\end{equation} 
where $t$ is the time evolution parameter, monotonically increasing from $0$, which corresponds to an unstretched and sound material state, namely $\vect{\epsilon}(t=0) = \vect{0}$ and $\sdam(t=0)=0$.

For proportional loading conditions, we assume
\begin{equation}
  \vect{\epsilon}(t) = \vect{\epsilon}_{\rm p}(t) = 
    t \bra{\dfrac{\cos\vartheta + \sin \vartheta}{2}\bra{\vect{e}_x \otimes \vect{e}_x }
    + 
    \dfrac{\cos\vartheta - \sin \vartheta}{2}\bra{\vect{e}_y \otimes \vect{e}_y }},
  \label{def_epsilon} 
\end{equation} 
where $\vartheta$ is the angle representative of the loading direction in the $\sepsv$--\thinspace$\sepss$ half-plane. For proportional loading conditions, the magnitude of the spherical and deviatoric parts of the homogeneous strain $\teps(t) = \vect{\epsilon}(t)$ compatible with the boundary condition are
\begin{equation}
  \sepsv(t) = t \cos\vartheta, \qquad \sepss(t) = \sqrt{2} \, \sepsd(t) = t\sin \vartheta.
  \label{eq_load_cond} 
\end{equation} 
In the normalized $\sepsvtilde$--\thinspace$\sepsdtilde$ half-plane, the angle~$\tilde\vartheta$ is related to $\vartheta$ by
\begin{equation}
  \tan \tilde\vartheta = \sqrt{\dfrac{\mu}{\kappa}} \, \tan \vartheta.
\end{equation} 

The constitutive, geometric, and model-specific parameters are reported in \Tab{tab_parAll}. Note that the parameters in \Tab{tab_par} are the same as those considered in~\citet{Vicentini2024}.

\begin{table}[h!]
\small
\centering
  \begin{subtable}[c]{0.4\linewidth}
    \centering
    \begin{tabular}{ccccccc}
      \toprule
        $E$ & $\nu$ & $\kappa$ & $\mu$ & $\Gf$ & $\ell$ & $D$\\
      \midrule
        100 & 0.3 & $\approx$ 71.43 & $\approx$ 38.46 & 0.06 & 0.04 & 1 \\
      \bottomrule
    \end{tabular}
    \caption{}
    \label{tab_par} 
  \end{subtable}
  \\[7mm]
  \begin{subtable}[c]{0.4\linewidth}
  \small
  \centering
  \renewcommand{\arraystretch}{1.2}
    \begin{tabular}{c|c|c}
      \toprule
        \texttt{Model} & \multicolumn{2}{c}{$f$ \textbf{parameters}} \\
        & \footnotesize proportional loading  & \footnotesize non-proportional loading 
      \\
      \midrule
        \texttt{M2} & 
        \multicolumn{2}{c}{
          $a^{\mathsmaller{+}} = 0.5$,\quad 
          $a^{\mathsmaller{-}} = 2$,\quad 
          $b = 1$,\quad 
          $c = 0$ 
        }
      \\
      \midrule
        \texttt{M3/M4} 
          & $a = 2$,\quad 
            $b = 0.75$ 
          & $a = 2$,\quad 
            $b = 1$ 
        \\
      \midrule
        \texttt{M5} 
          & $a = 1.75$,\quad 
            $b = 1.5$ 
          & $a = 0.5$,\quad 
            $b = 1$ 
        \\
      \bottomrule
    \end{tabular}
    \caption{}
    \label{tab_par_mod} 
  \end{subtable}
  \caption{
    (\subref{tab_par}) constitutive and geometric parameters and (\subref{tab_par_mod}) model-specific parameters for the homogeneous response analyses under proportional and non-proportional loading conditions and the numerical nucleation tests for the biaxial disk problem of~\Fig{fig_HR_domain}.
  }
  \label{tab_parAll} 
\end{table}

\begin{figure}[H]
\small
\centering
  \begin{minipage}{0.6\linewidth}
    \small
		\centering
		\includeinkscape[]{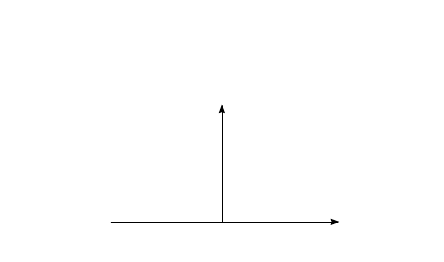}
  \end{minipage}
  \begin{minipage}{0.35\linewidth}
    \small
    \begin{itemize}
    \setlength\itemsep{2ex}
    \item $\vartheta = 0$, isotropic expansion
    \item $\vartheta = \pi/4$, uniaxial tension
    \item $\vartheta = \pi/2$, pure shear
    \item $\vartheta = 3\pi/4$, uniaxial compression
    \item $\vartheta = \pi$, isotropic contraction
    \end{itemize}
  \end{minipage}
  \caption{Illustration of paradigmatic loading directions and corresponding angles in the $\sepsv$--\thinspace$\sepss$ half-plane, employed in the proportional loading tests.}
  \label{fig_HR_load}
\end{figure}

In the homogeneous response analyses under proportional loading (\Sec{sec_HR_PL}), we consider as loading angles $\vartheta \in \cbra{0,\frac{\pi}{4},\frac{\pi}{2},\frac{3\pi}{4},\pi}$, corresponding to the paradigmatic loading conditions illustrated in~\Fig{fig_HR_load}. For the nucleation tests (\Sec{sec_NT}), we also include intermediate angles, so that $\vartheta \in \cbra{0,\frac{\pi}{8},\frac{\pi}{4},\frac{3\pi}{8},\frac{\pi}{2},\frac{5\pi}{8},\frac{3\pi}{4},\frac{7\pi}{8},\pi}$. 
For the analysis of homogeneous responses under non-proportional loading conditions, the strain tensor $\vect{\epsilon}(t)$ in~\eqref{eq_bc} is defined in terms of the auxiliary function
\begin{equation}
  \vect{\epsilon}(\gamma) = 
    \dfrac{3}{4}\bra{\sepsvcheck \sin\gamma + \sqrt{2}\,\sepsdcheck \bra{1 - \cos\gamma}}\bra{\vect{e}_x \otimes \vect{e}_x }
    + 
    \dfrac{3}{4}\bra{\sepsvcheck \sin\gamma - \sqrt{2}\,\sepsdcheck \bra{1 - \cos\gamma}}\bra{\vect{e}_y \otimes \vect{e}_y },
    \label{def_NPload}
\end{equation}
where $\gamma$ is a time-dependent parameter whose evolution is prescribed according to the specific loading path considered, as illustrated in \Fig{fig_NPHR}. Further details on the definition of the non-proportional loading paths considered are provided in \Sec{sec_HR_NPL}.

\subsection{Homogeneous responses}

\subsubsection{Proportional loading conditions}
\label{sec_HR_PL} 

The homogeneous response of each model is first derived for different proportional loading directions~$\vartheta$ by substituting the loading condition~\eqref{eq_load_cond} into the corresponding definition of the elastic domain. After the elastic limit is reached, damage is assumed to evolve so that the response remains on the strength surface. The critical values of the loading parameter $t$, corresponding to instants where the elastic limit is first reached, are reported in \Tab{tab_crit_t} for different values of the loading direction $\vartheta$.

\begin{table}[H]
\small
\centering
\renewcommand{\arraystretch}{1.2}
 \begin{tabular}{cccccccccc}
   \toprule
      \texttt{model} 
        & $ \vartheta = 0$ 
        & $ \vartheta = \pi/8$
        & $ \vartheta = \pi/4$
        & $ \vartheta = 3\pi/8$
        & $ \vartheta = \pi/2$
        & $ \vartheta = 5\pi/8$
        & $ \vartheta = 3\pi/4$
        & $ \vartheta = 5\pi/8$
        & $ \vartheta = \pi$ \\
    \midrule
      \texttt{M1} 
        & 0.145 & 0.15 & 0.165 & 0.186 & 0.197 & 0.186 & 0.165 & 0.15 & 0.145
        \\
    \midrule
      \texttt{M2} 
        & 0.0725 & 0.0775 & 0.0962 & 0.142 & 0.197 & 0.206 & 0.231 & 0.268 & 0.29 
        \\
    \midrule
      \texttt{M3/M4} 
        & 0.29 & 0.173 & 0.139 & 0.132 & 0.148 & 0.203 & 0.428 & -- & -- \\
    \midrule
      \texttt{M5} 
        & 0.254 & 0.259 & 0.272 & 0.289 & 0.296 & 0.321 & 0.419 & 0.774 & -- \\
    \bottomrule
 \end{tabular}
 \caption{Critical values of the loading parameter $t$, corresponding to instants where the elastic limit is first reached, approximated to the third decimal digit for different values of the loading angle.}
\label{tab_crit_t} 
\end{table}

\begin{table}[H]
\centering
\small
\renewcommand{\arraystretch}{1.5}
  \begin{tabular}{c>{\small}c}
    \toprule
      \texttt{Model} 
        & {\small$t$ {\footnotesize during the damaging stage}} \\
    \midrule
      \texttt{M1} 
        & 
        $\sqrt{\frac{\Gf}{\ell\bra{1-\sdam}\bra{\kappa(\cos\vartheta)^2 + \mu(\sin\vartheta)^2}}}$
      \\[1ex]
    \midrule
      \texttt{M2} 
        & 
        $ \frac{
            \sqrt{\Gf}\bra{
              b^2\,c\,\sqrt{\kappa}\,\cos\vartheta
              +\sqrt{(a^\mathsmaller{\pm})^2\,b^2\,(1-\sdam)\bra{\kappa\,b^2(\cos\vartheta)^2 + \mu\bra{(a^\mathsmaller{\pm})^2-c^2(1-\sdam)^3}(\sin\vartheta)^2}}
              }
          }{
            \sqrt{\ell}\bra{1-\sdam}\bra{\kappa\,b^2(\cos\vartheta)^2 + \mu\,(a^\mathsmaller{\pm})^2(\sin\vartheta)^2}
          }
        $
      \\[1ex]
    \midrule
      \texttt{M3/M4} 
        & 
        $ \frac{
            a\,b\,\sqrt{\Gf}
          }{
            \sqrt{\ell\,\bra{1-\sdam}}\bra{b\sqrt{\kappa}\,\cos\vartheta + a\sqrt{\mu}\sin\vartheta}
          }
        $
      \\[1ex]
    \midrule
      \texttt{M5} 
        & 
        $\begin{cases}
           \text{\footnotesize the same as \texttt{M2}},
             & \text{\footnotesize if $\sepsvtilde \geq 0$ }
             \\
             \frac{
                b\,\sqrt{\Gf}
              }{
            \sin\vartheta\sqrt{\ell\,\bra{1-\sdam} \mu} 
          }
             & \text{\footnotesize otherwise }\\
         \end{cases}
         $
      \\[1ex]
      \bottomrule
  \end{tabular}
  \caption{Time parameter $t$ during the damaging stage as a function of the phase-field variable~$\sdam$, the loading angle~$\vartheta$, and constitutive parameters for all considered models. The parameter $a^\mathsmaller{\pm}$ in \texttt{M2} equals $a^\mathsmaller{+}$ if $\vartheta\leq \pi/2$ and  $a^\mathsmaller{-}$ otherwise.}
\label{tab_t_crit} 
\end{table}

\begin{figure}[h!]
  \centering
  \small
  \vspace{5mm}
  \raisebox{23mm}{$\sepsdtilde$}\hspace{2mm}
    \begin{overpic}[width=10cm]{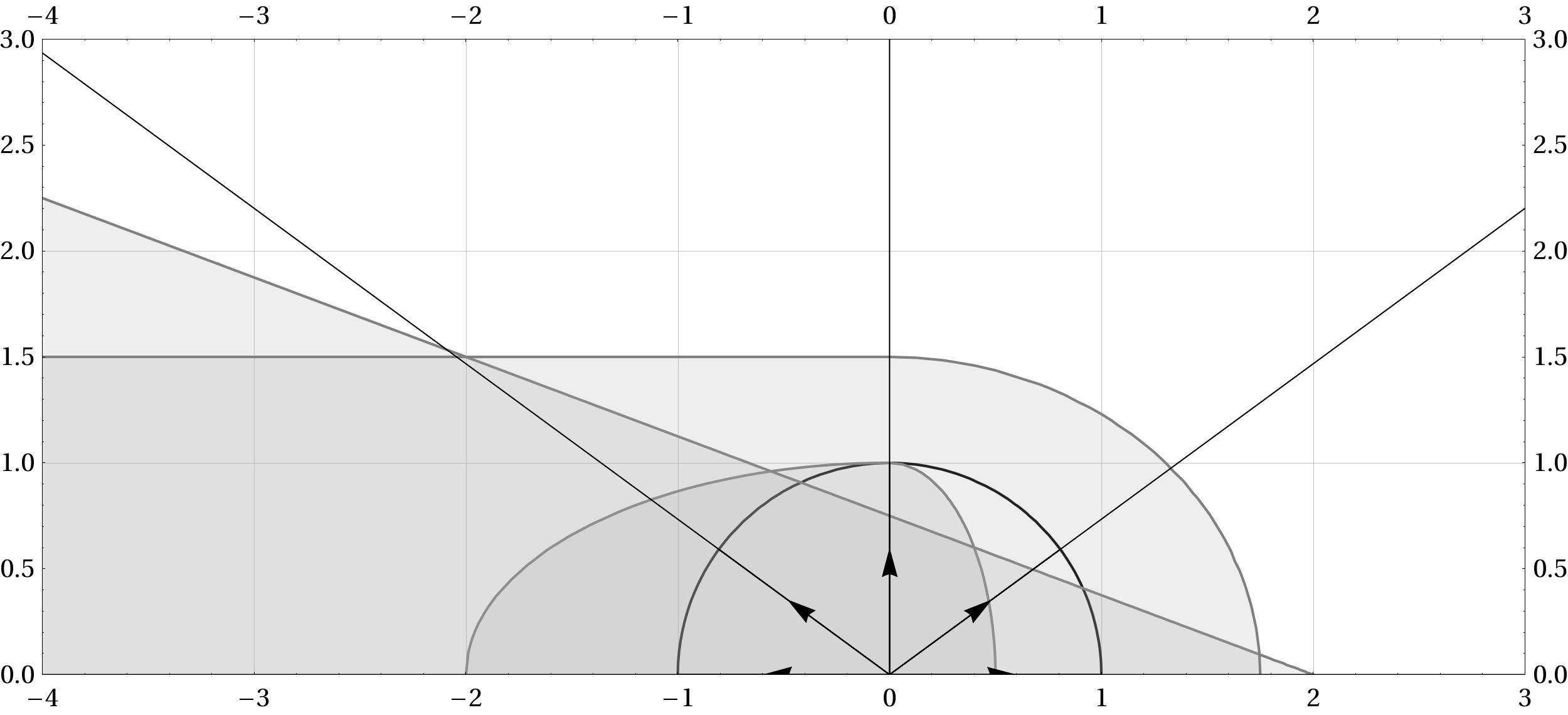}
      \put(65,14){$\Rm_{\texttt{1}}$}
      \put(30,12){$\Rm_{\texttt{2}}$}
      \put(21,27){$\Rm_{\texttt{3}}$ and $\Rm_{\texttt{4}}$}
      \put(70,20){$\Rm_{\texttt{5}}$}
      \put(88,4){$\vartheta = 0$}
      \put(80,29){$\vartheta = \pi/4$}
      \put(58,40){$\vartheta = \pi/2$}
      \put(14,36){$\vartheta = 3\pi/4$}
      \put(4,4){$\vartheta = \pi$}
    \end{overpic} 
  \\
  \hspace{19mm} $\sepsvtilde$
  \\[5mm]
  \renewcommand{\arraystretch}{1.2}
  \fcolorbox{black}{white}{
    \begin{minipage}{11.5cm}
      \footnotesize
      initial elastic domain: \quad \includegraphics[scale=0.05,trim=1000 100 0 0, clip]{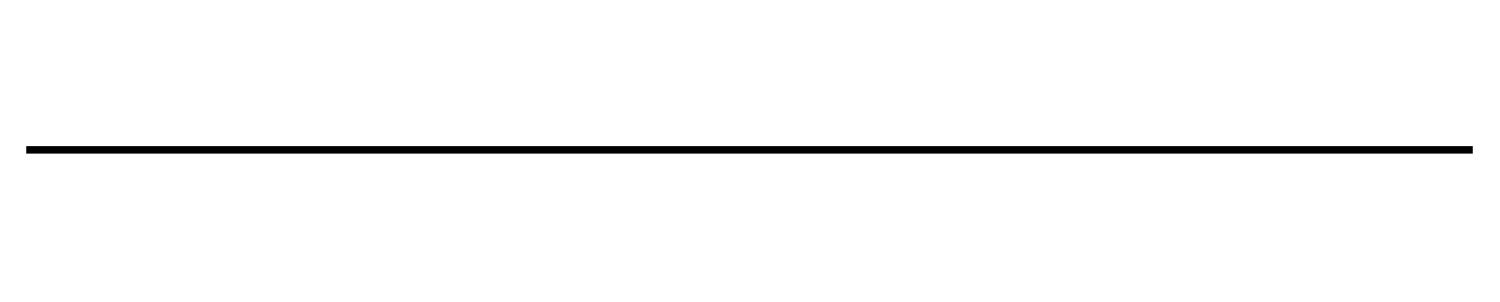}
      \\[1ex]
      damaging responses for different loading directions:
      \\
      \includegraphics[scale=0.05,trim=1000 100 0 0, clip]{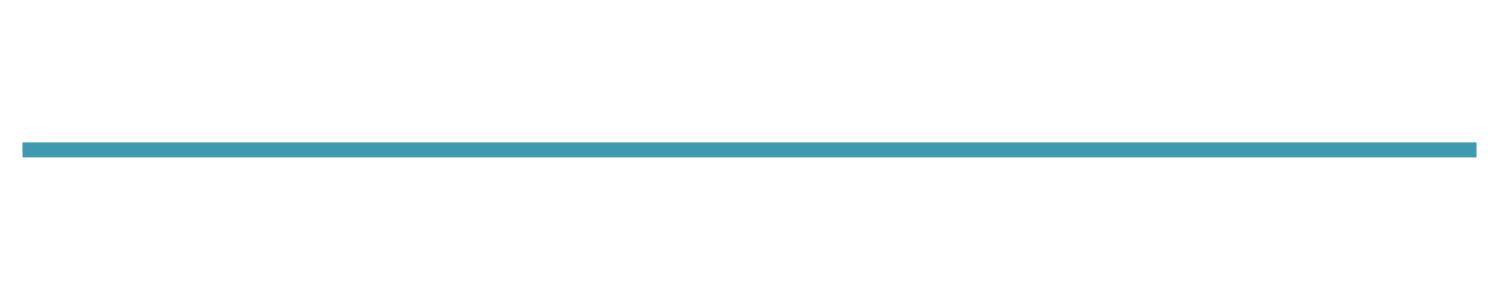} $ \vartheta = 0$,
      \quad 
      \includegraphics[scale=0.05,trim=1000 100 0 0, clip]{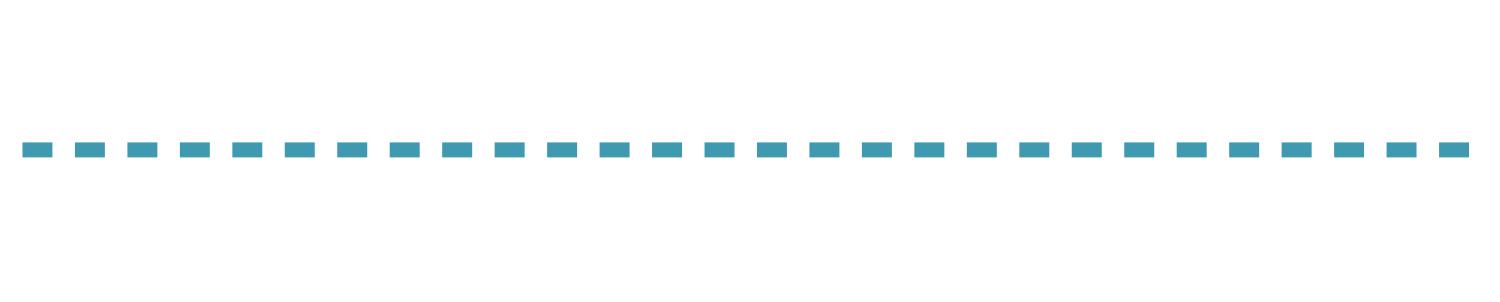} $ \vartheta = \pi/4$,
      \quad 
      \includegraphics[scale=0.05,trim=1000 100 0 0, clip]{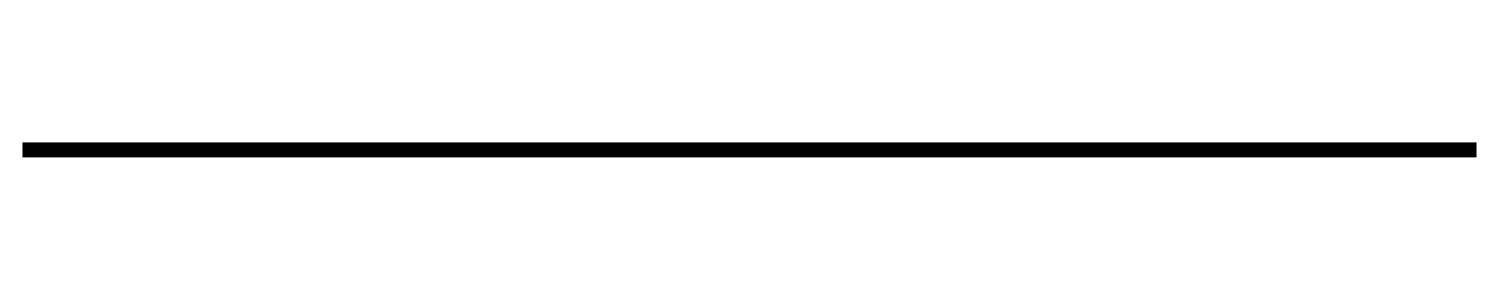} $ \vartheta = \pi/2$,
      \quad 
      \includegraphics[scale=0.05,trim=1000 100 0 0, clip]{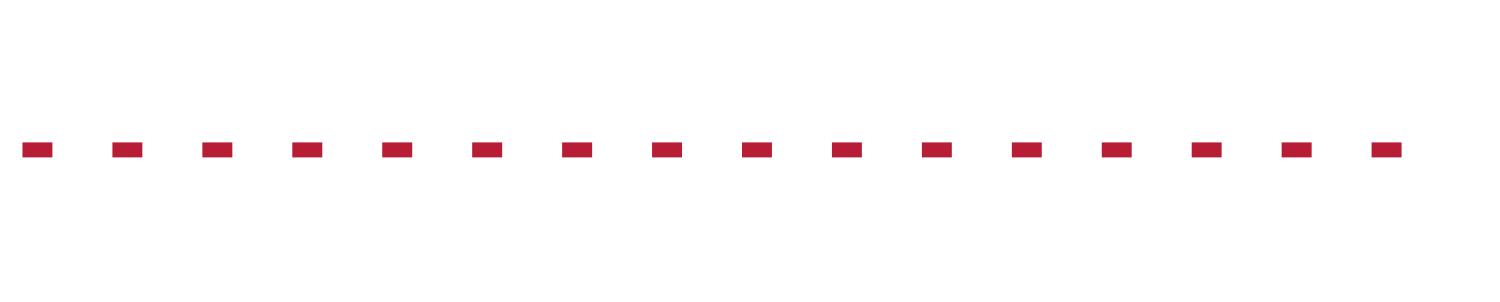} $ \vartheta = 3\pi/4$,
      \quad 
      \includegraphics[scale=0.05,trim=1000 100 0 0, clip]{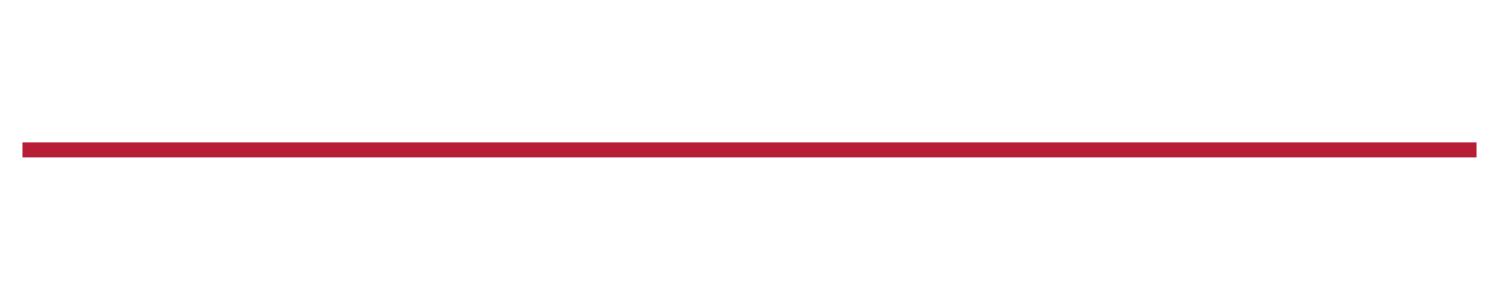} $ \vartheta = \pi$
    \end{minipage}
  }
  \caption{Elastic domains for all considered models in the $\sepsvtilde$--\thinspace$\sepsdtilde$ half-plane and proportional loading paths.}
  \label{fig_domains-HR}
\end{figure}

Relations linking the time parameter $t$ to the phase-field variable~$\sdam$, the loading angle~$\vartheta$, and the constitutive parameters during the damaging stage are reported in \Tab{tab_t_crit} for all considered models. These relations allow us to explicitly derive the corresponding homogeneous responses. \Fig{fig_domains-HR} illustrates the elastic domains in the $\sepsvtilde$--\thinspace$\sepsdtilde$ strain half-plane  at $\sdam = 0$ for all considered models and loading directions. The corresponding analytical homogeneous responses are shown in \Fig{fig_HR}.

Model \texttt{M1} exhibits the standard damaging response described in detail in~\citet{Marigo2016}, with symmetric behavior representing isotropic expansion/contraction and tension/compression. Compared to \texttt{M1}, model \texttt{M2} instead entails a non-symmetric behavior in the $\sepsvtilde$--\thinspace$\sstrhtilde$ half-plane. The non-symmetry is also reflected in the $\sepsdtilde$--\thinspace$\sstrdtilde$ half-plane, where the response in uniaxial tension differs from that in uniaxial compression.

Unlike the previous models, the mechanical response of \texttt{M3} and \texttt{M4} for $\vartheta = \pi$ never reaches the elastic limit. It is interesting to note that, although both models share the same elastic domain, no bulk modulus degradation occurs in \texttt{M4}, as is evident from the mechanical responses in the $\sepsvtilde$--\thinspace$\sstrhtilde$ half-plane. The volumetric response thus remains elastic despite the evolution of damage. In the $\sepsdtilde$--\thinspace$\sstrdtilde$ half-plane, instead, the mechanical responses are exactly the same. 

Model \texttt{M5} exhibits a non-symmetric degradation behavior in the $\sepsvtilde$--\thinspace$\sstrhtilde$ half-plane, qualitatively similar to \texttt{M3} for positive volumetric strains and to \texttt{M4} for negative volumetric strains. Moreover, in the $\sepsvtilde$--\thinspace$\sstrhtilde$ half-plane, the mechanical response is unaffected by the loading direction for negative volumetric strains, with the exception of~$\vartheta=\pi$. The model proposed and investigated in~\citet{Zolesi2024} appears to coincide with \texttt{M5}. However, in~\citet{Zolesi2024}, the positive volumetric limit strain (or hydrostatic limit stress) and the deviatoric limit strain (or deviatoric limit  stress), can be tuned separately only by choosing two different degradation functions for the volumetric and deviatoric energy parts. In \texttt{M5}, instead, the degradation behavior is completely separate from the shape of the elastic domain, thereby allowing them to be tuned~independently.

\Fig{subfig_EnDisH} shows the dissipated work density~$\disswork$ plotted against the phase-field variable. For homogeneous responses, the dissipated work density is given by 
$
\disswork\bra{\teps,\sdam;t} \coloneqq \intargt{\disspot(\sdamdt,\vect{0};\teps,\sdam,\vect{0})}.
$
As expected from the state-dependent nature of the dissipation potential, the dissipated work is independent of the loading direction only for model~\texttt{M1}. For all the other models, the dissipated work varies with the loading direction, indicating that the energy dissipated during the evolution may differ from one loading mode to another. This feature suggests that the proposed framework can potentially accommodate a mode-dependent fracture toughness. It is also worth noting that, as expected, no energy is dissipated by models~\texttt{M3}--\texttt{M5} along the loading direction $\vartheta=\pi$, since the corresponding strength surface is never reached and damage is therefore not activated. Finally, an interesting feature of model~\texttt{M5} is that the loading directions $\vartheta=\pi/2$ and $\vartheta=3\pi/4$ (actually any load direction $\pi/2 <= \vartheta < \pi$) lead to the same dissipated work.

\begin{landscape}

\newlength{\myFigLen}
\setlength{\myFigLen}{0.15\linewidth}

\newcommand{\mySubFig}[2]{
  \begin{minipage}{0.17\linewidth}
    \centering
    \includegraphics[width=0.9\linewidth]{#1}
    \\ \hspace{0mm} #2
  \end{minipage}
}

\begin{figure}[h!]
  \centering
  \small
    \begin{subfigure}{\linewidth}
      \begin{tabular}{p{2mm}ccccc}
        & \texttt{M1} & \texttt{M2} & \texttt{M3} & \texttt{M4} & \texttt{M5} \\[5mm]
          \rotatebox{90}{$\sstrhtilde$} 
          & \mySubFig{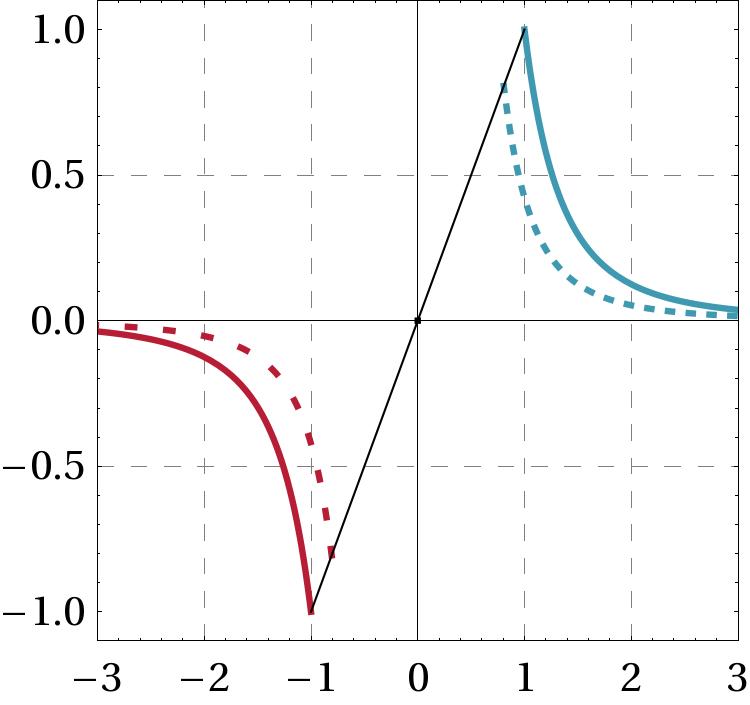}{$\sepsvtilde$}
          & \mySubFig{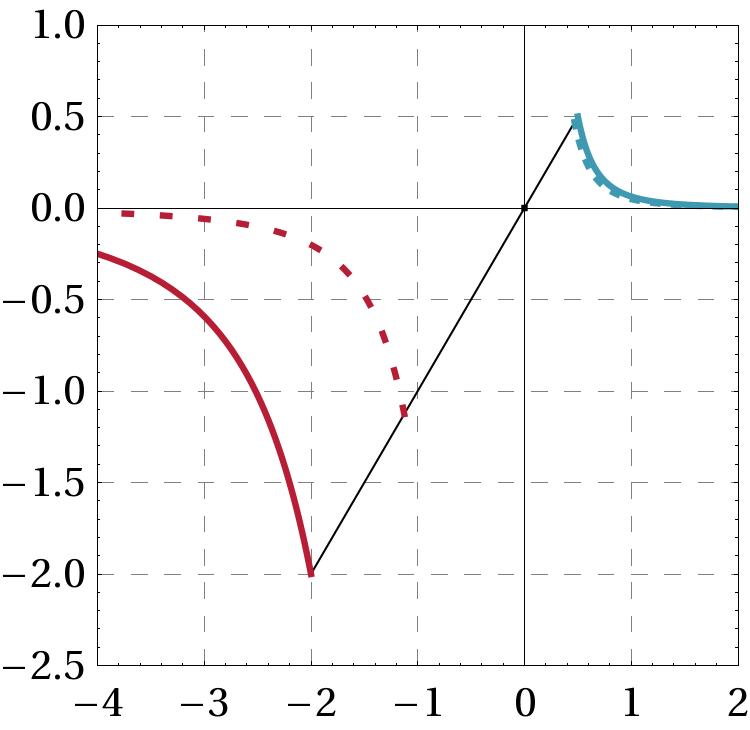}{$\sepsvtilde$}
          & \mySubFig{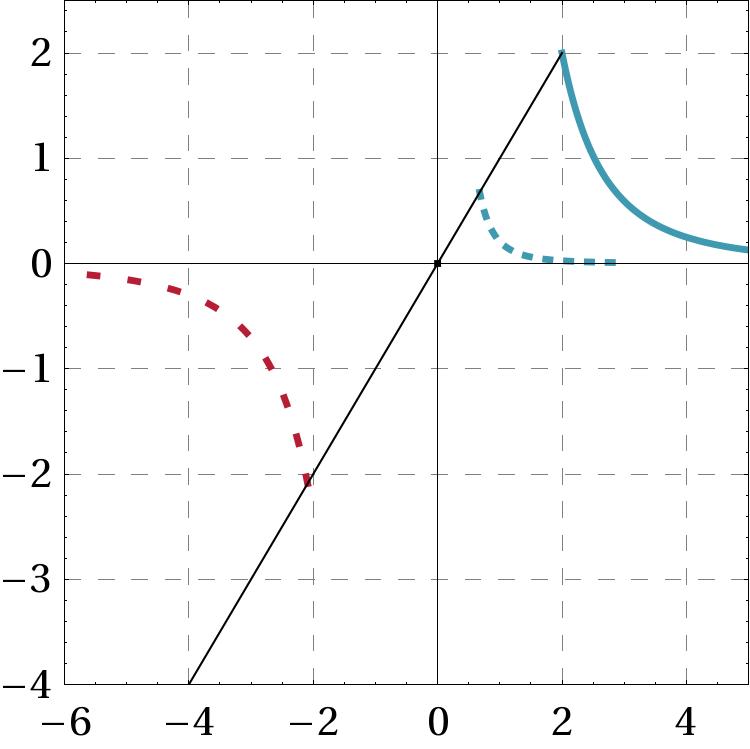}{$\sepsvtilde$}
          & \mySubFig{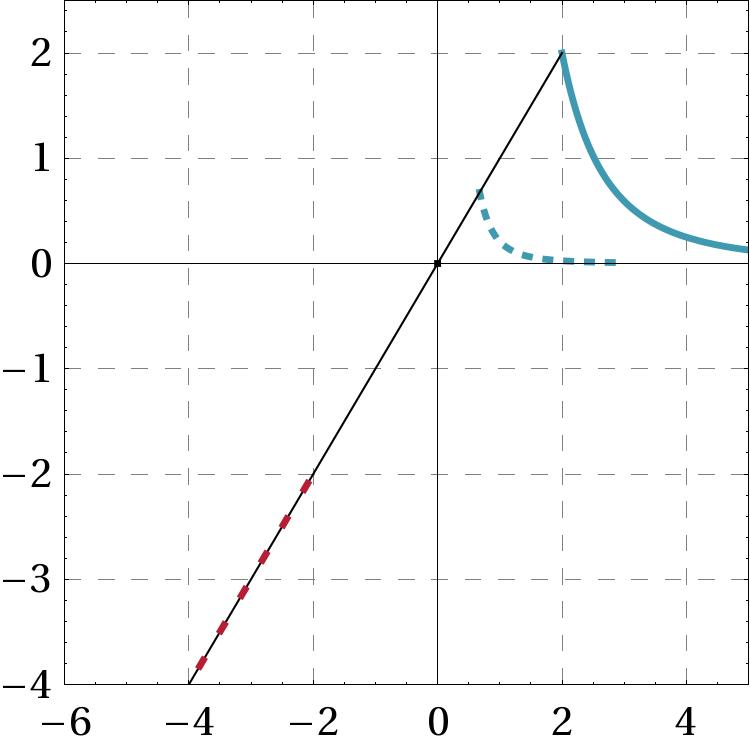}{$\sepsvtilde$}
          & \mySubFig{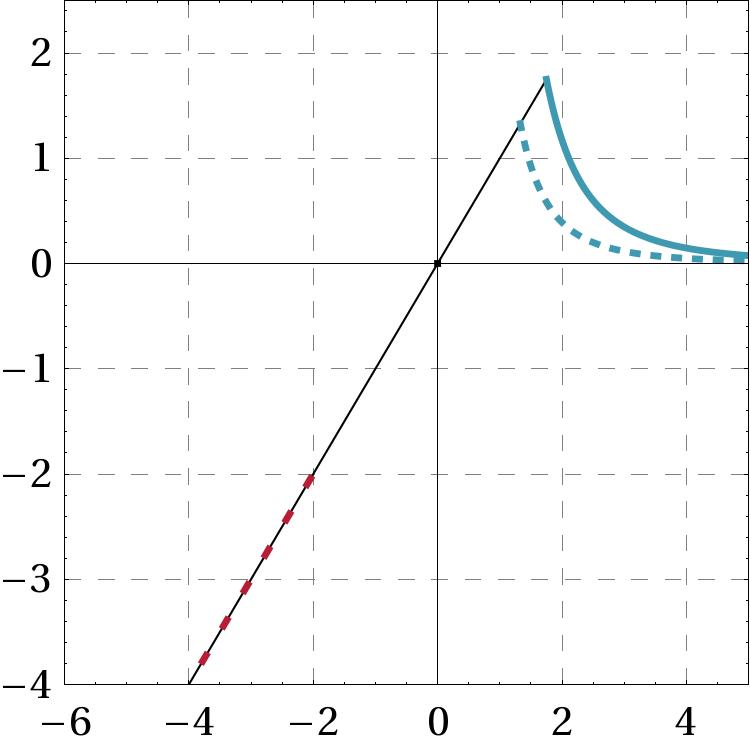}{$\sepsvtilde$}
        \end{tabular}
      \caption{}
      \label{subfig_Mv} 
    \end{subfigure}  
    \\[5mm]
    \begin{subfigure}{\linewidth}
      \begin{tabular}{p{2mm}ccccc}
          \rotatebox{90}{$\sstrdtilde$}
          & \mySubFig{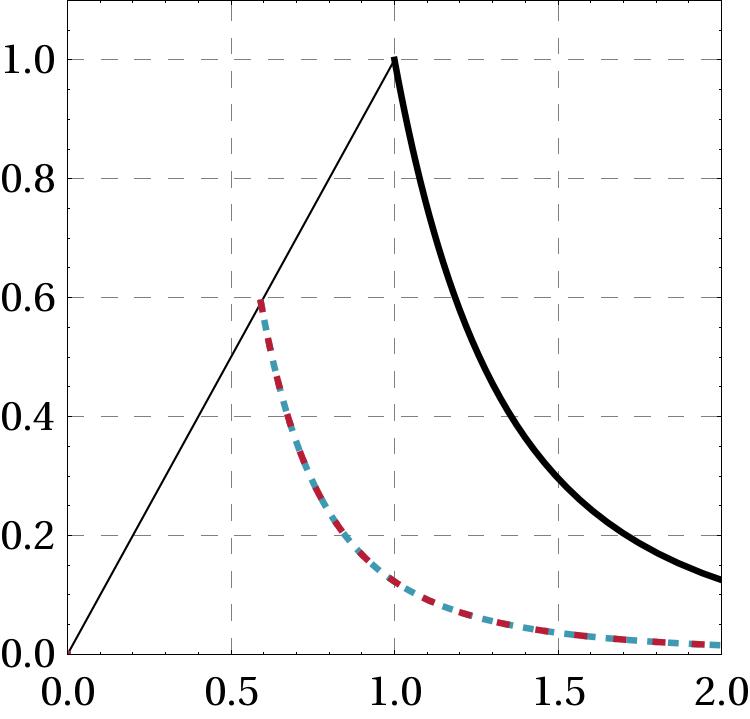}{$\sepsdtilde$}
          & \mySubFig{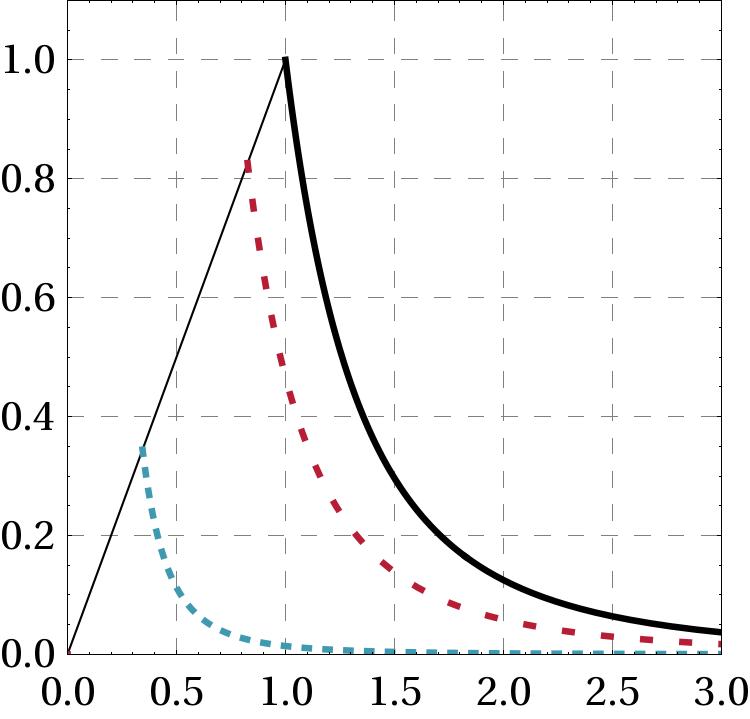}{$\sepsdtilde$}
          & \mySubFig{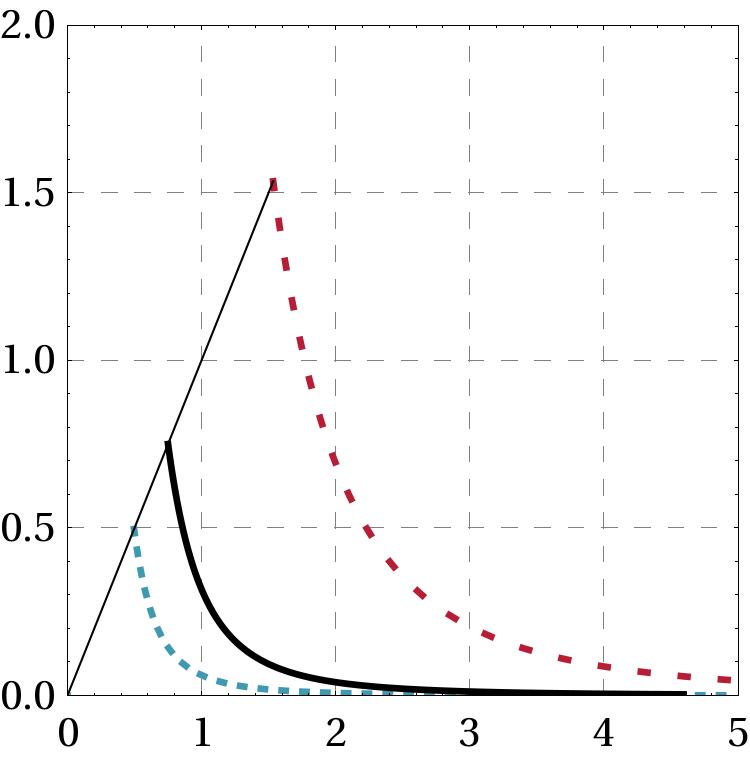}{$\sepsdtilde$}
          & \mySubFig{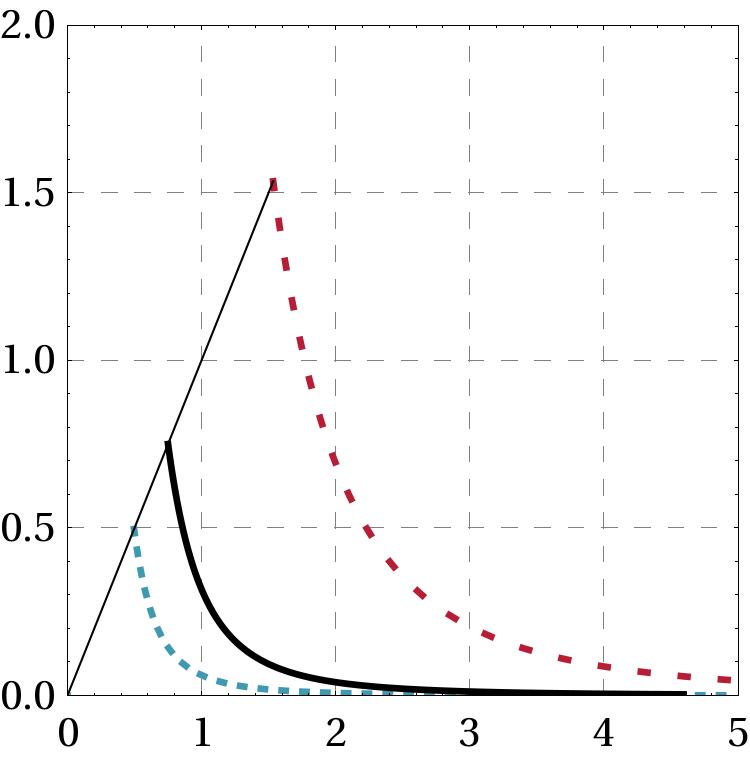}{$\sepsdtilde$}
          & \mySubFig{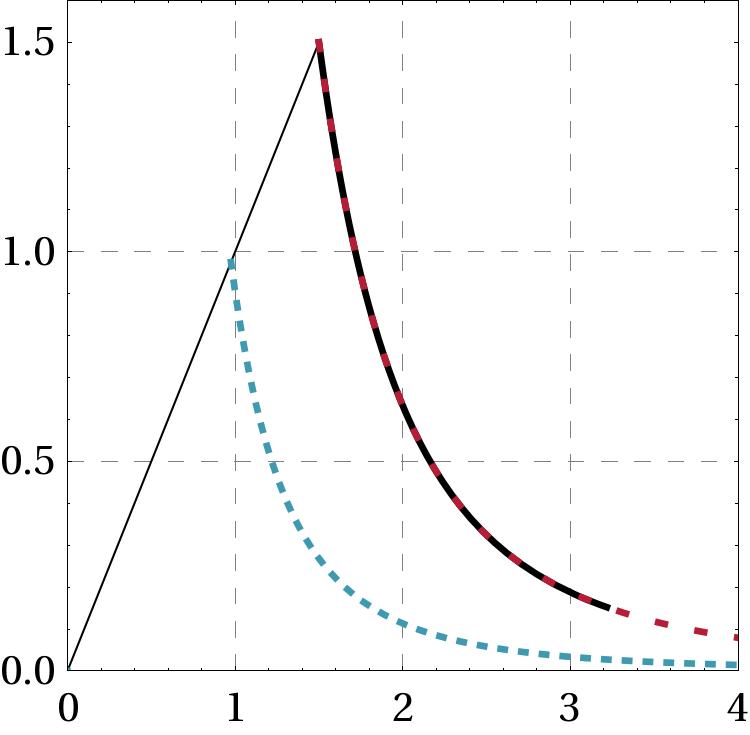}{$\sepsdtilde$}
      \end{tabular}
      \caption{}
      \label{subfig_Md} 
    \end{subfigure}  
    \\[5mm]
    \begin{subfigure}{\linewidth}
      \begin{tabular}{p{2mm}ccccc}
          $\disswork$
          & \mySubFig{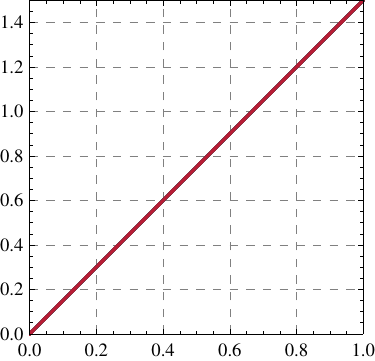}{$\sdam$}
          & \mySubFig{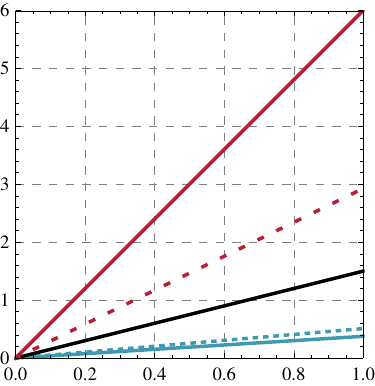}{$\sdam$}
          & \mySubFig{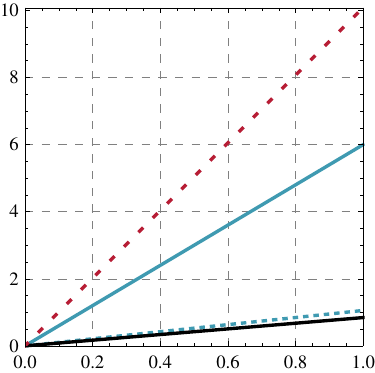}{$\sdam$}
          & \mySubFig{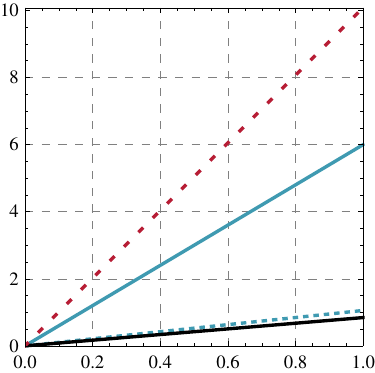}{$\sdam$}
          & \mySubFig{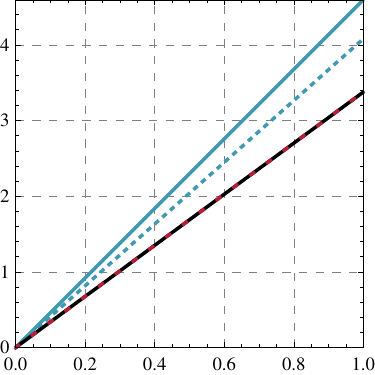}{$\sdam$}
      \end{tabular}
      \caption{}
      \label{subfig_EnDisH} 
    \end{subfigure}  
  \caption{
    Homogeneous responses for the proportional loading directions $\vartheta \in \cbra{0,\frac{\pi}{4},\frac{\pi}{2},\frac{3\pi}{4},\pi}$ of \Fig{fig_domains-HR} and for all considered models: 
    (\subref{subfig_Mv}) and (\subref{subfig_Md}) stress--strain responses. The initial elastic response is represented by a thin gray line;
    (\subref{subfig_EnDisH}) dissipated work density plotted against the phase-field variable.
    }
  \label{fig_HR} 
\end{figure}

\end{landscape}

\subsubsection{Non-proportional loading}
\label{sec_HR_NPL} 

To highlight the effects of the path-dependent dissipated work arising from a state-dependent dissipation potential, we compare the proportional loading path associated with a pure shear test ($\vartheta=\pi/2$) with two non-proportional loading paths. \Fig{fig_NPHR} shows the three loading paths in the $\sepsvtilde$--\thinspace$\sepsdtilde$ half-plane. They all start from the unloaded state $A$ and terminate at the same final state $B$, given by $(\sepsvtilde,\sepsdtilde)=(0,3)$ and corresponding to $t_{\l{B}}=3\,t_{\rm c}$, where $t_{\rm c}=0.197$ is the critical value of the loading parameter for model~\texttt{M1} under proportional pure shear loading ($\vartheta=\pi/2$), see \Tab{tab_crit_t}.
It is worth emphasizing that the loading path does not affect the strength criterion itself, but only the subsequent dissipative evolution.

As reported in \Tab{tab_par_mod}, for the present non-proportional loading analyses, the models are calibrated to share the same critical value along the proportional loading path by setting $b=1$ in the fracture perturbation function~$f$. The corresponding initial elastic domains are also presented in \Fig{fig_NPHR}.

The two strain functions $\vect{\epsilon}$ in \eqref{eq_bc} associated to the two non-proportional loading paths are
\begin{subequations}
  \begin{align}
    \vect{\epsilon}_1(t) &= \vect{\epsilon}_{\rm np}(-\beta(t)), &\hspace{-30mm} \text{(path 1)} \hspace{30mm} 
    \\
    \vect{\epsilon}_2(t) &= \vect{\epsilon}_{\rm np}(\beta(t)),  &\hspace{-30mm} \text{(path 2)} \hspace{30mm} 
  \end{align}
\end{subequations}
where
$\vect{\epsilon}_{\rm np}$ is the auxiliary function provided in \eqref{def_NPload} 
and $\beta(t)=\pi\,t/(3\,t_{\rm c})$. With this parametrization, the normalized arc-length is the same for all three loading paths and given by $\varsigma=t/(3t_c)$. Accordingly, in the $\sepsvtilde$--\thinspace$\sepsdtilde$ half-plane, the two non-proportional paths can be written as (\Fig{fig_NPHR})
\begin{subequations}
  \begin{align}
    &\text{path 1:} \quad
      \begin{cases}
        \sepsvtilde = (3/2) \, \sin(-\beta(t)), \\
        \sepsdtilde = (3/2) \bra{1 - \cos(\beta(t))},
      \end{cases}
    \\[1ex]
    &\text{path 2:} \quad
      \begin{cases}
        \sepsvtilde = (3/2) \, \sin(\beta(t)), \\
        \sepsdtilde = (3/2) \bra{1 - \cos(\beta(t))}.
      \end{cases}
  \end{align}
\end{subequations} 

\begin{figure}[h!]
  \centering
  \small
  \raisebox{26mm}{$\sepsdtilde$}\hspace{2mm}
  \begin{overpic}[width=10cm]{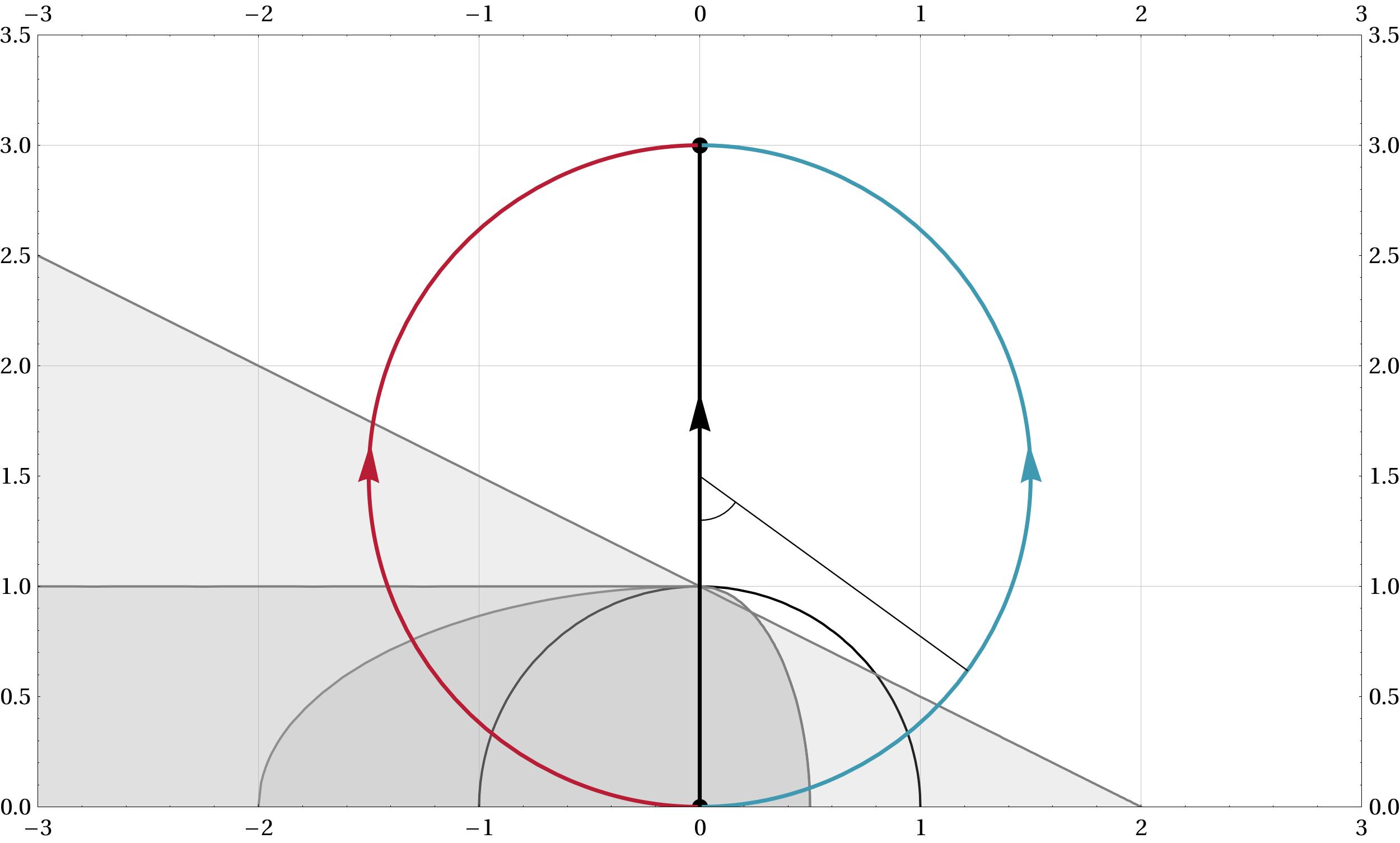}
    \put(51,20){$\beta$}
    \put(40,10){$\Rm_{\texttt{1}}$}
    \put(18,12){$\Rm_{\texttt{2}}$}
    \put(4,43){$\Rm_{\texttt{3}}$ and $\Rm_{\texttt{4}}$}
    \put(10,20){$\Rm_{\texttt{5}}$}
    \put(51,4){$A$}
    \put(51,52){$B$}
    \put(31,35){\textcolor{myLightRed}{path 1}}
    \put(74,35){\textcolor{myLightBlue}{path 2}}
    \put(51,35){
      $\substack{
        \text{\footnotesize prop.~load.}
        \\ 
        \text{\footnotesize  path ($\pi/2$)}
        }$}
  \end{overpic} 
  \\
  \hspace{10mm} $\sepsvtilde$
  \\[5mm]
   \fcolorbox{black}{white}{
      \footnotesize
      \begin{tabular*}{65mm}{@{}c@{\:}l@{}}
        \multicolumn{2}{l}{\hspace{-3mm} loading paths:} \\
        \includegraphics[scale=0.05,trim=1000 100 0 0, clip]{3-mod} 
          & \, proportional loading path ($\vartheta = \pi/2$) \\
        \includegraphics[scale=0.05,trim=1000 100 0 0, clip]{5-mod} 
          & \, non-proportional loading path 1 \\
        \includegraphics[scale=0.05,trim=1000 100 0 0, clip]{1-mod} 
          & \, non-proportional loading path 2
      \end{tabular*}\hspace{-5mm}
  }
  \caption{Examined proportional and non-proportional laoding paths from state $A$ to state $B$ together with the initial elastic domains for all considered models in the $\sepsvtilde$--\thinspace$\sepsdtilde$ half-plane.}
  \label{fig_NPHR} 
\end{figure}

Since the three loading paths considered share the same initial and final states, comparing their material response allows us to directly assess the influence of the loading path on the dissipated work. The result is shown in \Fig{fig_NPrespb}.
Specifically, \Fig{subfig_sa} shows the phase-field evolution as a function of the normalized path arc-length. Although the final phase-field values are nearly identical across all models and loading conditions, their evolution along the loading paths differs significantly.
Clearly, the phase-field evolution along paths 1 and 2 is exactly the same for the standard phase-field model \texttt{M1}.
Moreover, for models~\texttt{M2}--\texttt{M5}, the final phase-field value attained along path~2 is slightly larger than those obtained along path~1 and the proportional loading path. This difference results from the irreversibility condition imposed on the phase-field variable. Indeed, along path~2, $\sdam$ reaches its maximum value before the final state, at a point with $\beta=\hat{\beta}\,(<\pi)$ (\Fig{fig_NPHR}), and irreversibility prevents it from subsequently decreasing. Thus, $\sdam$ remains constant from $\hat{\beta}$ to the final state.

\Fig{subfig_sEn} reports the evolution of the free energy density~$\enfree$ (dashed curves) and the dissipated work density~$\endiss$ (solid curves) as functions of the normalized path arc-length. Several relevant features of the models emerge from these plots. In particular, although the final value of the phase-field variable is nearly the same for all models and loading conditions, the corresponding dissipated work may differ substantially. This clearly highlights the strong path dependence of the dissipation, which need not be \emph{directly} correlated with the phase-field evolution. For instance, for model~\texttt{M2}, the phase-field evolutions along path~1 and the proportional loading path are quite similar (\Fig{subfig_sa}), whereas the corresponding dissipated works are markedly different. Moreover, although the phase field evolves considerably faster along path~2 than along the other paths, the associated dissipated work is significantly lower. The opposite trend is observed along path~1 for models~\texttt{M3}--\texttt{M4}, where a slower phase-field evolution results in a substantially larger final dissipated work. These features are further highlighted in \Fig{subfig_aEnDiss}, where the dissipated work density is plotted directly against the phase-field variable.

For models~\texttt{M3}--\texttt{M4}, it is also worth noting that, along path~1, which involves negative volumetric strains, the free-energy density decreases much more rapidly during the phase-field evolution for model~\texttt{M3} than for model~\texttt{M4}. This difference is consistent with the distinct degradation functions entering the free-energy density~\eqref{def_enfree}. The same feature is reflected in the $\sepsvtilde$--\thinspace$\sstrhtilde$ responses shown in \Fig{subfig_v}: along path~1, model~\texttt{M4} exhibits no degradation of the volumetric elastic response, similarly to model~\texttt{M5}, consistent with the underlying degradation model~\FED. For completeness, the corresponding $\sepsdtilde$--\thinspace$\sstrdtilde$ responses are reported in \Fig{subfig_d}. These curves also show that all models exhibit the same response along the proportional, pure-shear loading path.

\begin{landscape}
\small

\newcommand{\mySubFigI}[1]{
  \begin{minipage}{0.17\linewidth}
    \centering
    \includegraphics[width=0.9\linewidth]{#1}
    \\ \hspace{0mm} $\varsigma$
  \end{minipage}
}

\newcommand{\mySubFigII}[1]{
  \begin{minipage}{0.17\linewidth}
    \centering
    \includegraphics[width=0.9\linewidth]{#1}
    \\ \hspace{0mm} $\sdam$
  \end{minipage}
}

\begin{figure}[h!]
  \centering
  \small
  \begin{subfigure}{\linewidth}
    \begin{tabular}{p{2mm}ccccc}
      & \texttt{M1} & \texttt{M2} & \texttt{M3} & \texttt{M4} & \texttt{M5} \\[5mm]
        \rotatebox{90}{$\sdam$} 
        & \mySubFigI{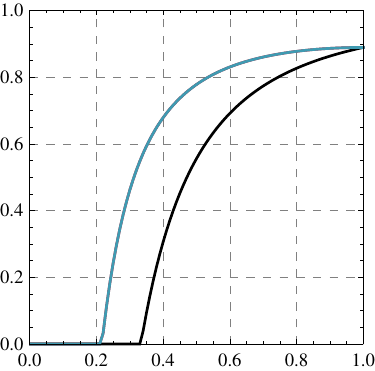}
        & \mySubFigI{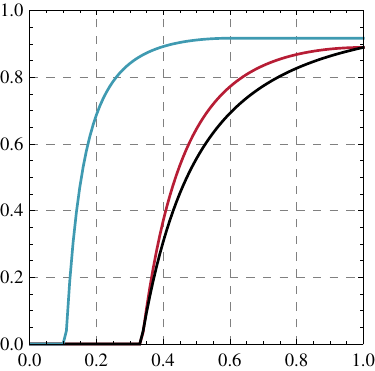}
        & \mySubFigI{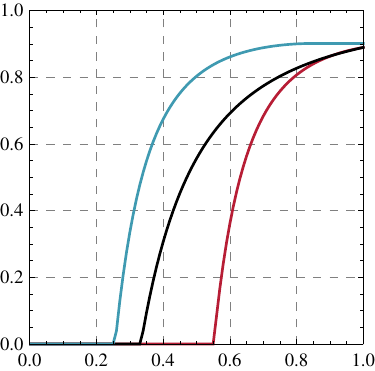}
        & \mySubFigI{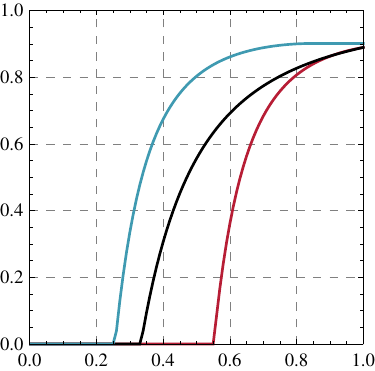}
        & \mySubFigI{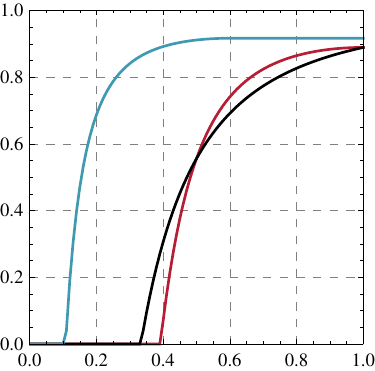}
      \end{tabular}
    \caption{}
    \label{subfig_sa} 
  \end{subfigure}  
  \\[5mm]
  \begin{subfigure}{\linewidth}
    \begin{tabular}{p{2mm}ccccc}
        \rotatebox{90}{$\enfree$, $\endiss$}
        & \mySubFigI{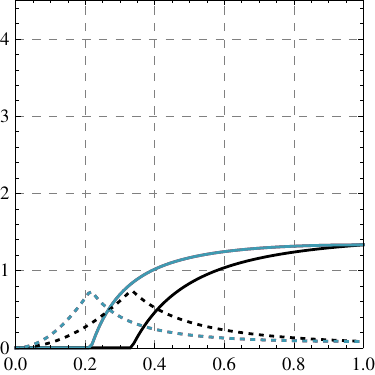}
        & \mySubFigI{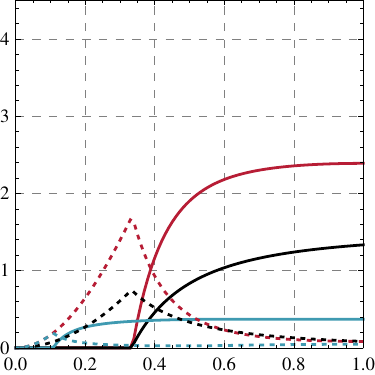}
        & \mySubFigI{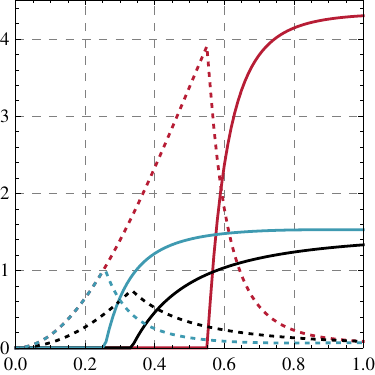}
        & \mySubFigI{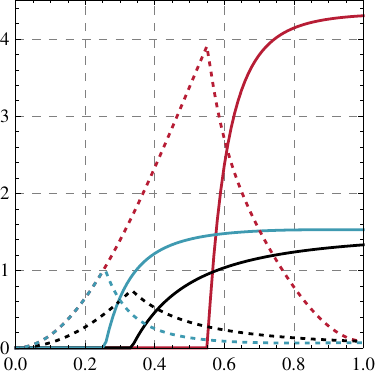}
        & \mySubFigI{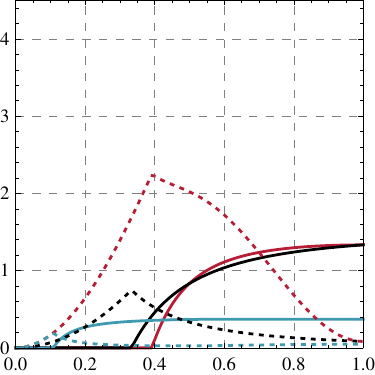}
      \end{tabular}
    \caption{}
    \label{subfig_sEn} 
  \end{subfigure}  
  \\[5mm]
  \begin{subfigure}{\linewidth}
    \begin{tabular}{p{2mm}ccccc}
        \rotatebox{90}{$\endiss$}
        & \mySubFigII{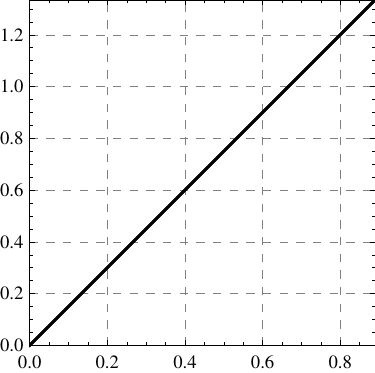}
        & \mySubFigII{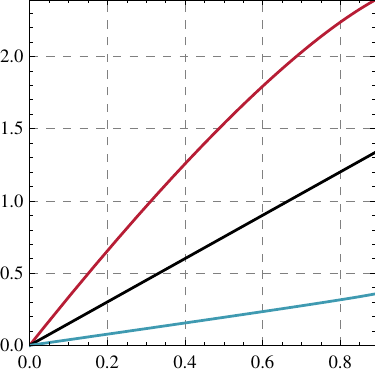}
        & \mySubFigII{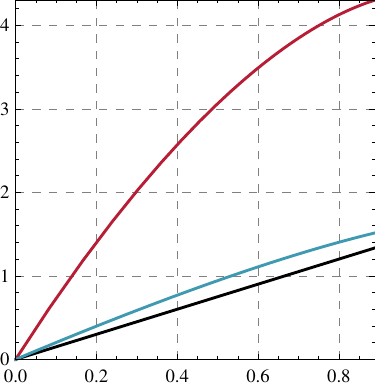}
        & \mySubFigII{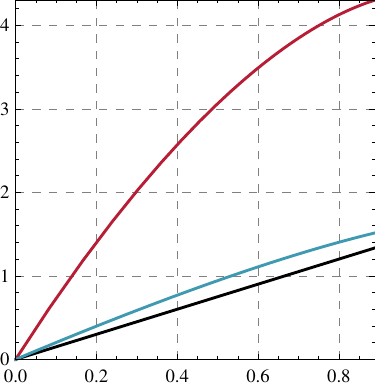}
        & \mySubFigII{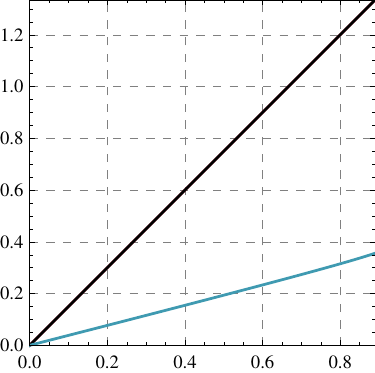}
    \end{tabular}
    \caption{}
    \label{subfig_aEnDiss} 
  \end{subfigure}  
  \caption{\emph{Continued on next page.}}
  \label{fig_NPrespa} 
\end{figure}
\end{landscape}

\begin{landscape}
\small

\newcommand{\mySubFigI}[1]{
  \begin{minipage}{0.17\linewidth}
    \centering
    \includegraphics[width=0.95\linewidth]{#1}
    \\ \hspace{0mm} $\sepsvtilde$
  \end{minipage}
}

\newcommand{\mySubFigII}[1]{
  \begin{minipage}{0.17\linewidth}
    \centering
    \includegraphics[width=0.95\linewidth]{#1}
    \\ \hspace{0mm} $\sepsdtilde$
  \end{minipage}
}

\begin{figure}[h!]
  \ContinuedFloat
  \centering
  \small
  \begin{subfigure}{\linewidth}
    \begin{tabular}{p{2mm}ccccc}
      & \texttt{M1} & \texttt{M2} & \texttt{M3} & \texttt{M4} & \texttt{M5} \\[5mm]
        \rotatebox{90}{$\sstrhtilde$} 
        & \mySubFigI{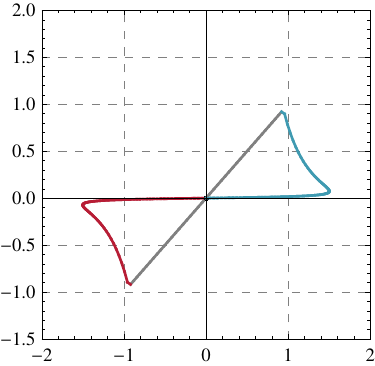}
        & \mySubFigI{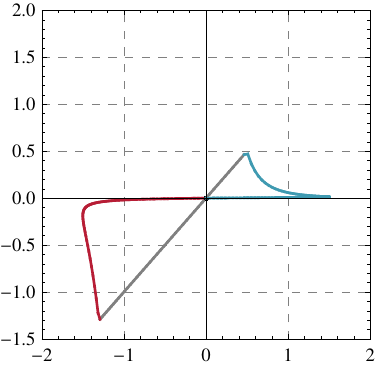}
        & \mySubFigI{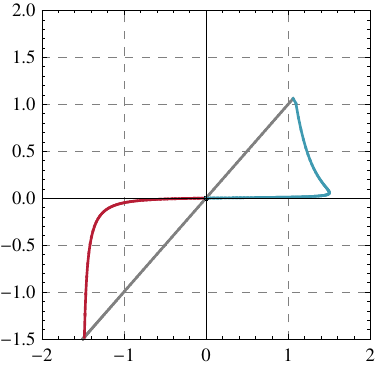}
        & \mySubFigI{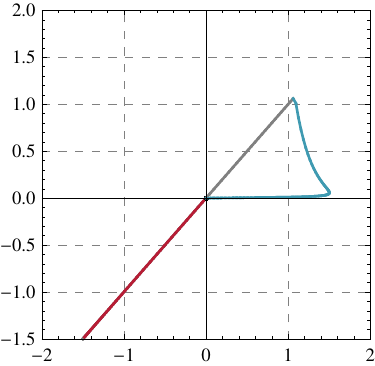}
        & \mySubFigI{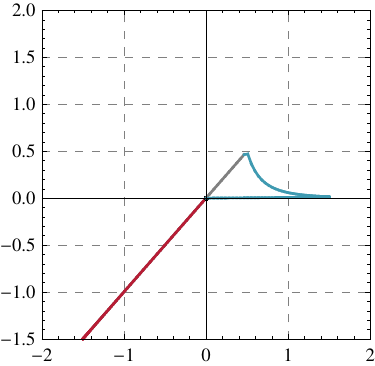}
      \end{tabular}
    \caption{}
    \label{subfig_v} 
  \end{subfigure}  
  \\[5mm]
  \begin{subfigure}{\linewidth}
    \begin{tabular}{p{2mm}ccccc}
        \rotatebox{90}{$\sstrdtilde$}
        & \mySubFigII{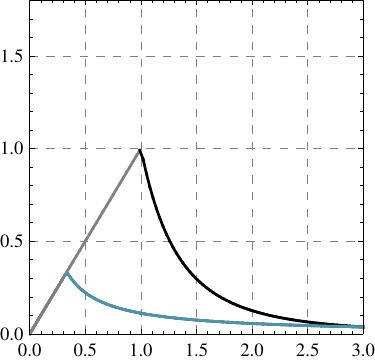}
        & \mySubFigII{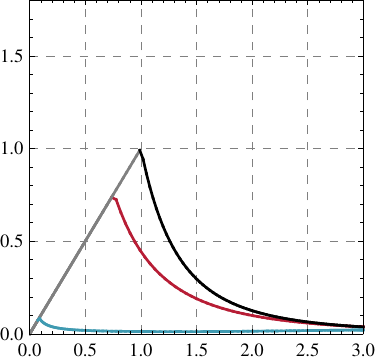}
        & \mySubFigII{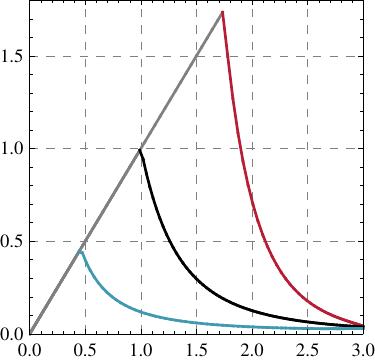}
        & \mySubFigII{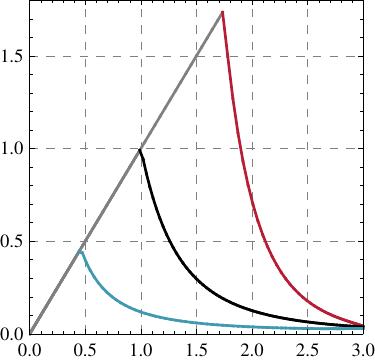}
        & \mySubFigII{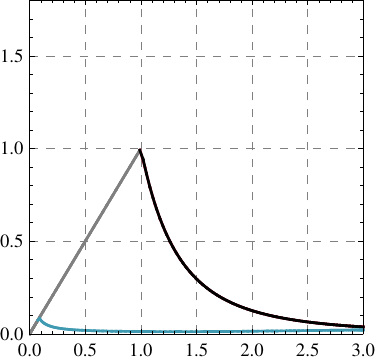}
      \end{tabular}
    \caption{}
    \label{subfig_d}
  \end{subfigure}  
  \caption{Homogeneous responses for the proportional and non-proportional laoding paths of \Fig{fig_NPHR} for all considered models: 
    (\subref{subfig_sa}) phase-field evolution plotted against the normalized path arc-length; 
    (\subref{subfig_sEn}) free energy density $\enfree$ (dashed curves) and dissipated work density $\endiss$ (solid curves) plotted against the normalized path arc-length;
    (\subref{subfig_aEnDiss}) dissipated work density plotted against the phase-field variable;
    (\subref{subfig_v}) and (\subref{subfig_d}) stress--strain responses. The initial elastic response is represented by a solid gray line ( \includegraphics[scale=0.1,trim=1250 25 0 0, clip]{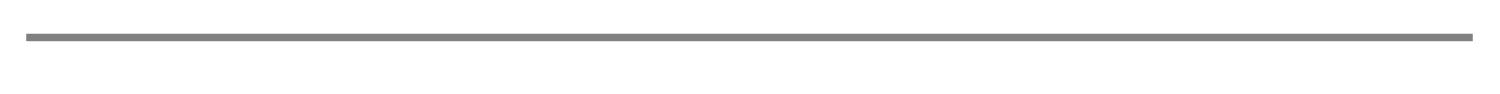}).
  }
\label{fig_NPrespb} 
\end{figure}
\end{landscape}

\subsection{Numerical multi-axial nucleation tests}
\label{sec_NT} 

\newlength{\mpw}
\setlength{\mpw}{30mm}
\newlength{\diskwidth}
\setlength{\diskwidth}{12mm}

\newcommand{\putDiskFig}[4]{
  \put(#1,#2){
    \makebox(0,0){
      \begin{minipage}{\mpw}
        \centering
        \footnotesize
        \vspace{-4mm}
        #3 \\
        \includegraphics[width=\diskwidth]{#4}
      \end{minipage}
    }}}
In this section, we perform numerical crack-nucleation tests for the biaxial disk problem using the numerical framework introduced in~\Sec{sec_num}. The specific purpose of these simulations is to assess the capability of the proposed model to predict fracture initiation consistently with the prescribed strength criterion, rather than to investigate the subsequent crack-propagation stage. Indeed, as discussed in~\citet[Sec.~5.1]{Vicentini2024}, once the boundary of the elastic domain is reached in this problem, crack nucleation is followed by an unstable propagation event. 
As discussed in~\Sec{sec_num}, the numerical treatment of such discontinuous evolutions requires particular care and lies beyond the scope of the present work. Both approximations for $\Gf$ in \Alg{algo_sol} were tested and found to capture the analytically predicted crack nucleation load. However, (\texttt{Approx-I}) exhibited numerical instabilities that prevented convergence immediately after crack nucleation. Accordingly, we employ (\texttt{Approx-E}) for the computational examples, as this approximation allowed the algorithm to converge without difficulty after nucleation.
Moreover, in addition to the displacement boundary conditions~\eqref{eq_bc}, we impose the Dirichlet condition $\sdam = 0$ on the whole boundary of the disk throughout the loading process.

The domain is discretized with standard triangular finite elements using an unstructured mesh. For models \texttt{M1}--\texttt{M3}, linear shape functions are considered with average mesh size $h_1 = \ell/5$. For models \texttt{M4}--\texttt{M5}, which adopt the \texttt{PED} model, quadratic shape functions are adopted for the displacement field with doubled average mesh size, namely $h_2 = 2 h_1 = 2\ell/5$, so as to keep the number of degrees of freedom approximately the same. This choice of shape functions is motivated by the fact that the \texttt{PED} model may exhibit volumetric locking problems, as examined in detail in~\citet{Alessi2020}, causing the damage localization bands to appear with excessive thickness, overestimating the fracture energy, and failing to predict correct crack-path geometries. These issues are mitigated by adopting selective reduced integration for the undamaged part of the elastic energy. Further implementation details can be found in~\citet{Alessi2020}.

The results of the considered models are reported in \Fig{fig_all_nuc}. In each model-specific subfigure, the prescribed elastic domain is represented by a gray region, while the red dots represent the numerically obtained strain states at damage initiation along the highlighted loading directions (black arrows and gray lines). For loading directions along which the strength surface is not reached, damage evolution is never observed. For each loading direction, we plot the phase-field variable at the time step corresponding to crack nucleation, whenever nucleation occurs. For loading directions that do not trigger nucleation, a zero-damage field is shown for completeness. This benchmark not only assesses nucleation behavior, but also offers insight into the incipient crack patterns associated with different loading conditions.

\begin{figure}[h!]
\centering
  \begin{subfigure}{\linewidth}
    \centering
    \small
    \raisebox{32mm}{$\sepsdtilde$}\hspace{2mm}
    \begin{overpic}[width=0.8\linewidth]{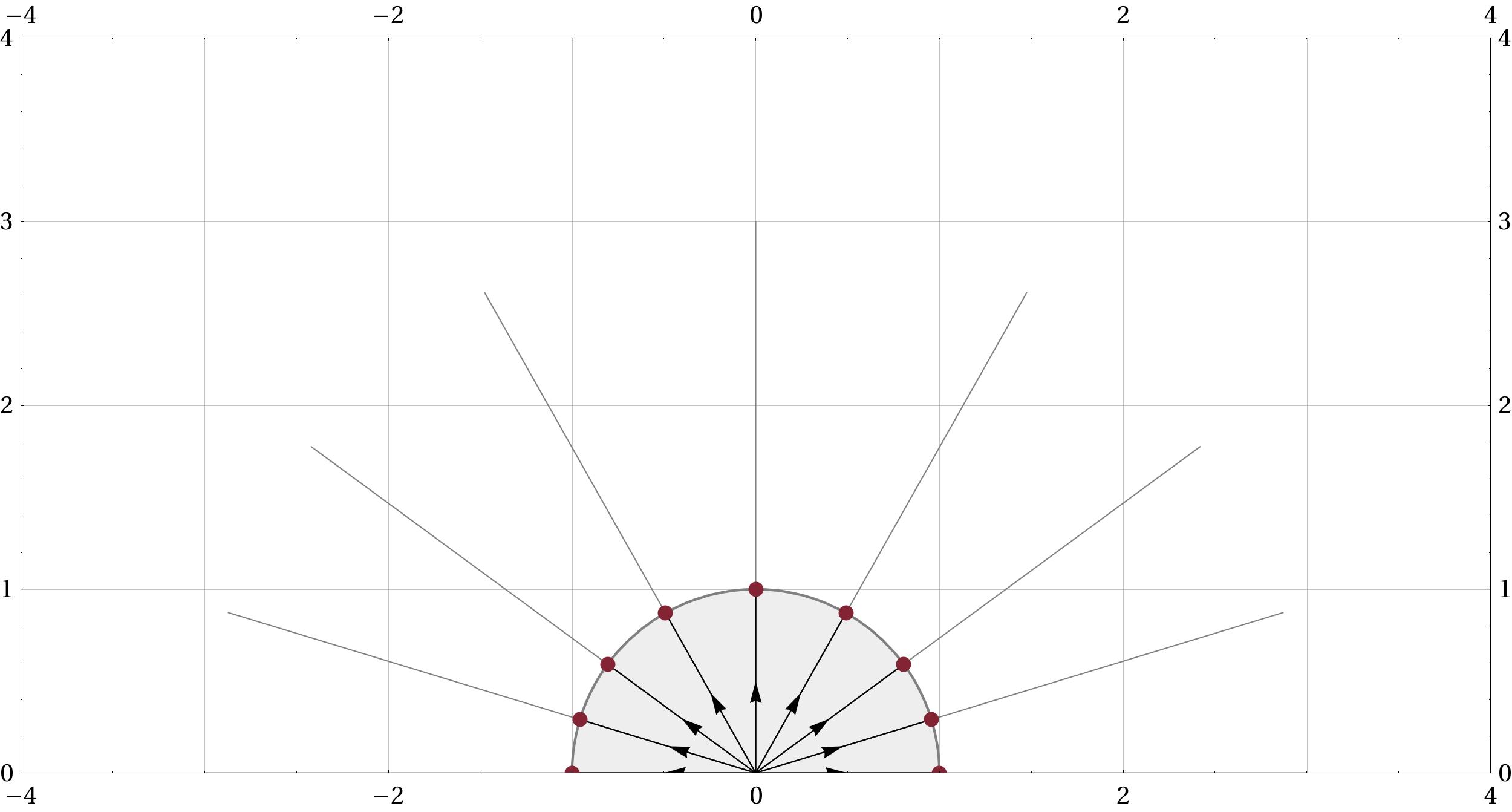}
      \putDiskFig{87}{2.5}{\hspace{15mm}$\theta = 0$}{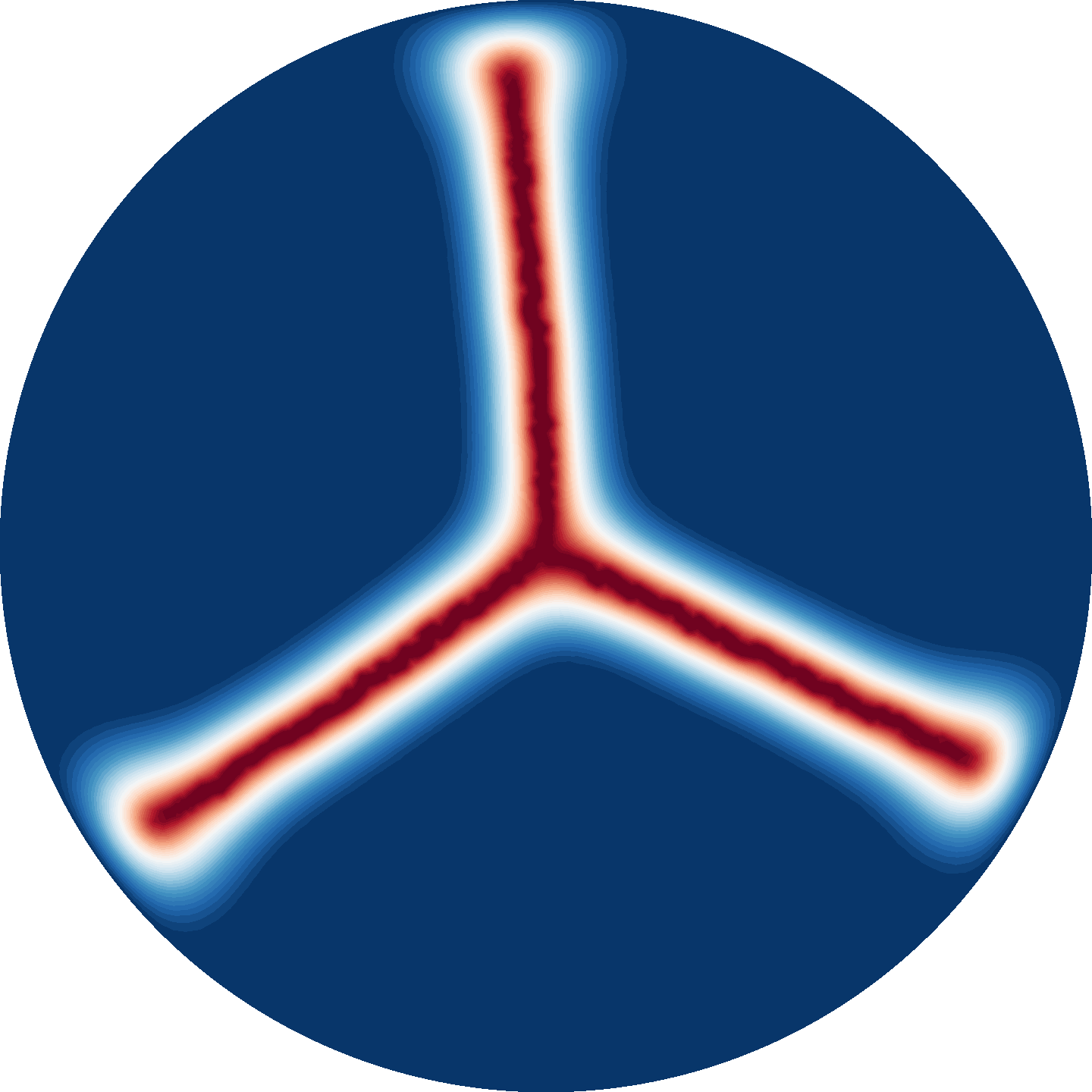}
      \putDiskFig{85}{13}{\hspace{8mm}$\theta = \pi/8$}{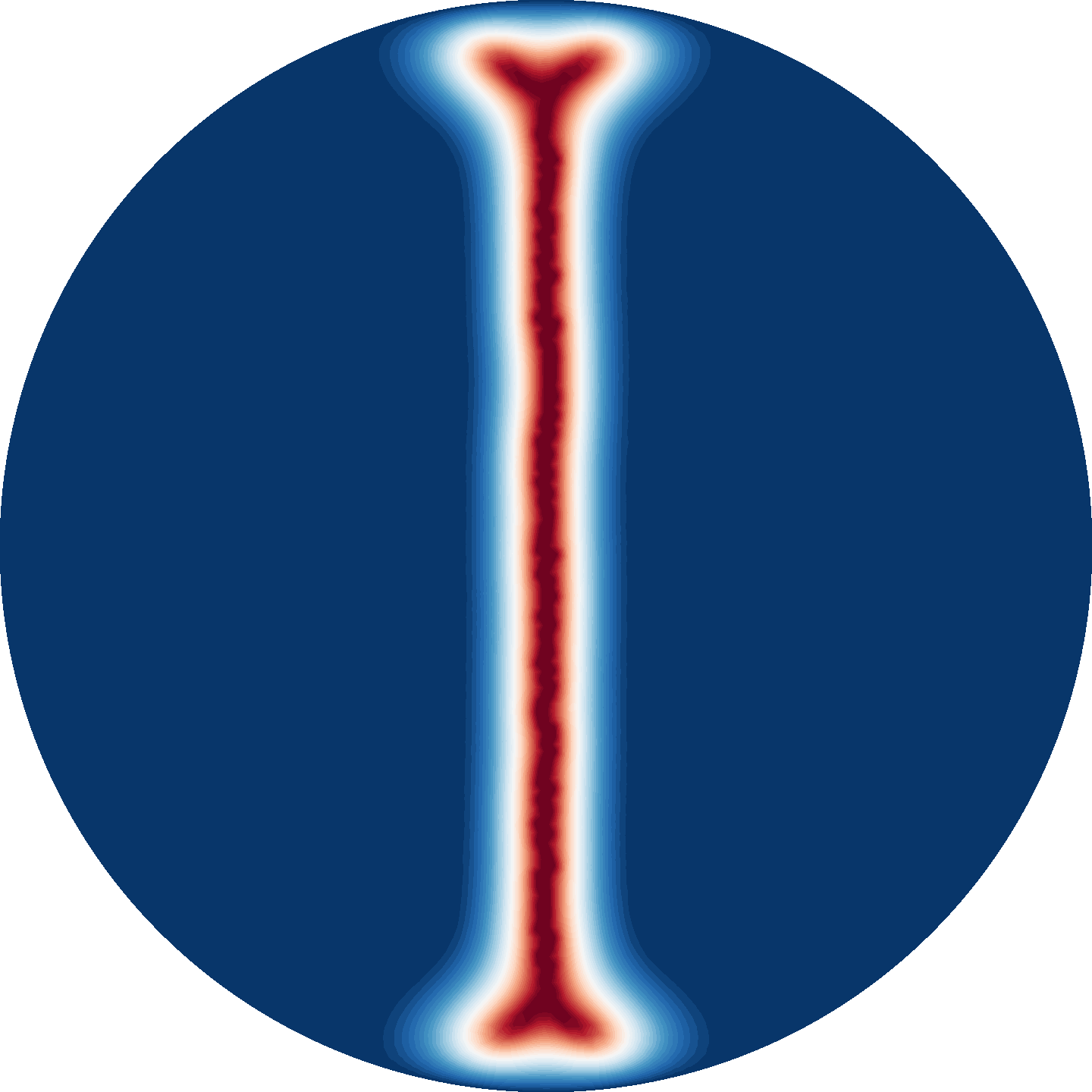}
      \putDiskFig{79}{24}{$\theta = \pi/4$}{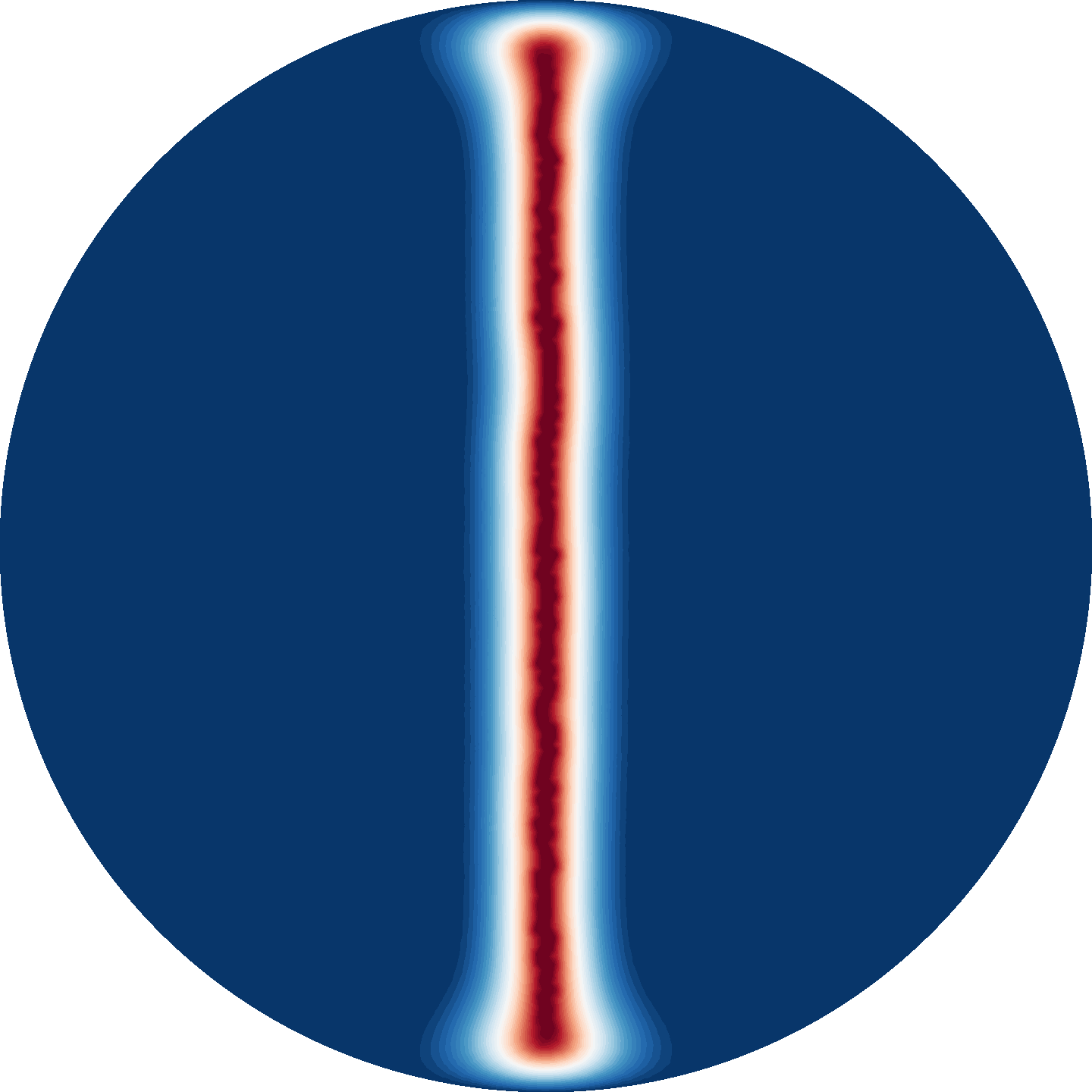}
      \putDiskFig{68}{34}{$\theta = 3\pi/8$}{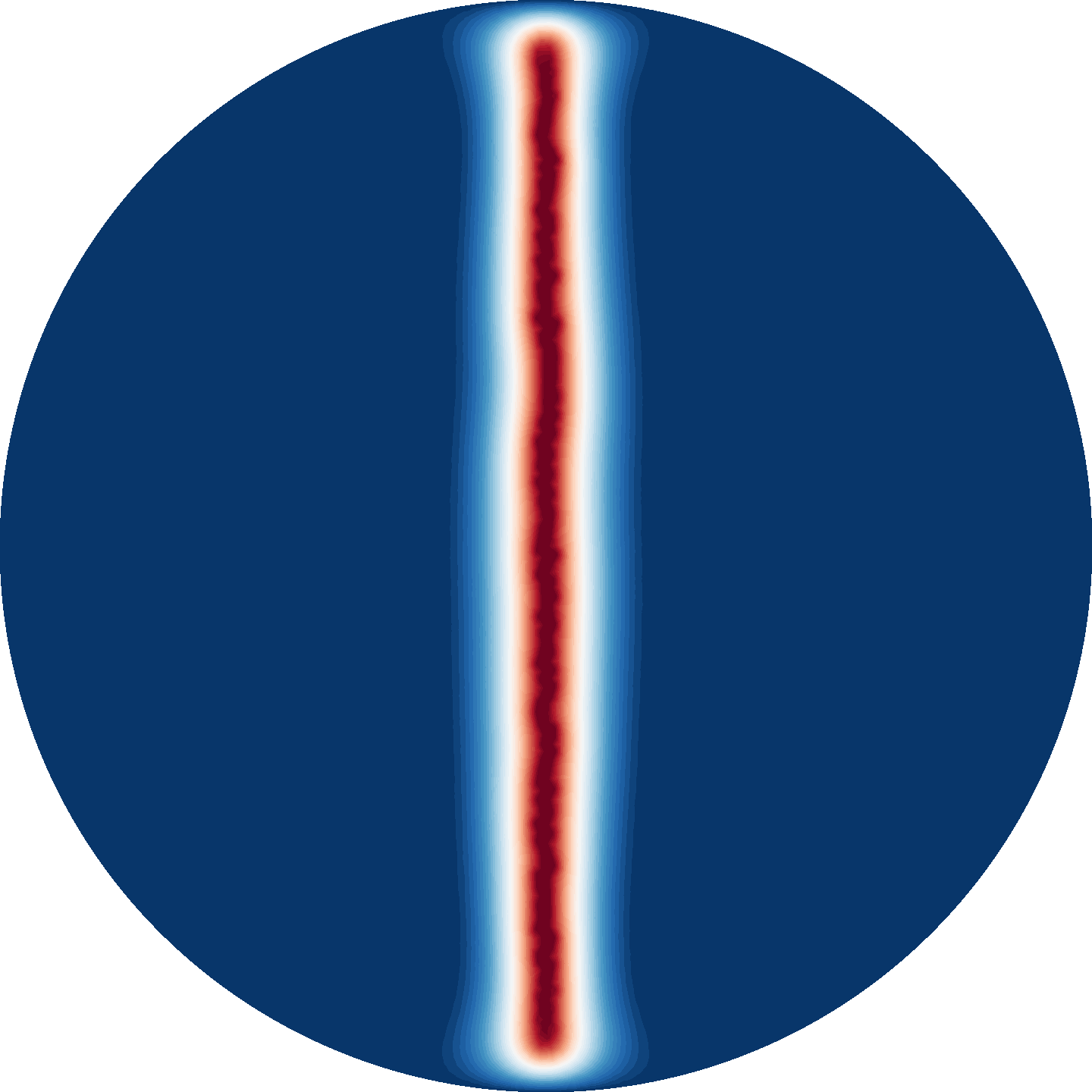}
      \putDiskFig{49}{38}{$\theta = \pi/2$}{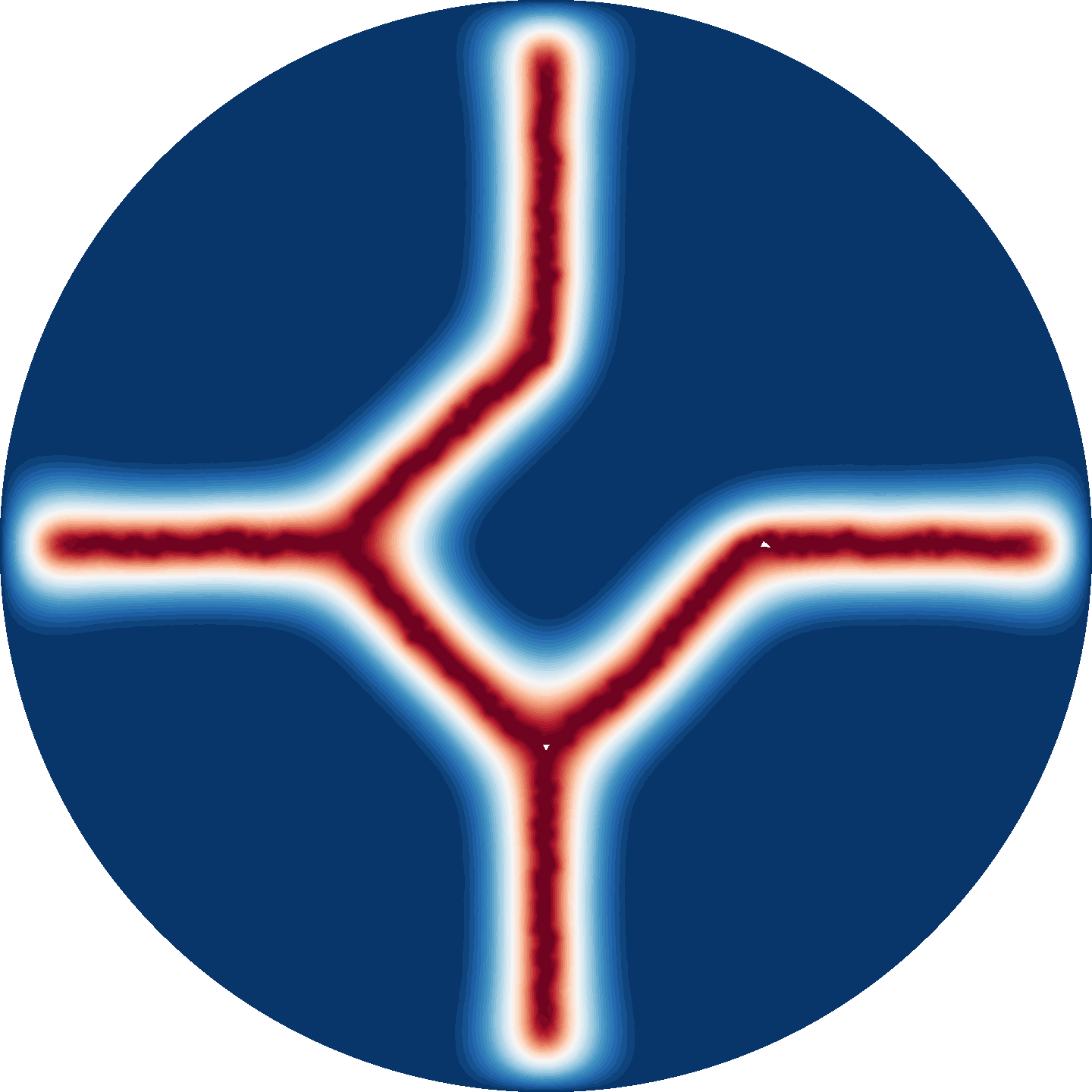}
      \putDiskFig{32}{34}{$\theta = 5\pi/8$}{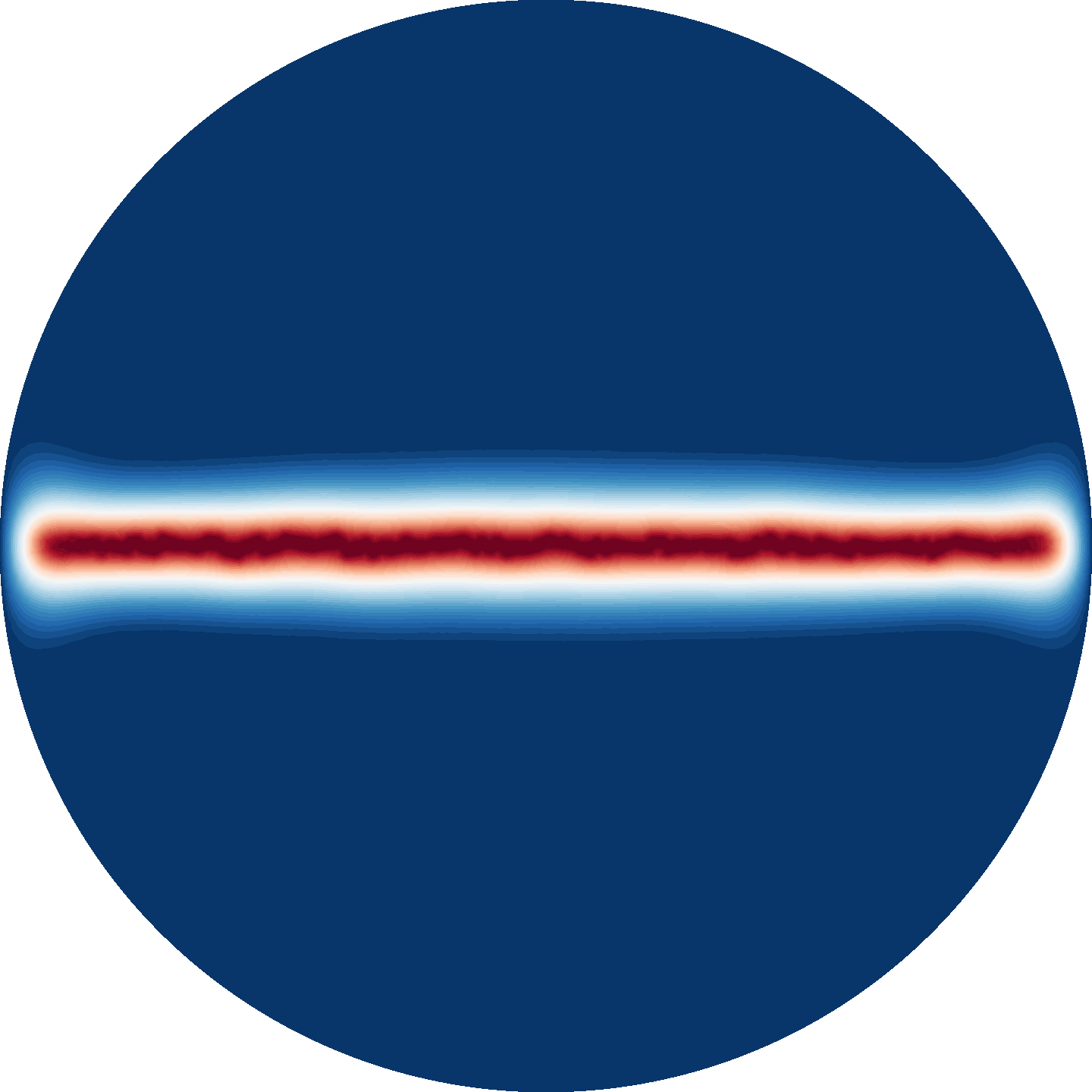}
      \putDiskFig{21}{24}{$\theta = 3\pi/4$}{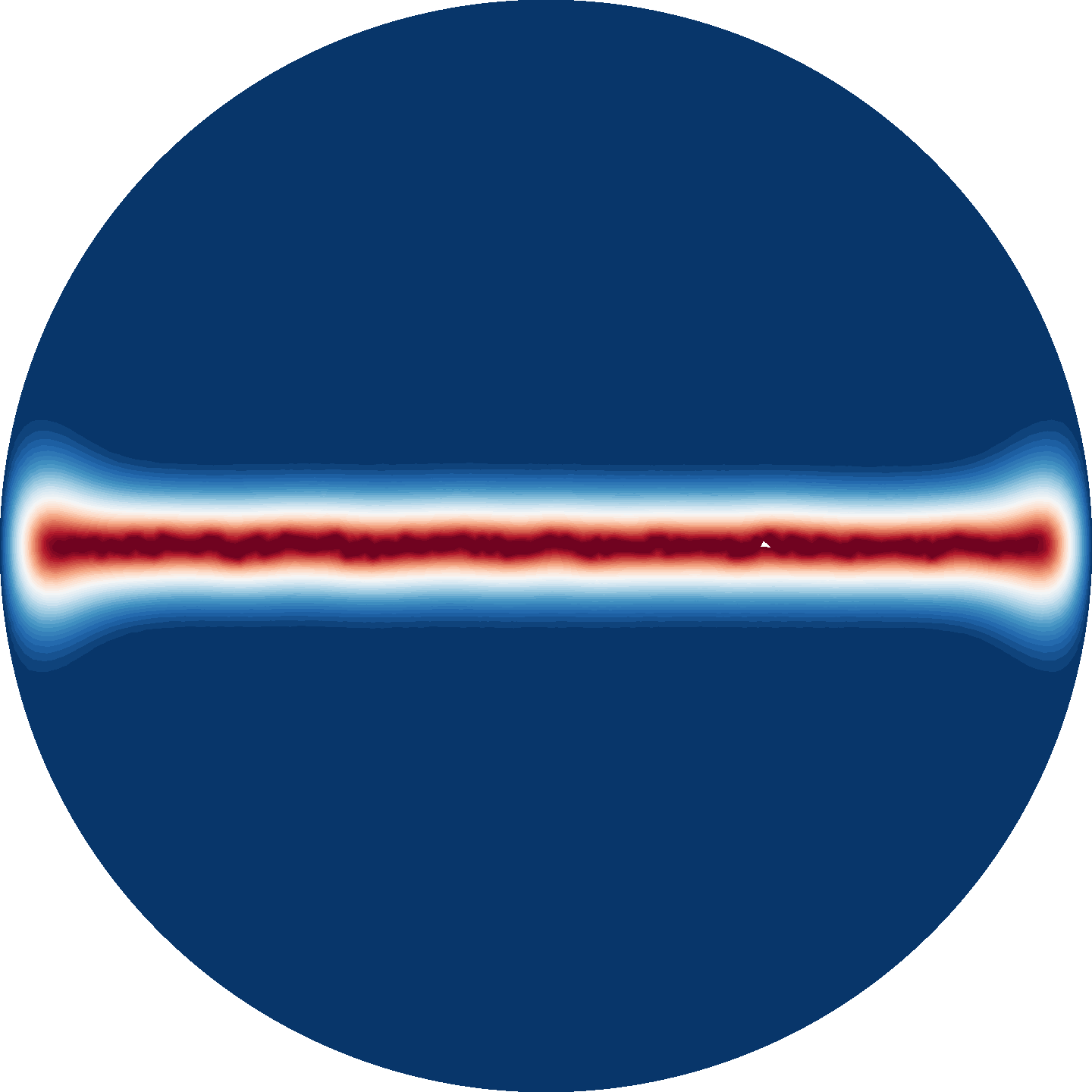}
      \putDiskFig{15}{13}{\hspace{-8mm}$\theta = 7\pi/8$}{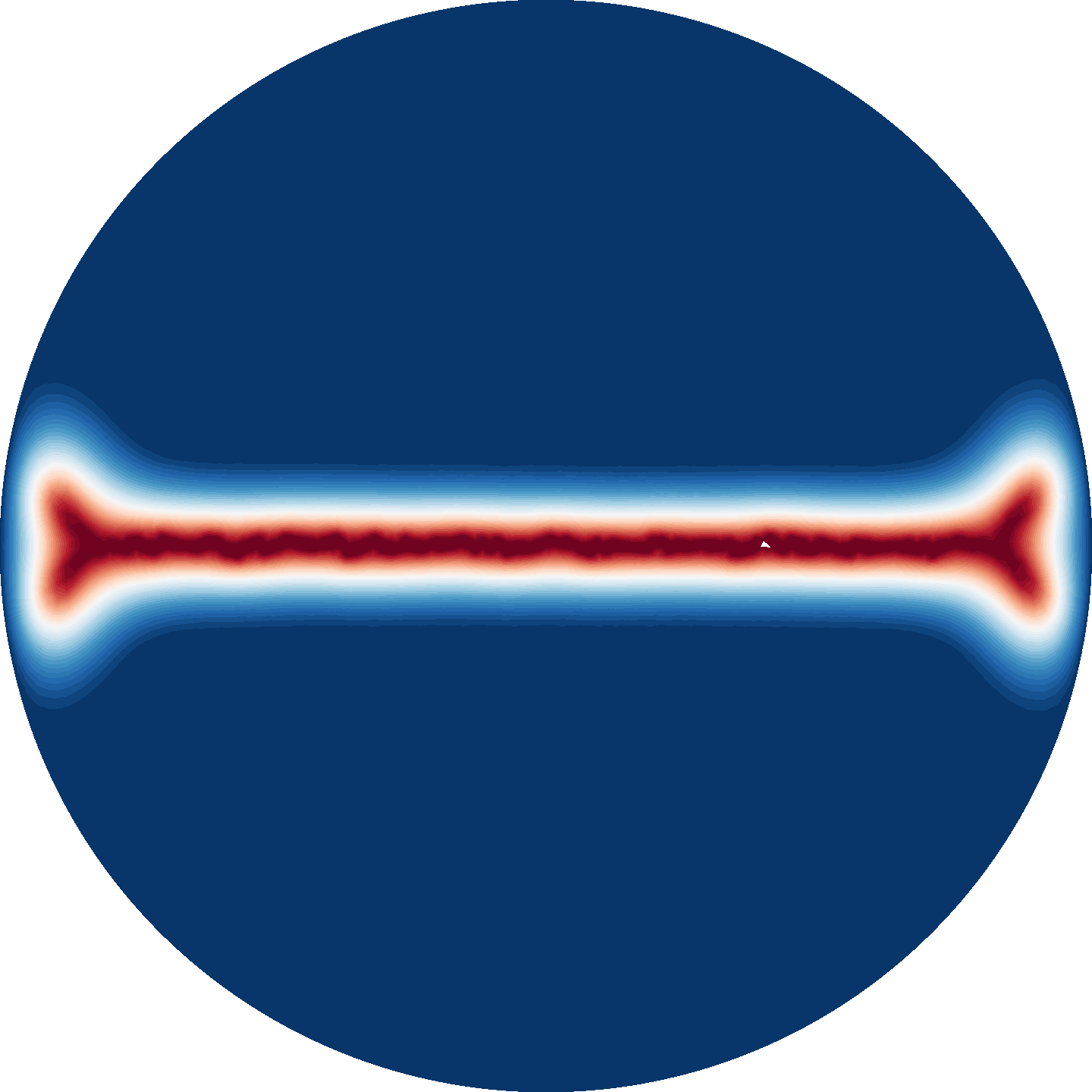}
      \putDiskFig{12}{2.5}{\hspace{-15mm}$\theta = \pi$}{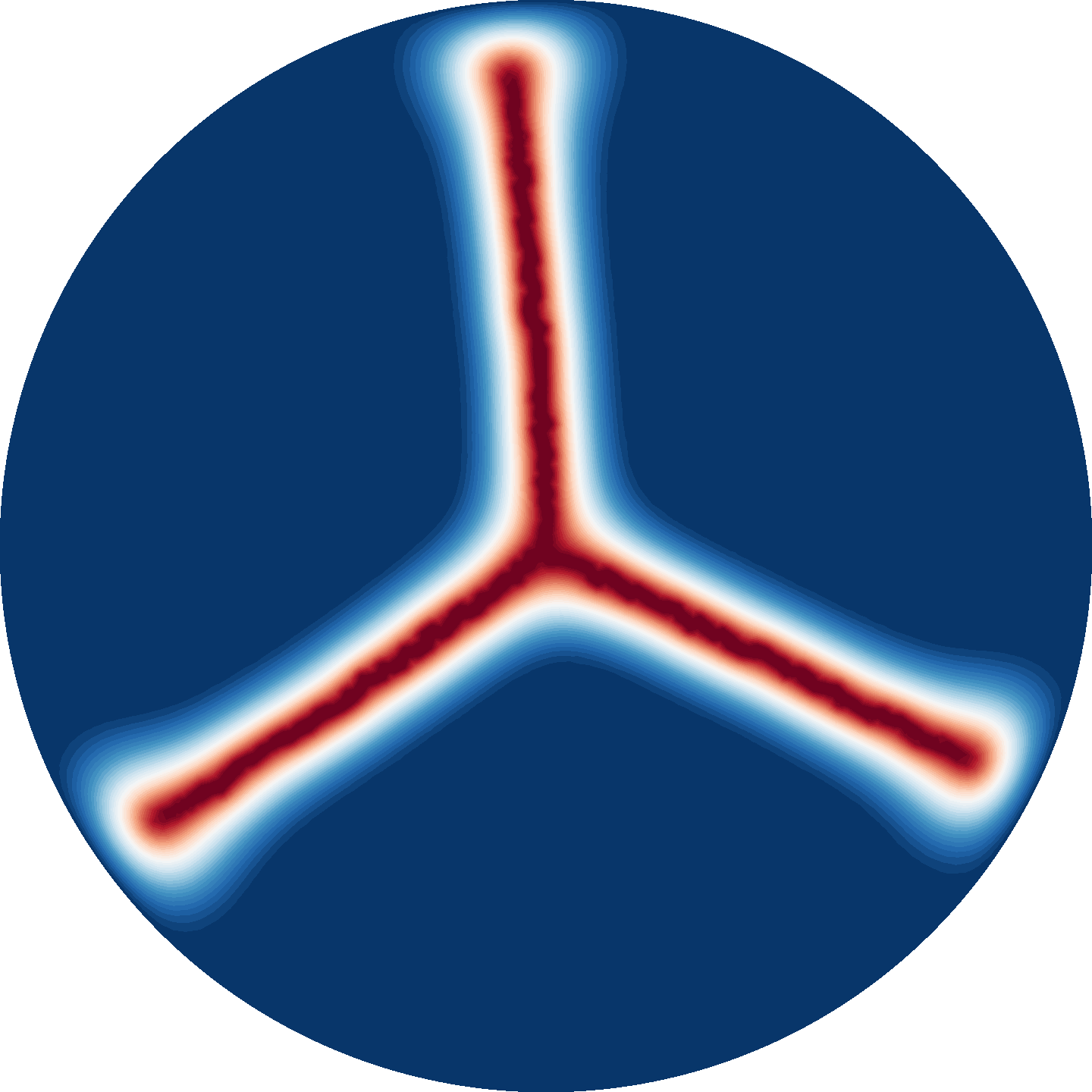}
    \end{overpic} 
    \\
    \hspace{7mm} $\sepsvtilde$
    \caption{Model \texttt{M1}.}
    \label{} 
  \end{subfigure}
  \\[2ex]
  \begin{subfigure}{\linewidth}
    \centering
    \small
    \raisebox{32mm}{$\sepsdtilde$}\hspace{2mm}
    \begin{overpic}[width=0.8\linewidth]{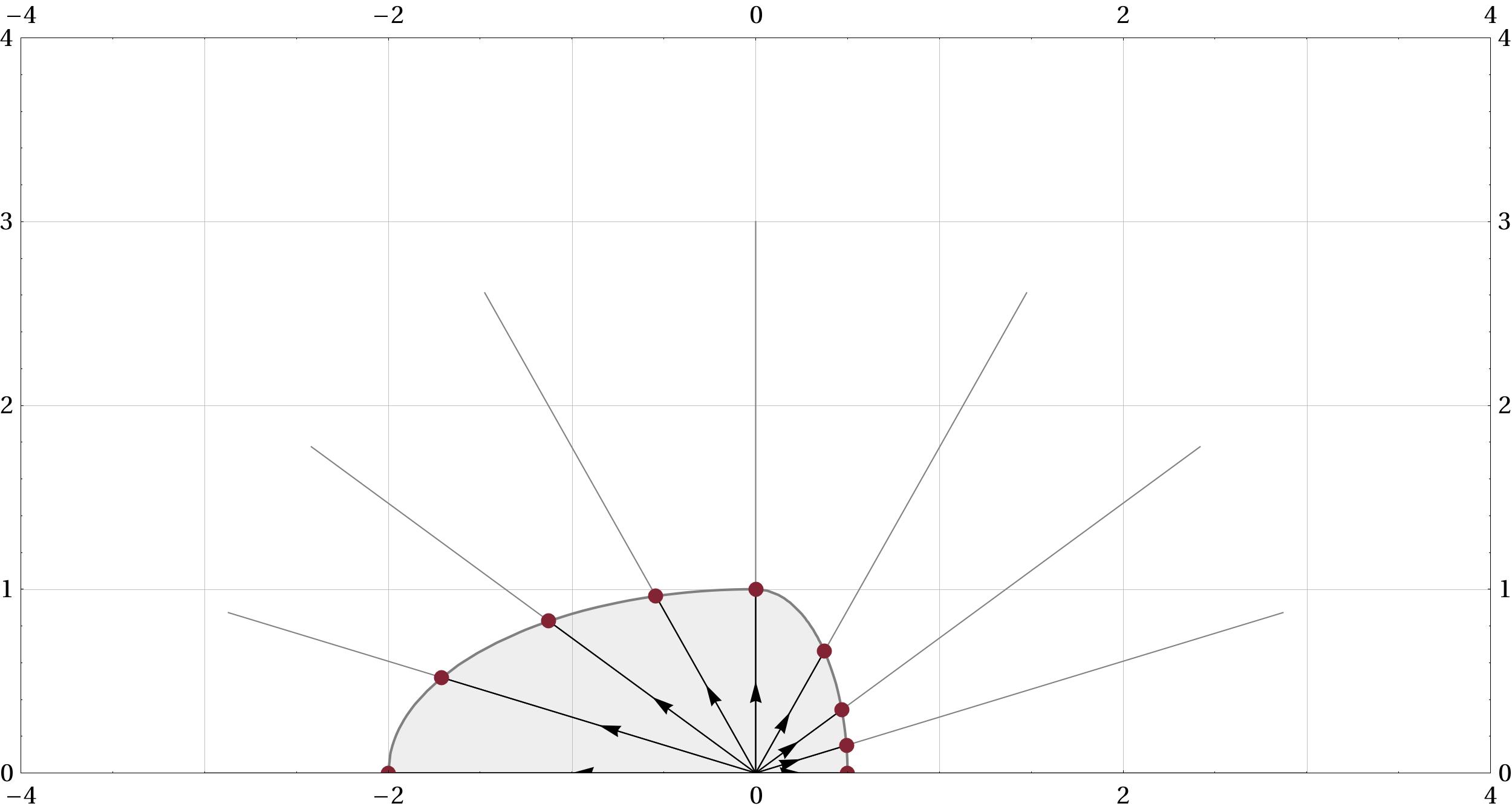}
      \putDiskFig{87}{2.5}{\hspace{15mm}$\theta = 0$}{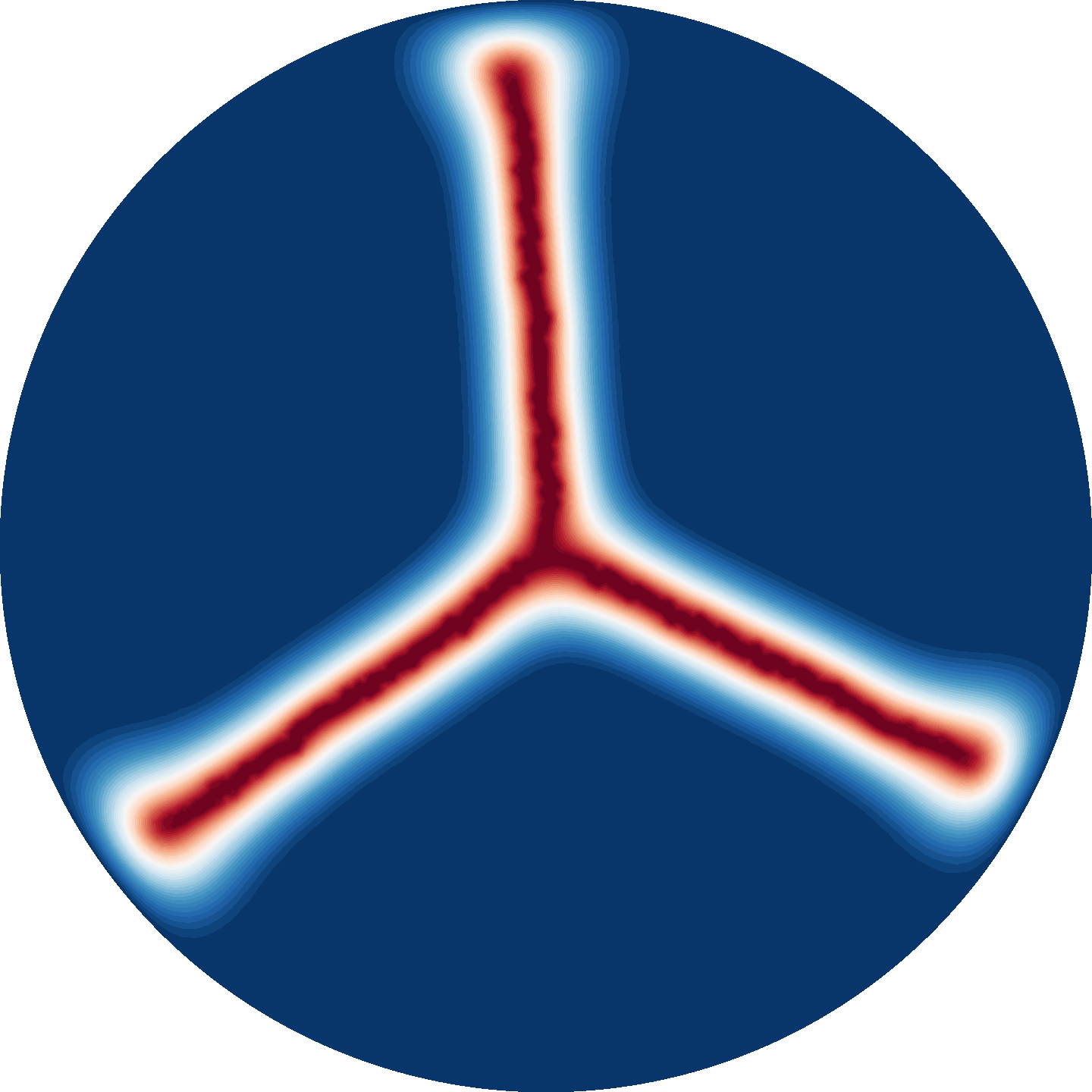}
      \putDiskFig{85}{13}{\hspace{8mm}$\theta = \pi/8$}{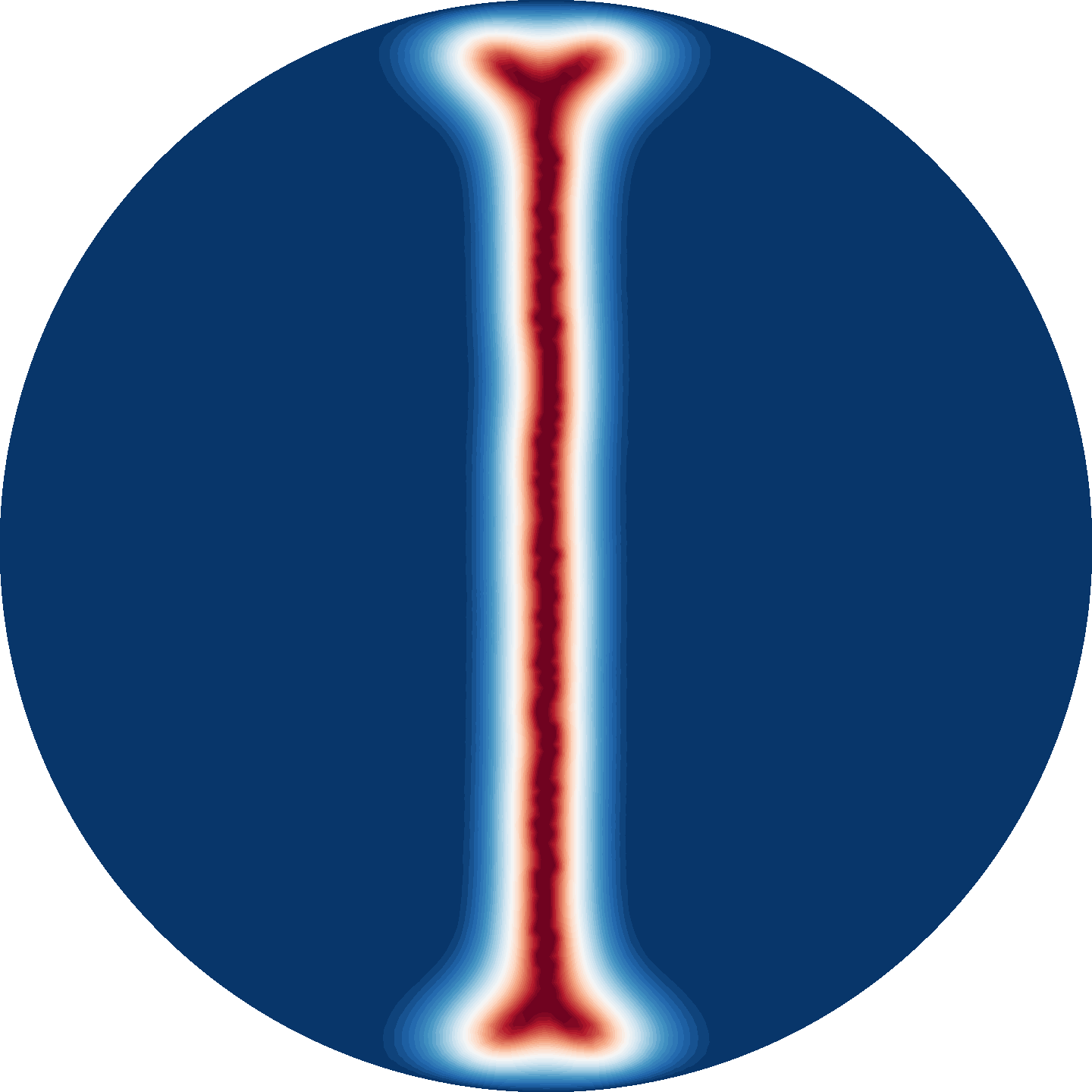}
      \putDiskFig{79}{24}{$\theta = \pi/4$}{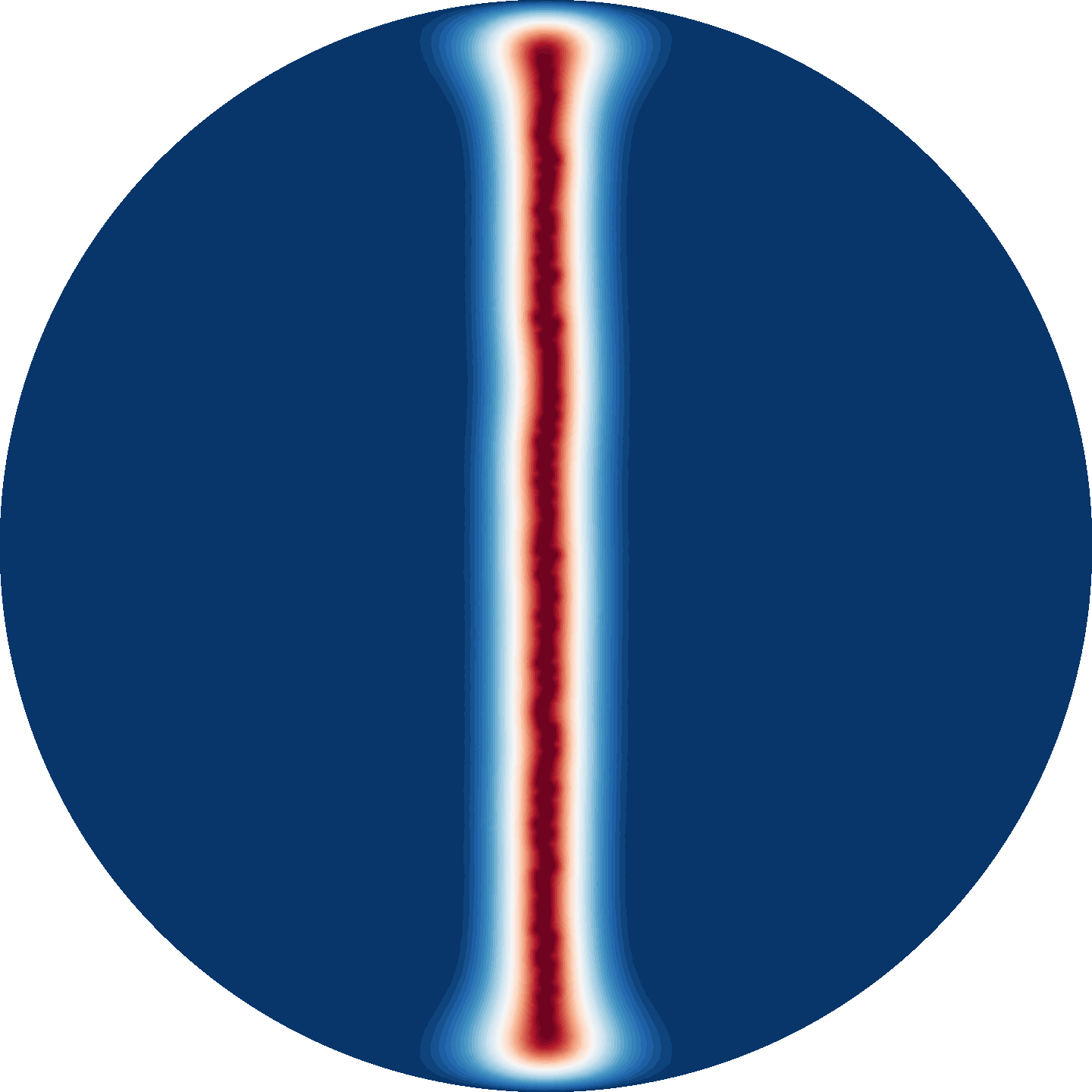}
      \putDiskFig{68}{34}{$\theta = 3\pi/8$}{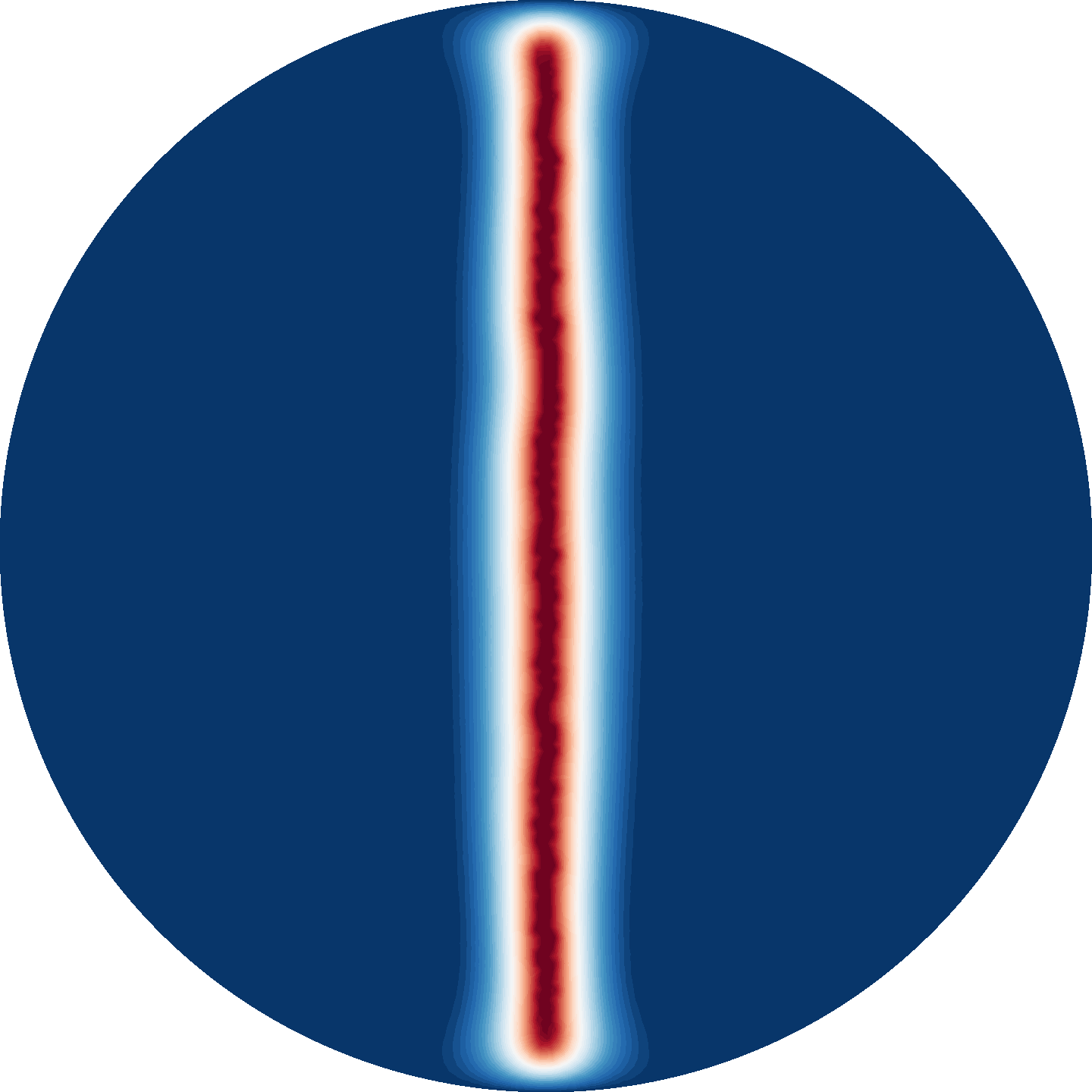}
      \putDiskFig{49}{38}{$\theta = \pi/2$}{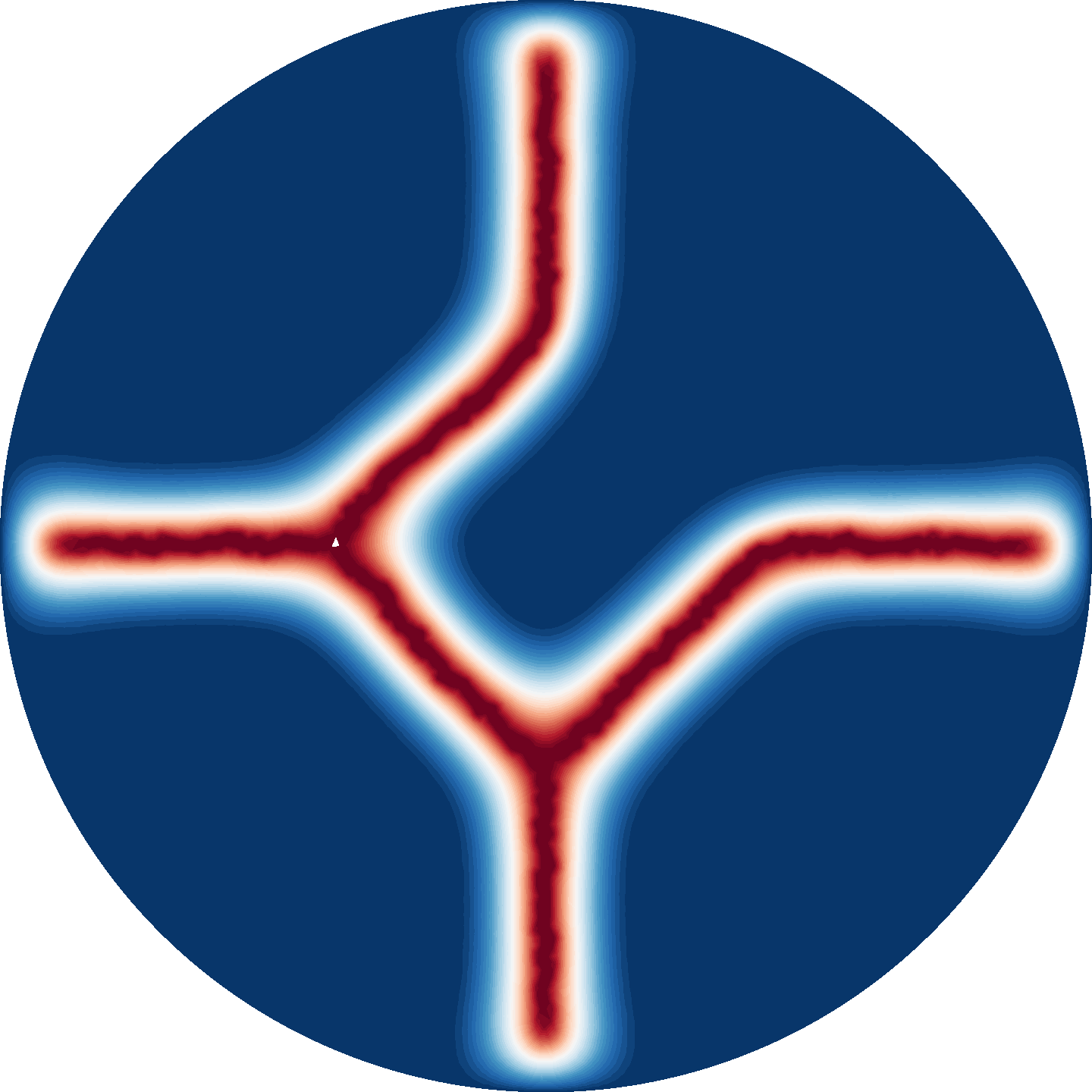}
      \putDiskFig{32}{34}{$\theta = 5\pi/8$}{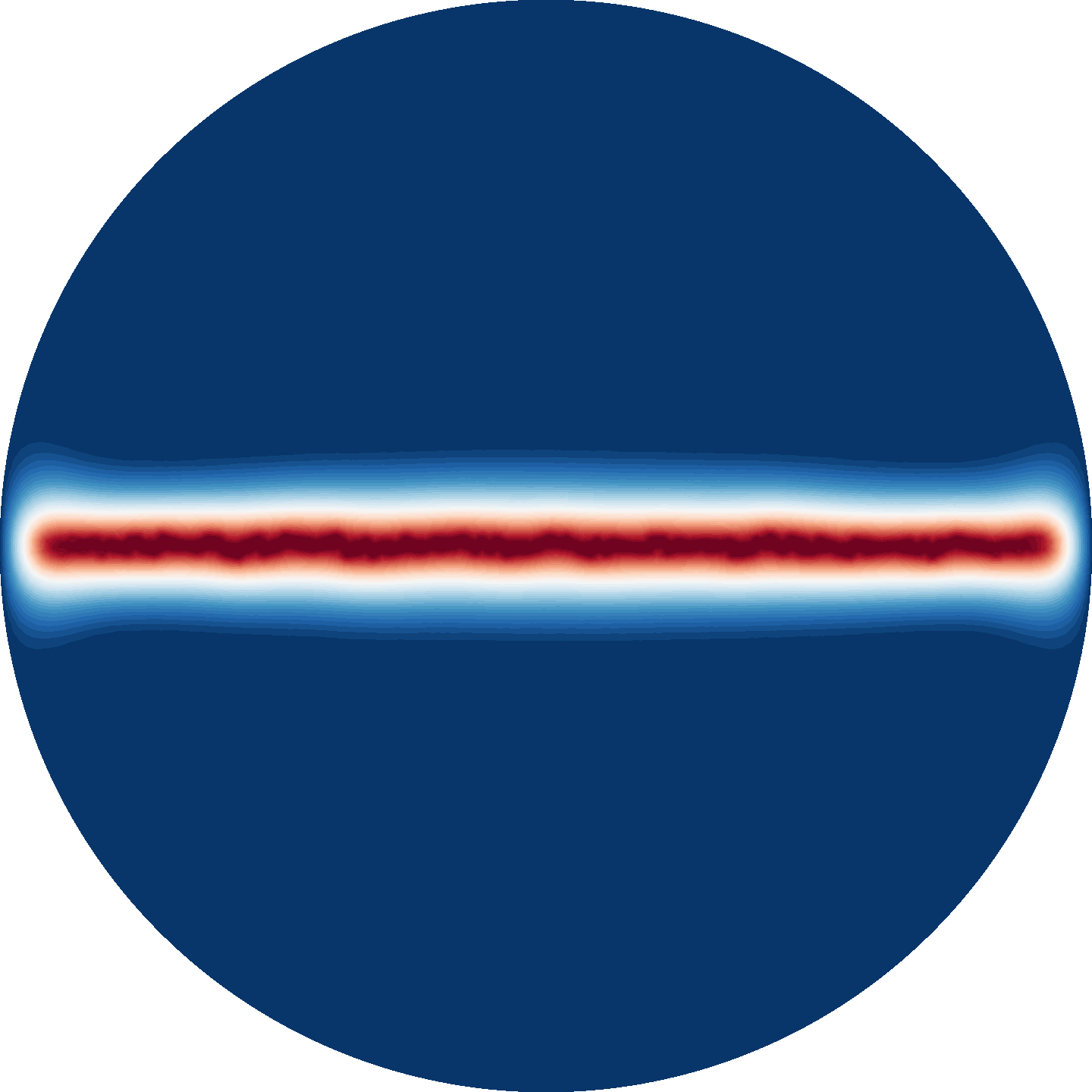}
      \putDiskFig{21}{24}{$\theta = 3\pi/4$}{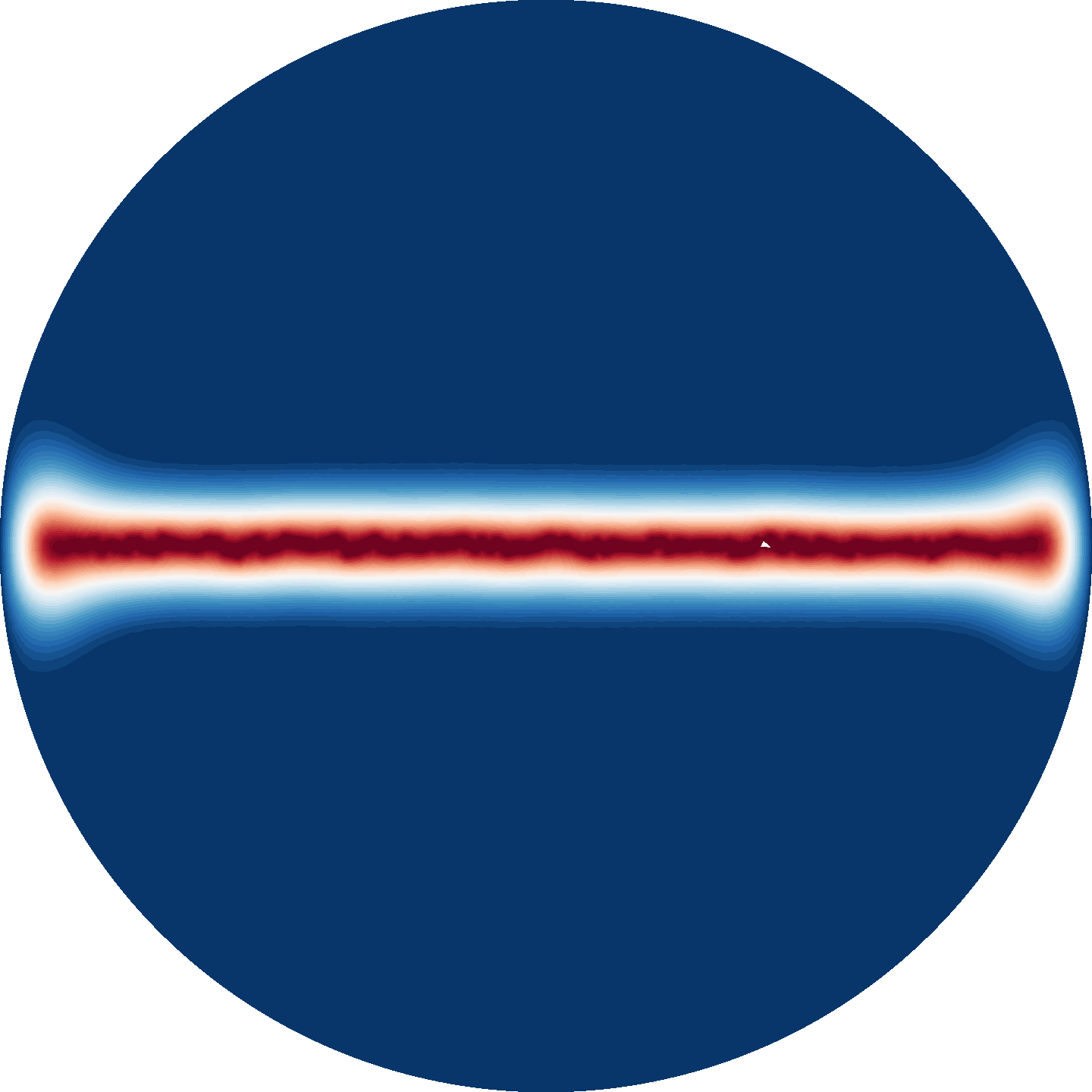}
      \putDiskFig{15}{13}{\hspace{-8mm}$\theta = 7\pi/8$}{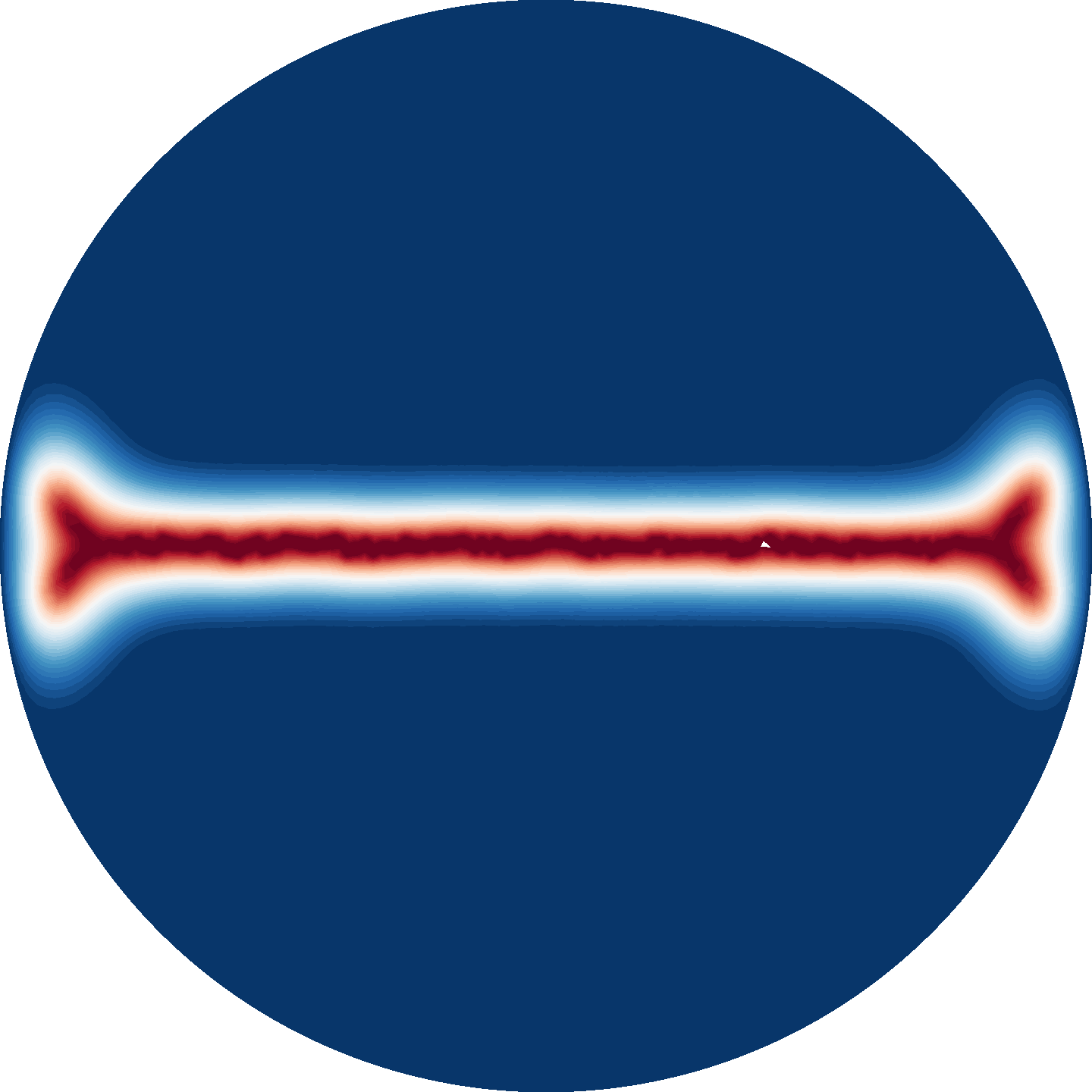}
      \putDiskFig{12}{2.5}{\hspace{-15mm}$\theta = \pi$}{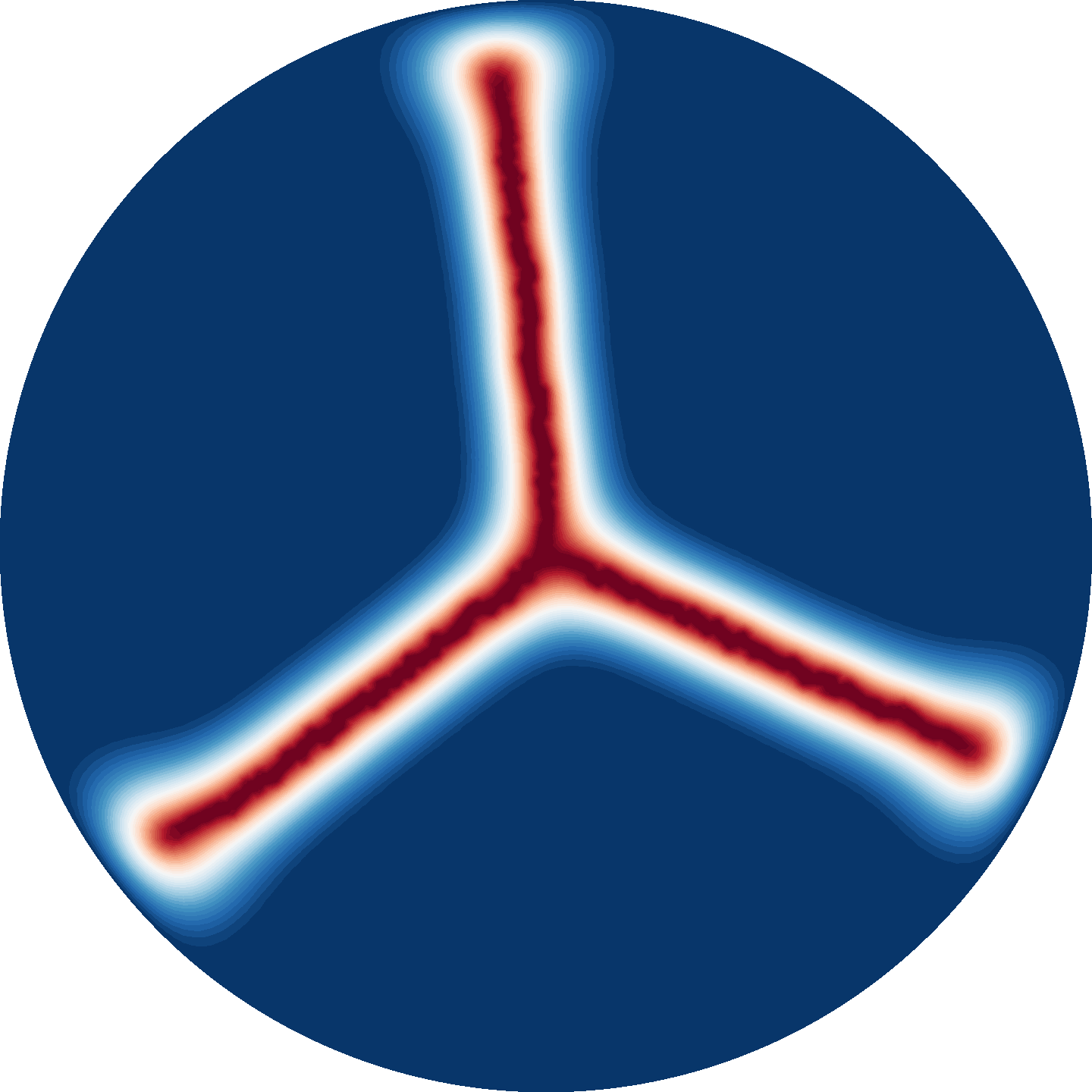}
    \end{overpic} 
    \\
    \hspace{7mm} $\sepsvtilde$
    \caption{Model \texttt{M2}.}
    \label{} 
  \end{subfigure}
  \\[2ex]
  \begin{subfigure}{\linewidth}
    \centering
    \small
    \raisebox{32mm}{$\sepsdtilde$}\hspace{2mm}
    \begin{overpic}[width=0.8\linewidth]{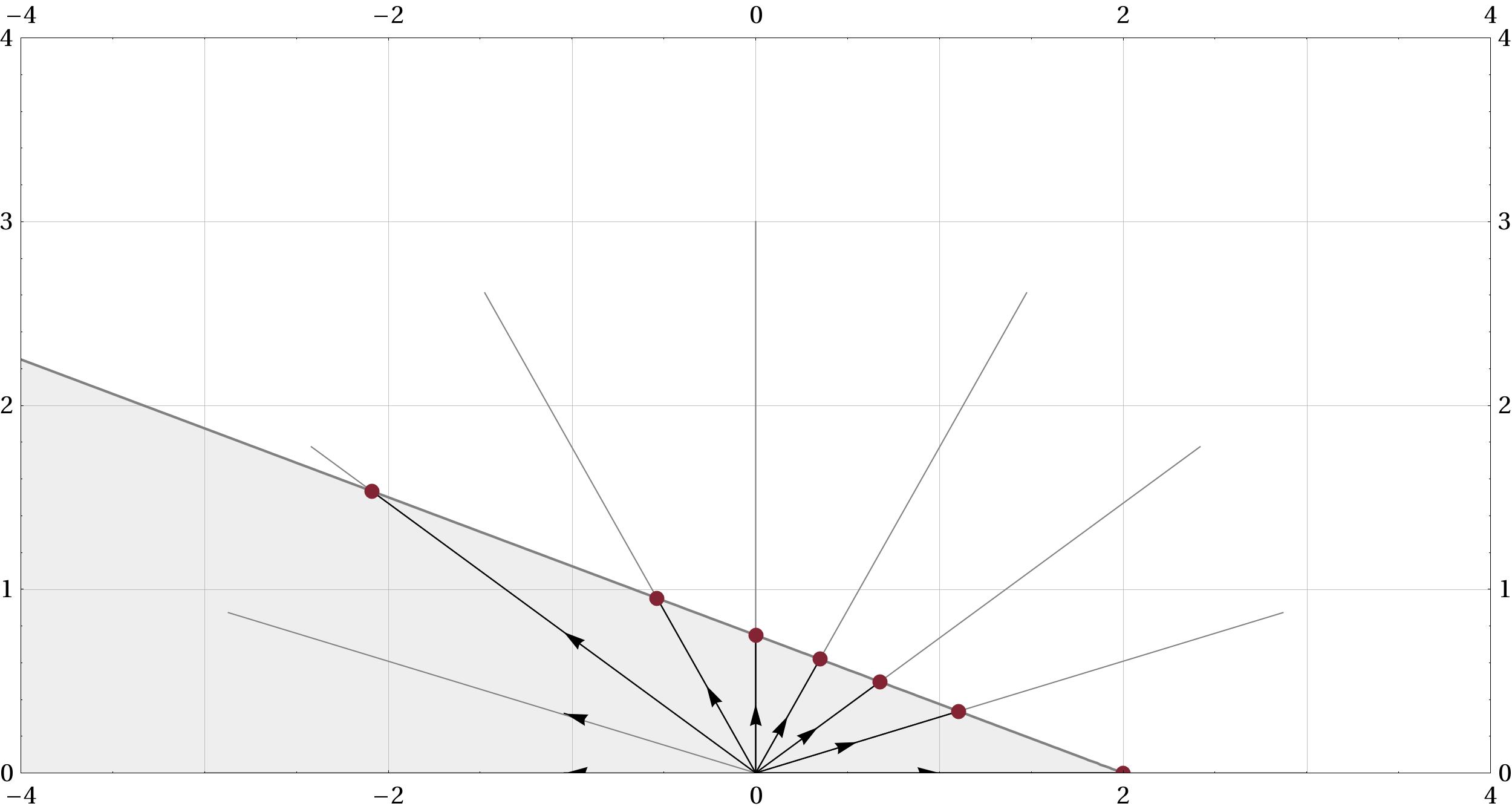}
      \putDiskFig{87}{2.5}{\hspace{15mm}$\theta = 0$}{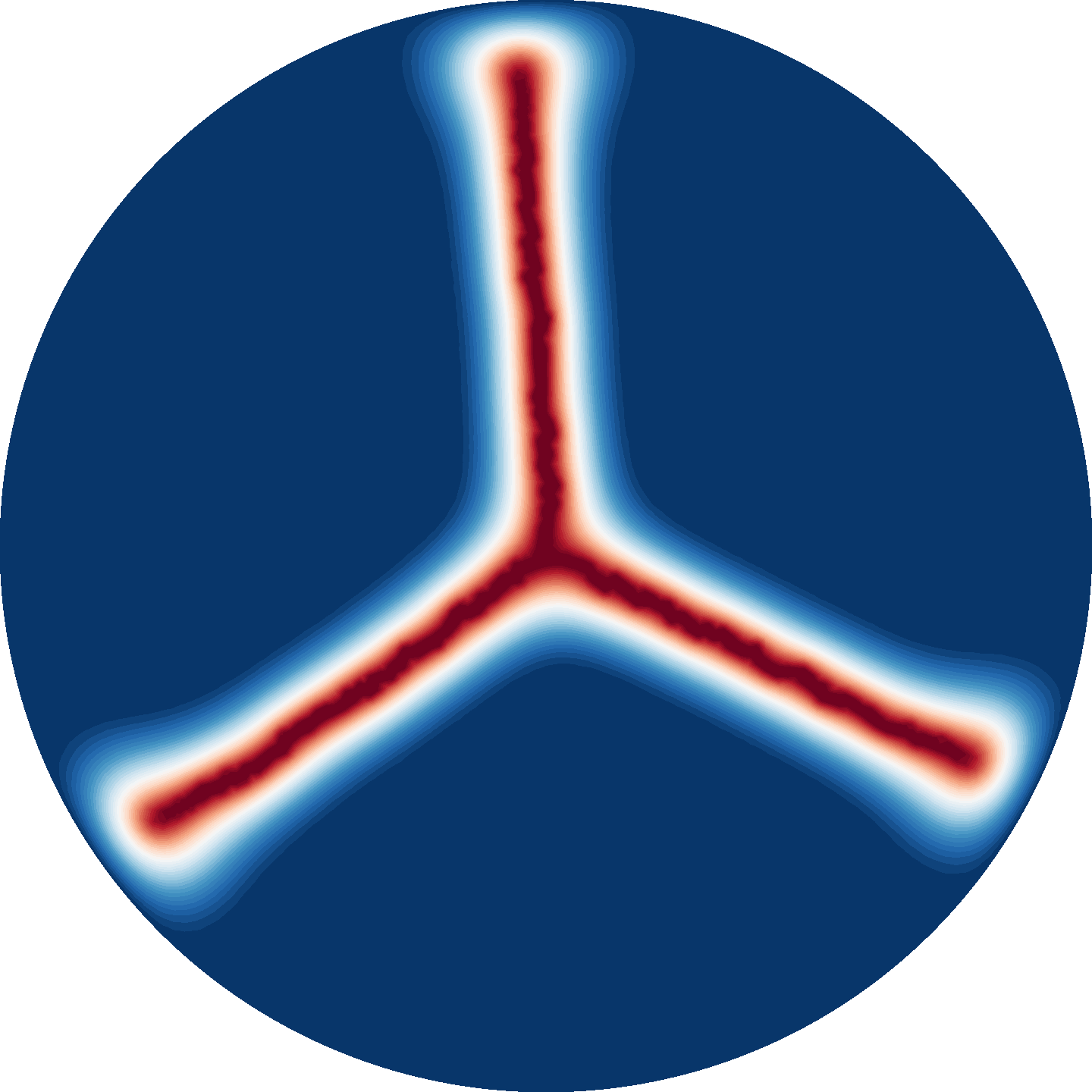}
      \putDiskFig{85}{13}{\hspace{8mm}$\theta = \pi/8$}{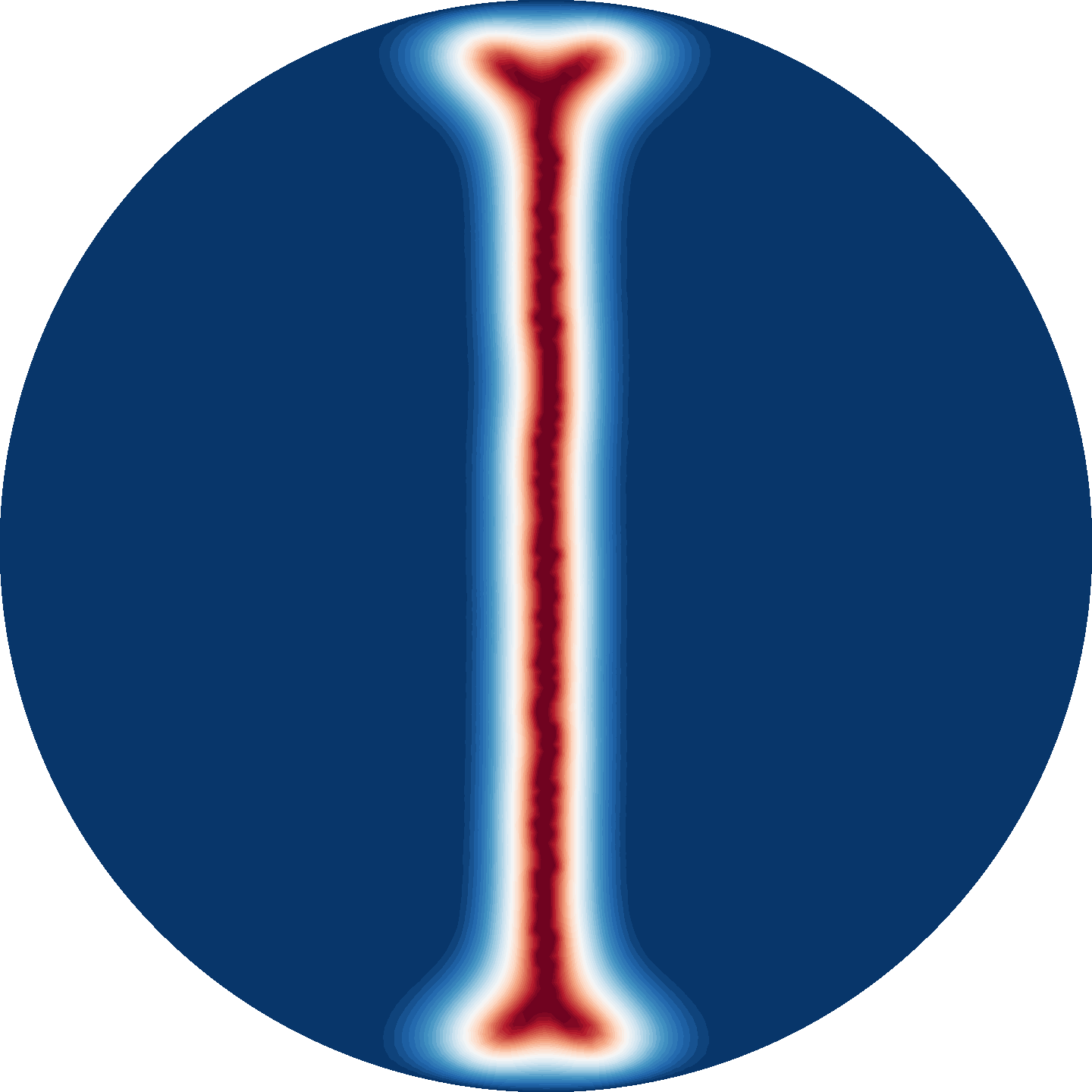}
      \putDiskFig{79}{24}{$\theta = \pi/4$}{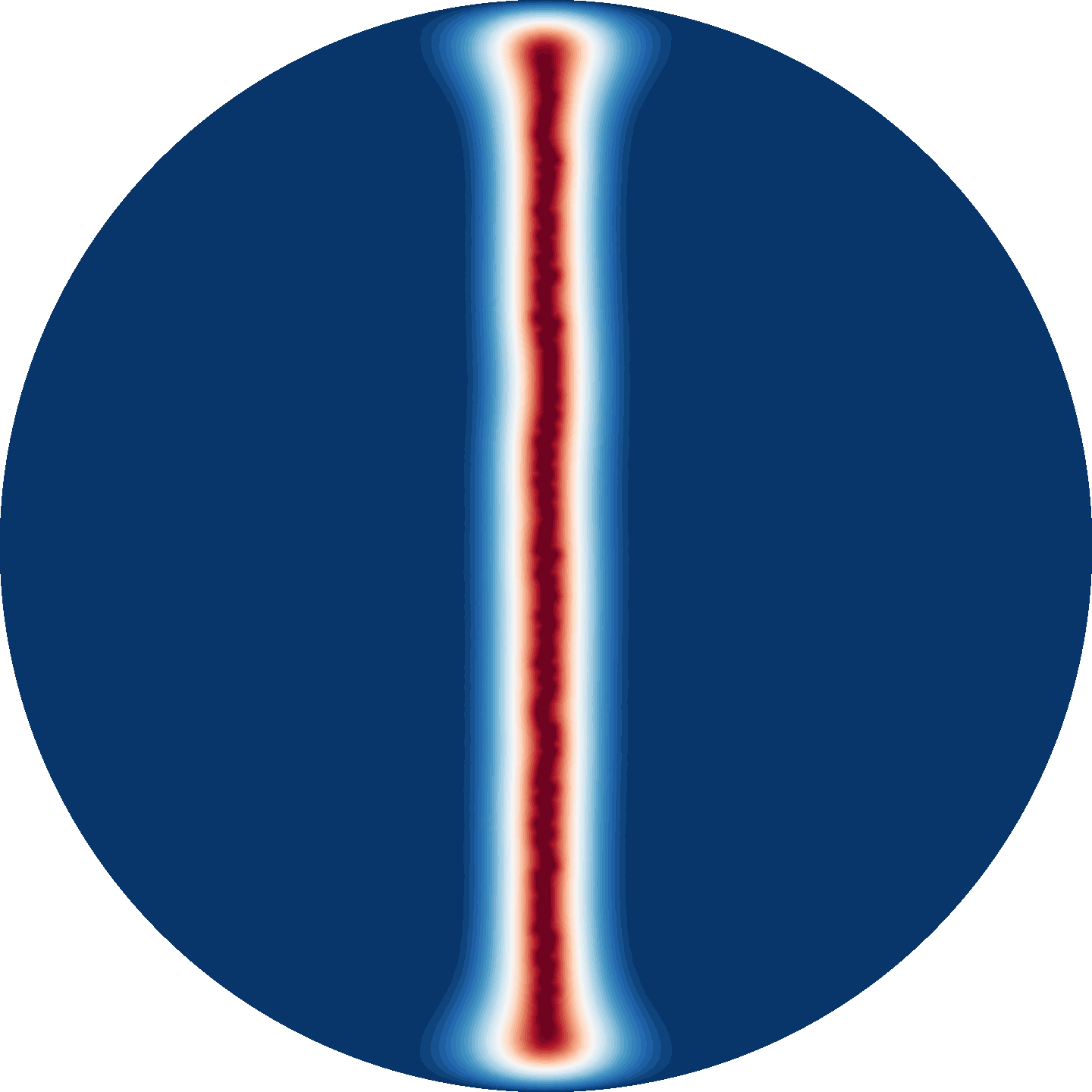}
      \putDiskFig{68}{34}{$\theta = 3\pi/8$}{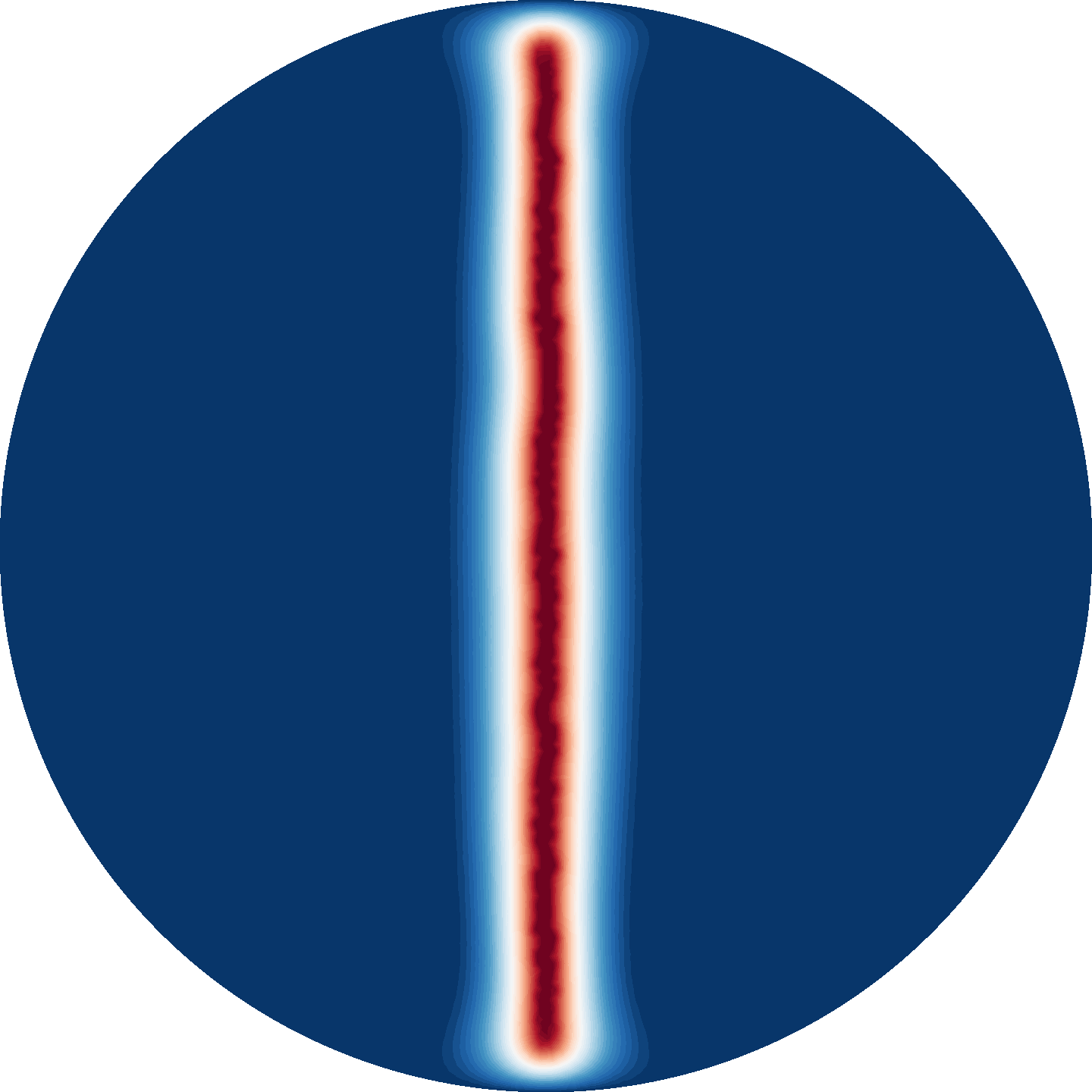}
      \putDiskFig{49}{38}{$\theta = \pi/2$}{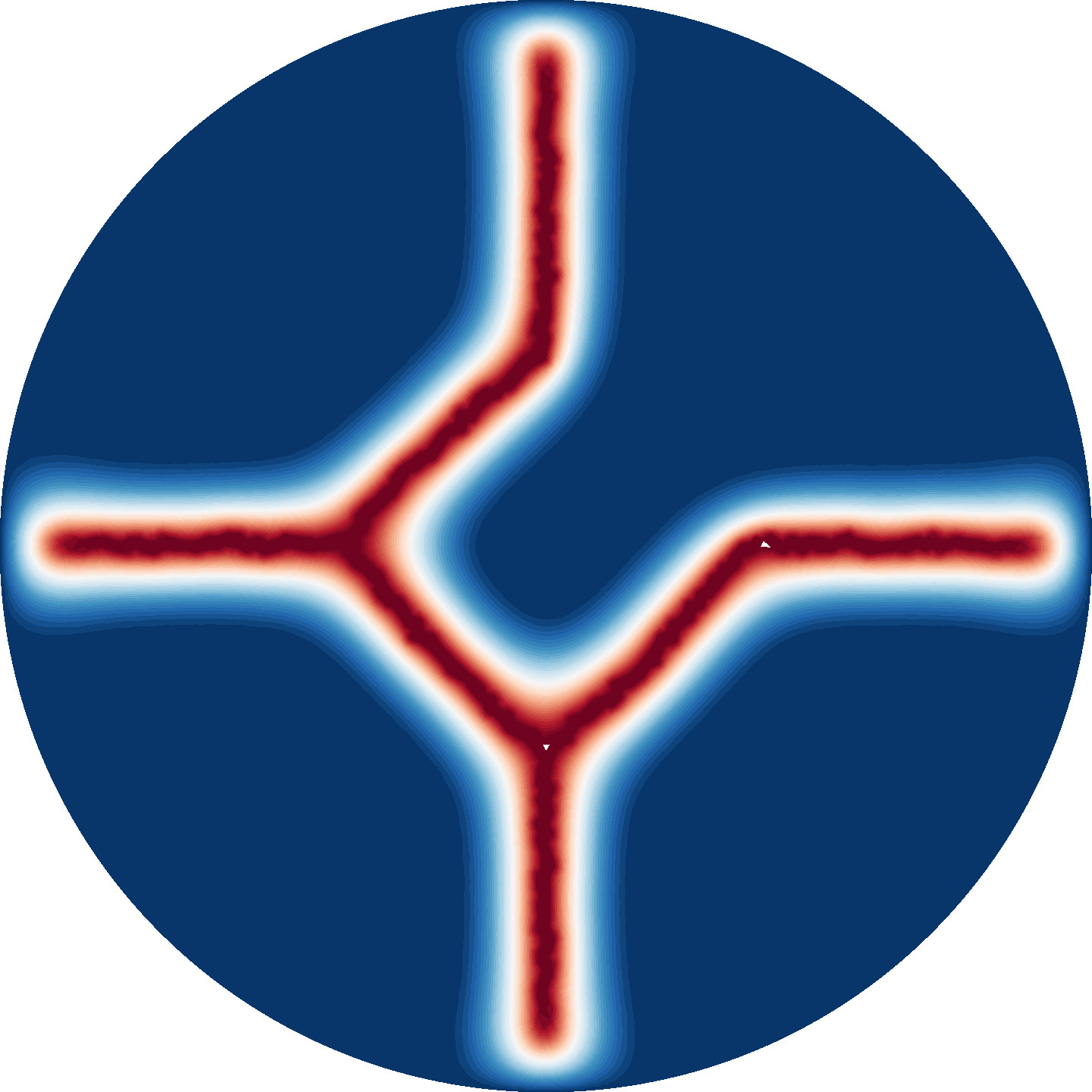}
      \putDiskFig{32}{34}{$\theta = 5\pi/8$}{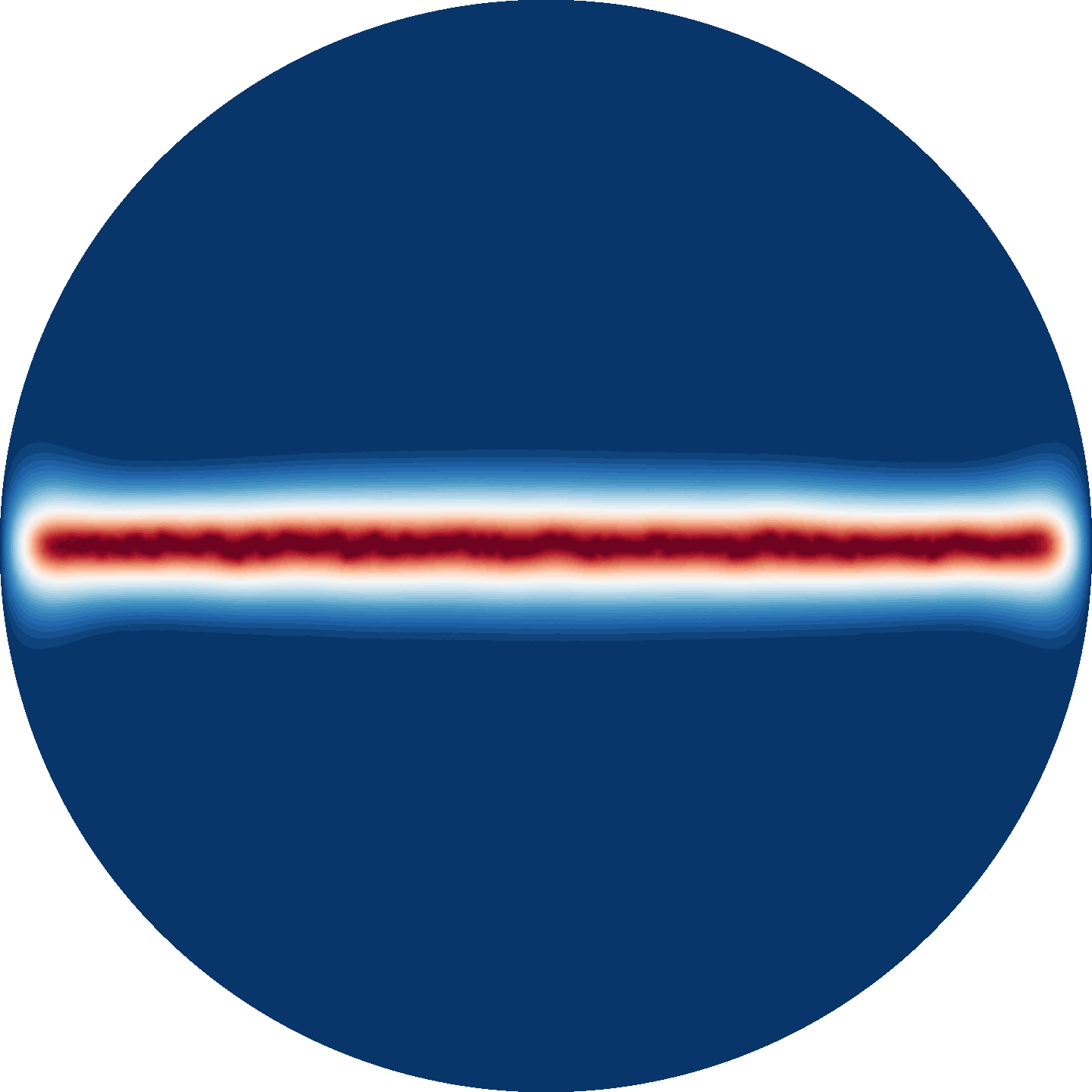}
      \putDiskFig{21}{24}{$\theta = 3\pi/4$}{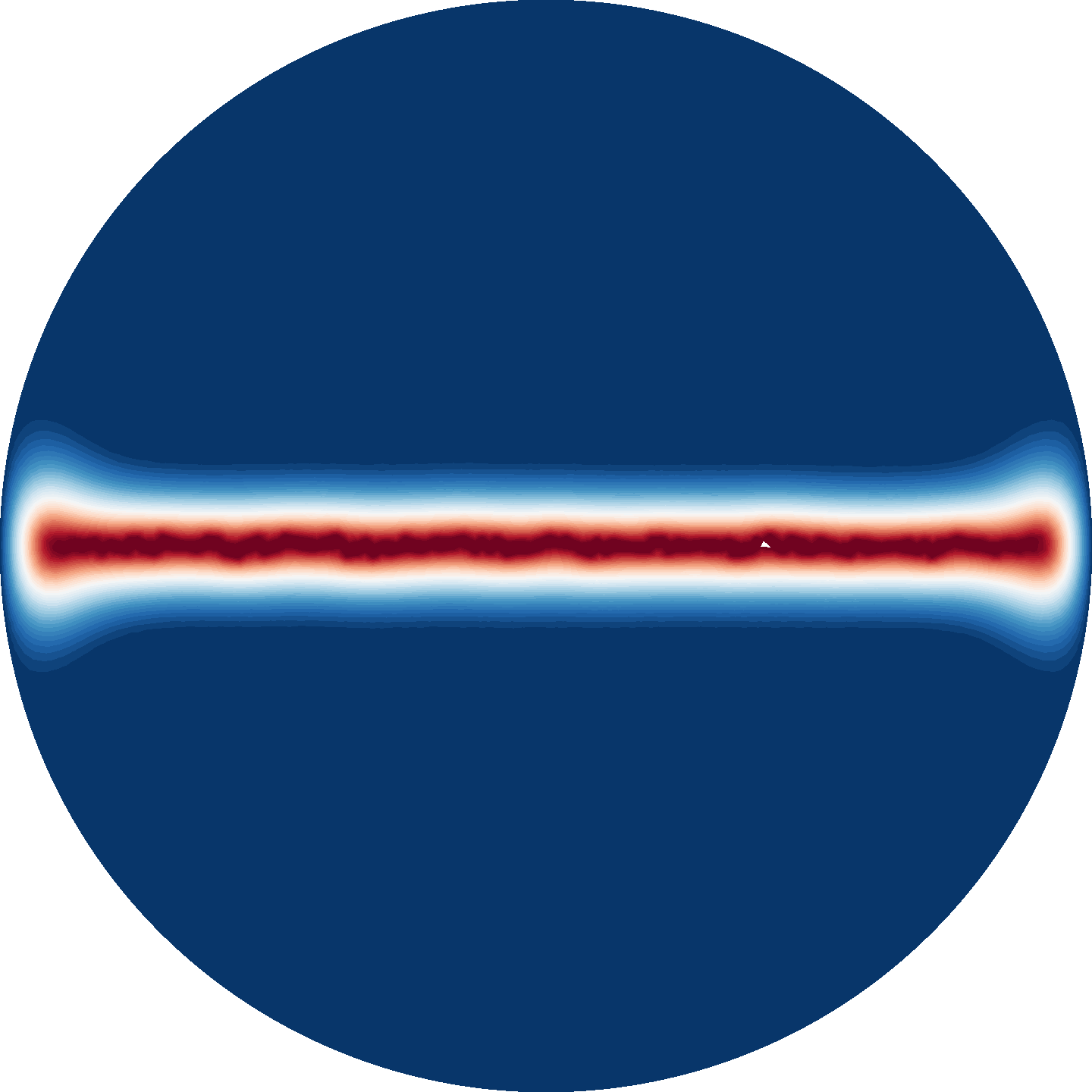}
      \putDiskFig{15}{13}{\hspace{-8mm}$\theta = 7\pi/8$}{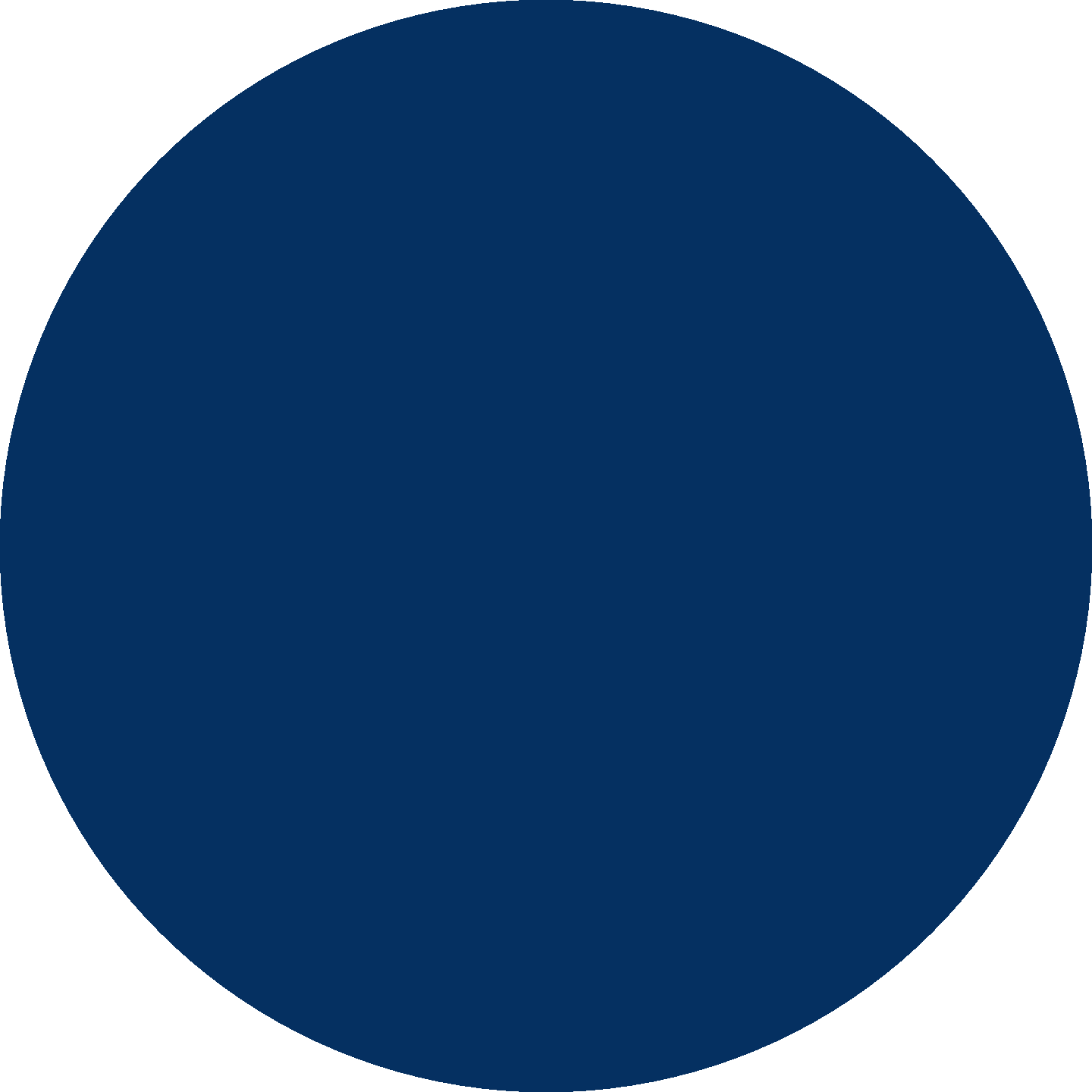}
      \putDiskFig{12}{2.5}{\hspace{-15mm}$\theta = \pi$}{M5-theta_180.00}
    \end{overpic} 
    \\
    \hspace{7mm} $\sepsvtilde$
    \caption{Model \texttt{M3}.}
    \label{} 
  \end{subfigure}  
  \caption{\emph{See next figure caption}.}
  \label{} 
\end{figure}

\begin{figure}[h!]
\ContinuedFloat
\centering
  \begin{subfigure}{\linewidth}
    \centering
    \small
    \raisebox{32mm}{$\sepsdtilde$}\hspace{2mm}
    \begin{overpic}[width=0.8\linewidth]{RepsPlotNumM3}
      \putDiskFig{87}{2.5}{\hspace{15mm}$\theta = 0$}{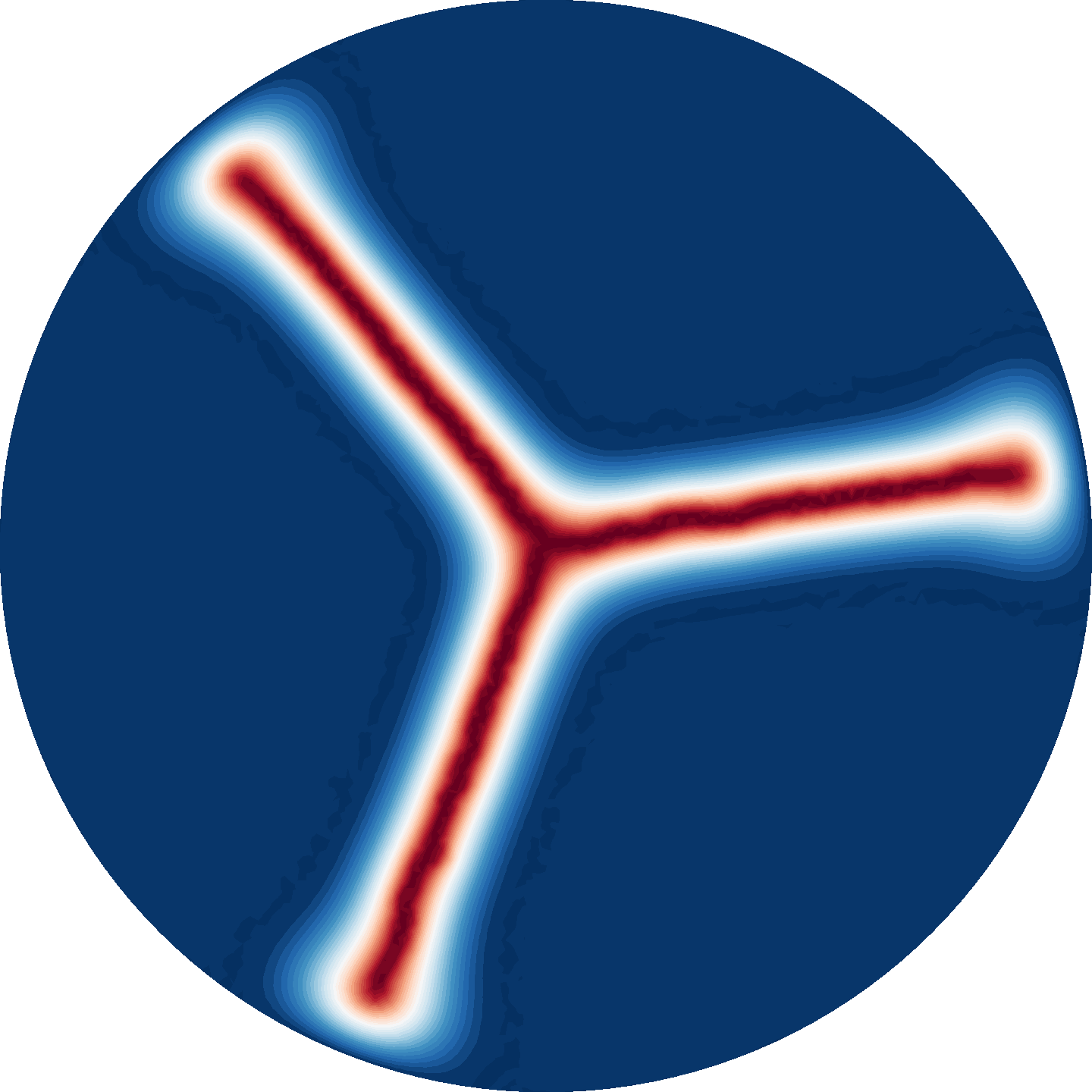}
      \putDiskFig{85}{13}{\hspace{8mm}$\theta = \pi/8$}{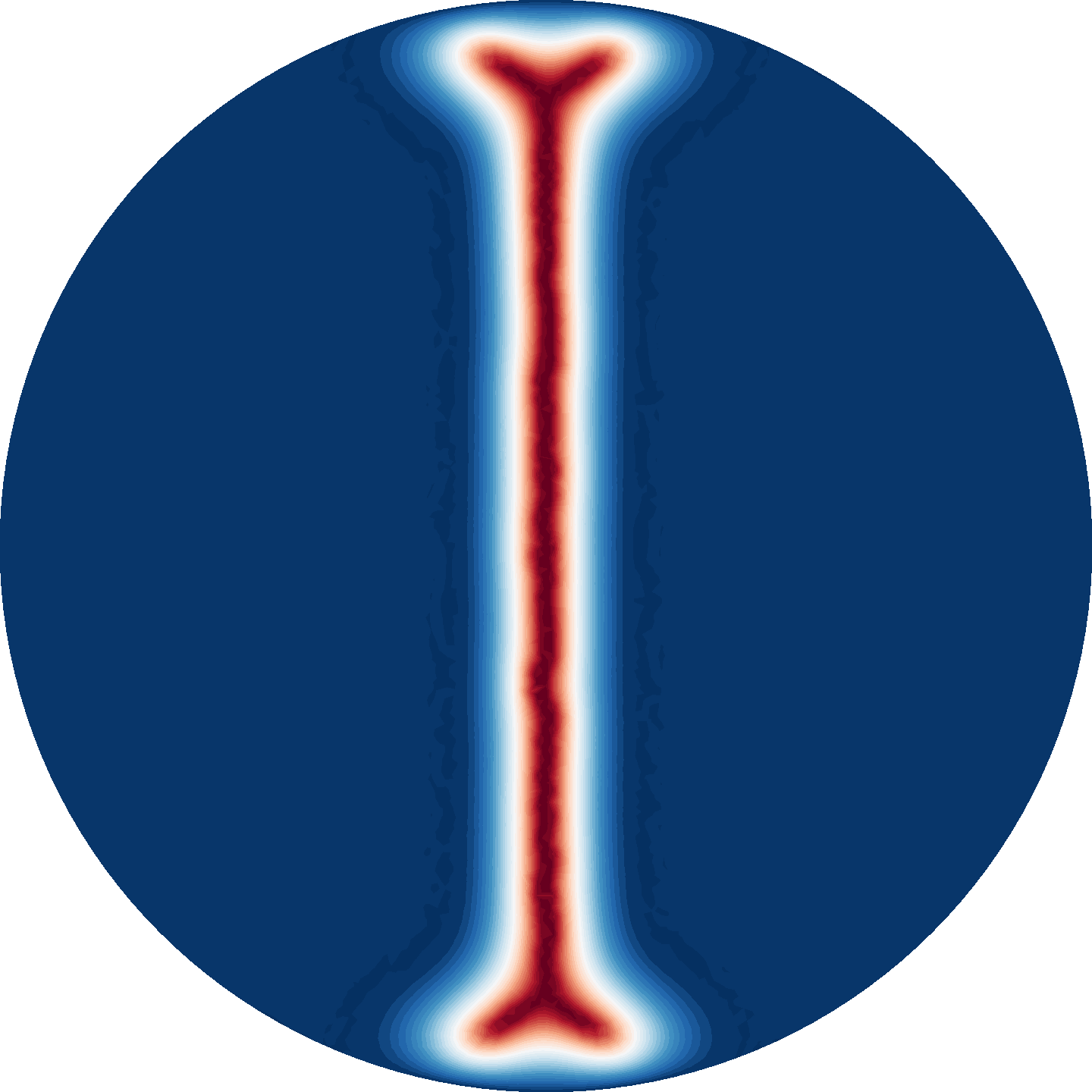}
      \putDiskFig{79}{24}{$\theta = \pi/4$}{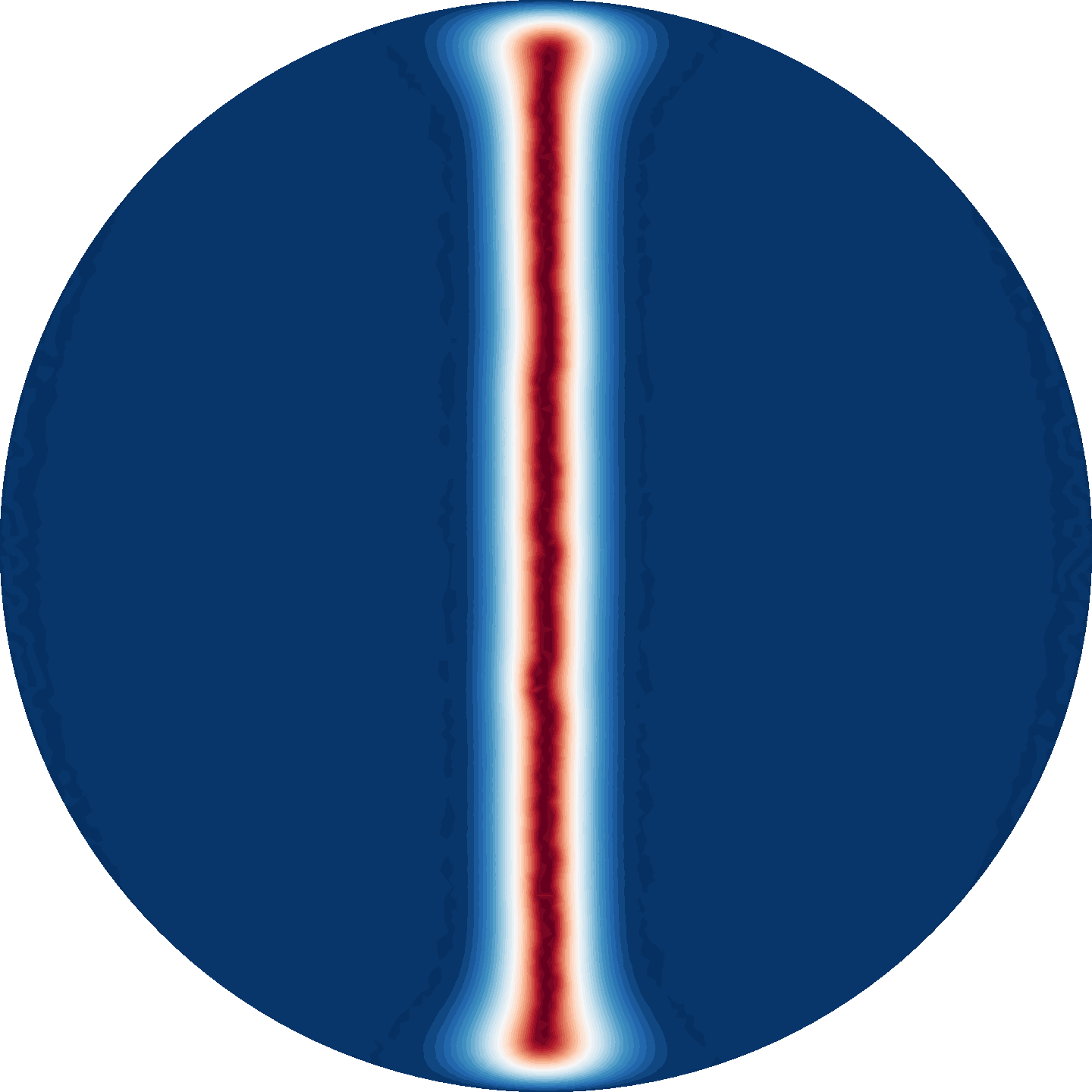}
      \putDiskFig{68}{34}{$\theta = 3\pi/8$}{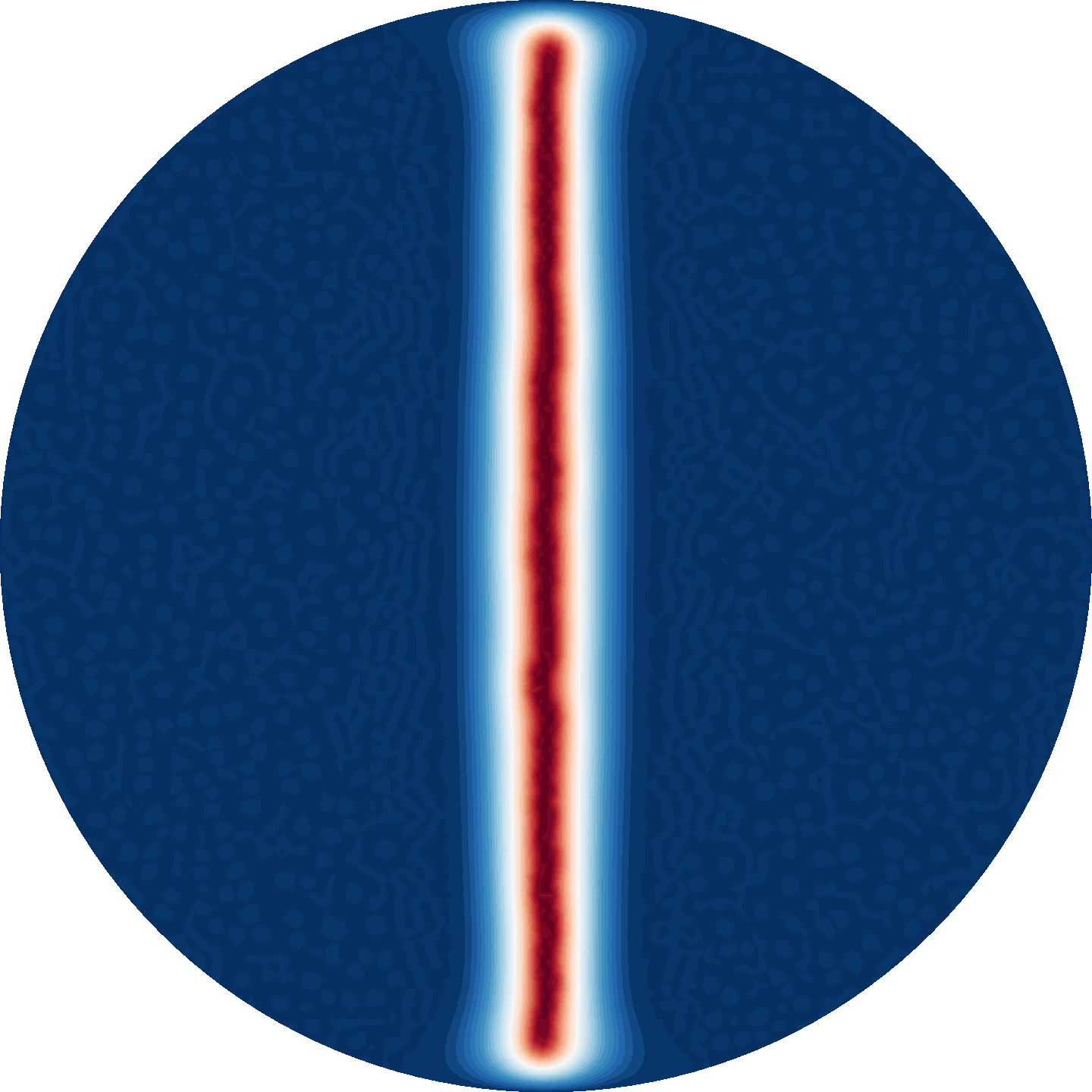}
      \putDiskFig{49}{38}{$\theta = \pi/2$}{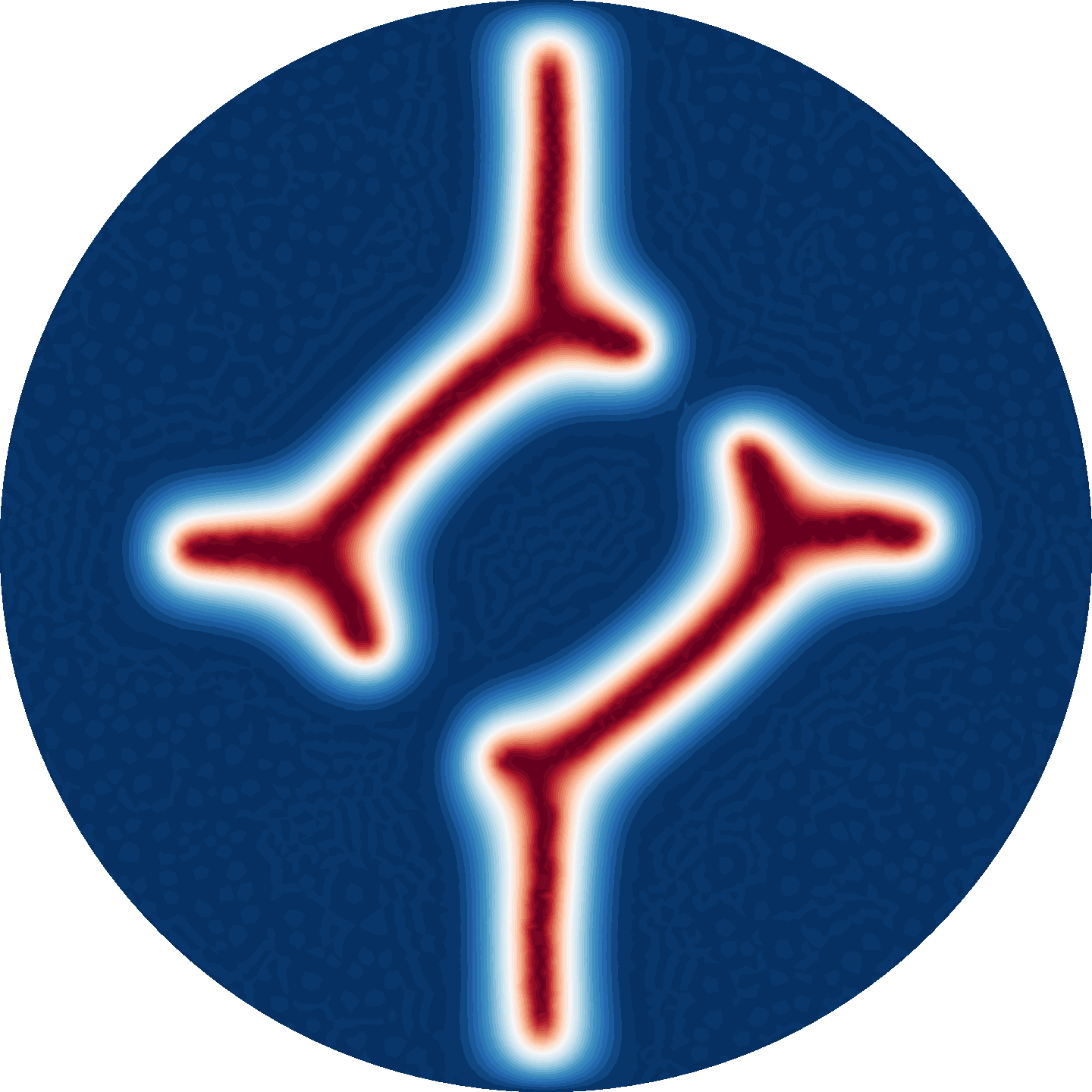}
      \putDiskFig{32}{34}{$\theta = 5\pi/8$}{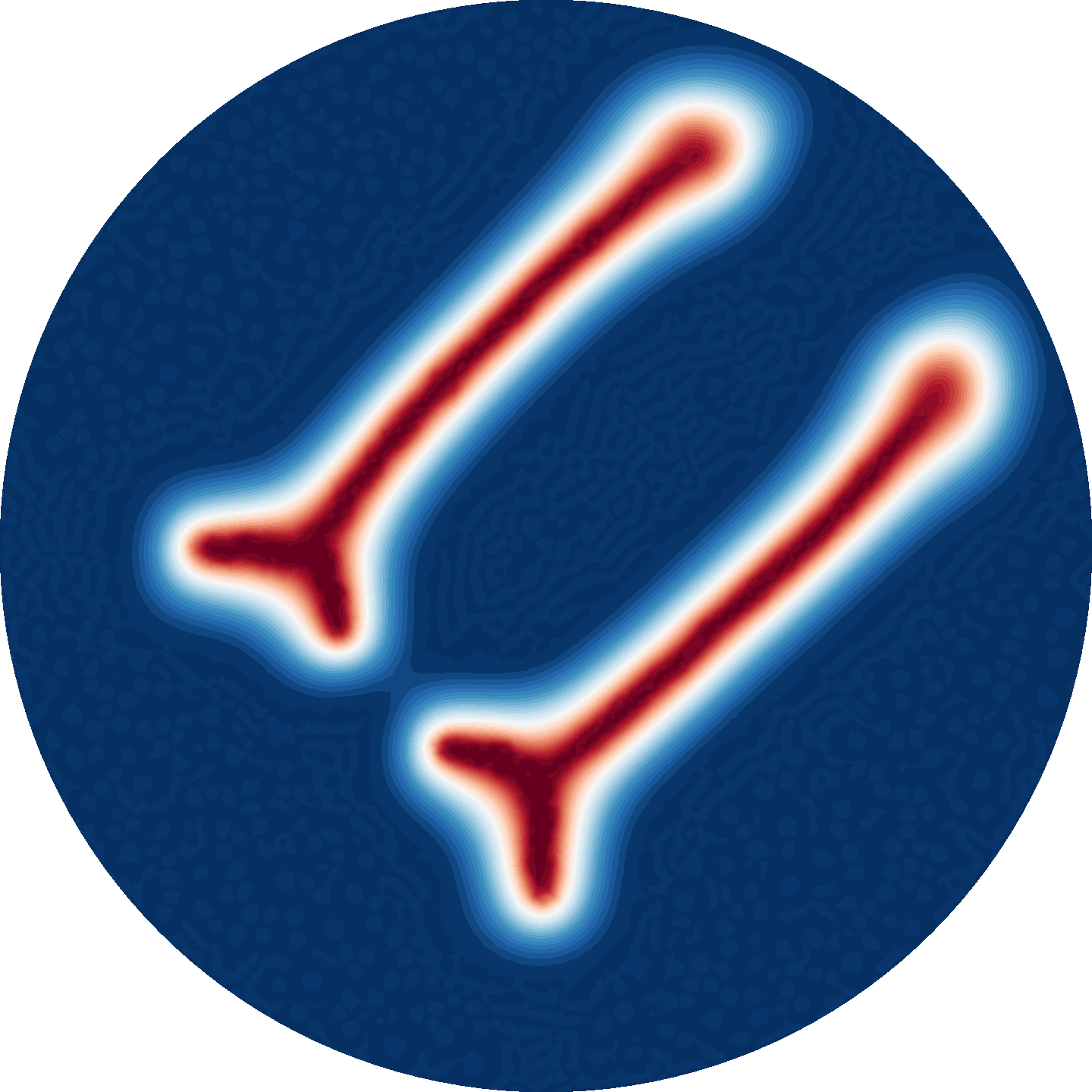}
      \putDiskFig{19}{26}{$\theta = 3\pi/4$}{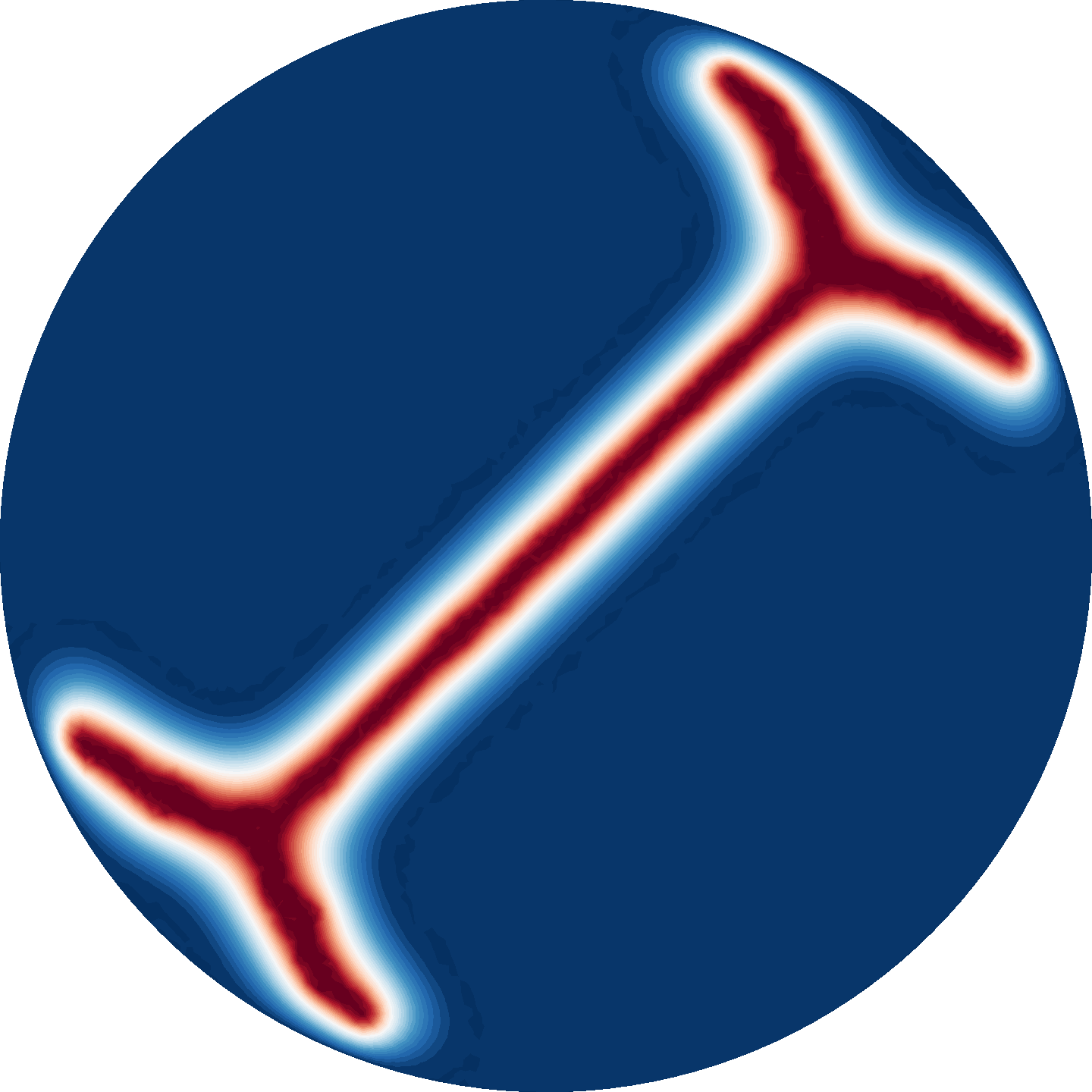}
      \putDiskFig{13}{15}{\hspace{-8mm}$\theta = 7\pi/8$}{M5-theta_180.00}
      \putDiskFig{11}{2.5}{\hspace{-15mm}$\theta = \pi$}{M5-theta_180.00}
    \end{overpic} 
    \\
    \hspace{7mm} $\sepsvtilde$
    \caption{Model \texttt{M4}.}
    \label{} 
  \end{subfigure}
  \\[2ex]
  \begin{subfigure}{\linewidth}
    \centering
    \small
    \raisebox{32mm}{$\sepsdtilde$}\hspace{2mm}
    \begin{overpic}[width=0.95\linewidth]{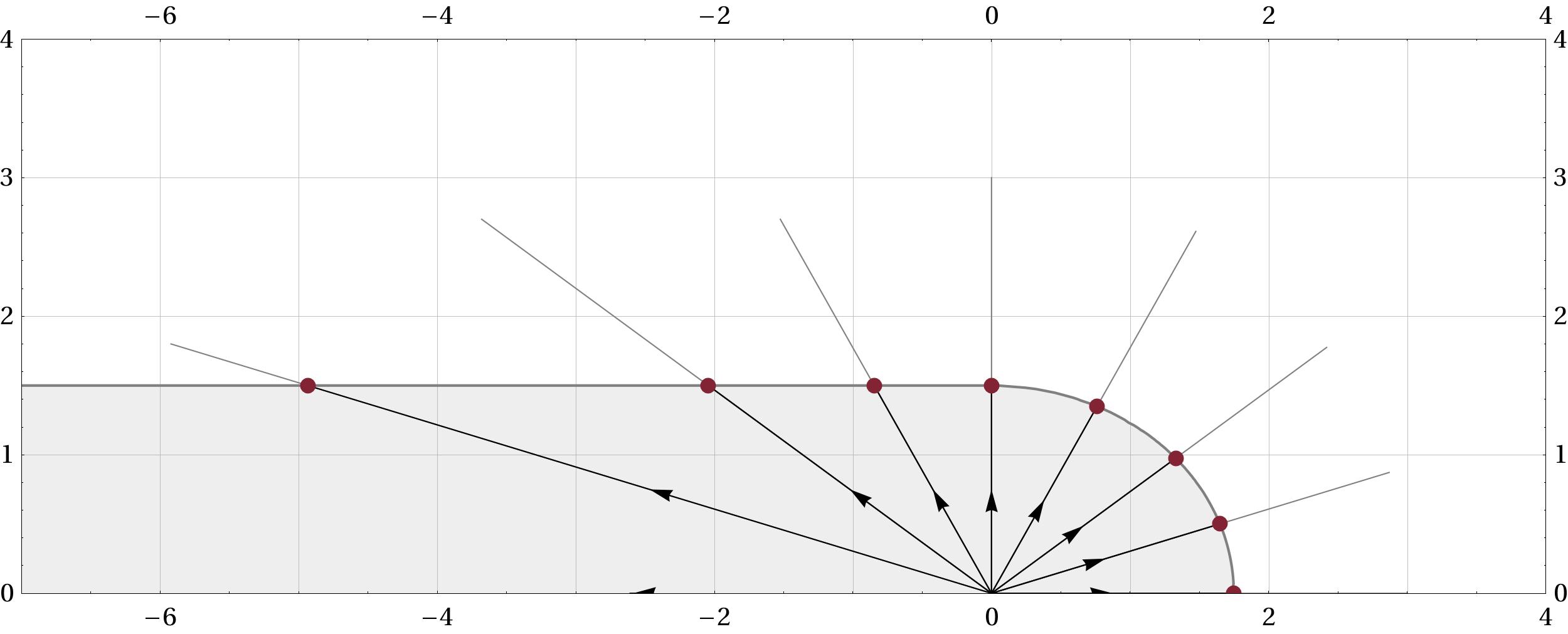}
      \putDiskFig{90}{2.5}{\hspace{15mm}$\theta = 0$}{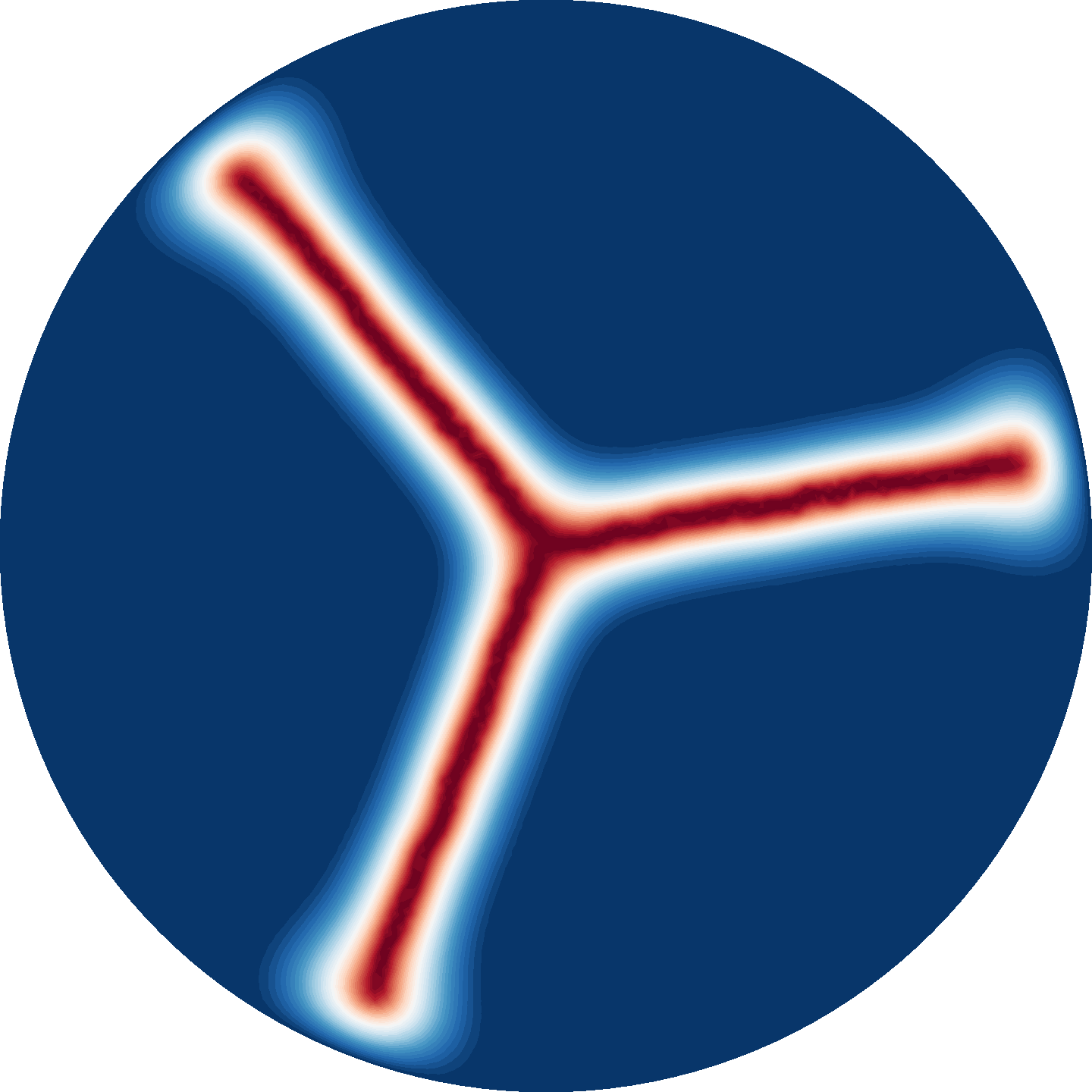}
      \putDiskFig{88}{11}{\hspace{8mm}$\theta = \pi/8$}{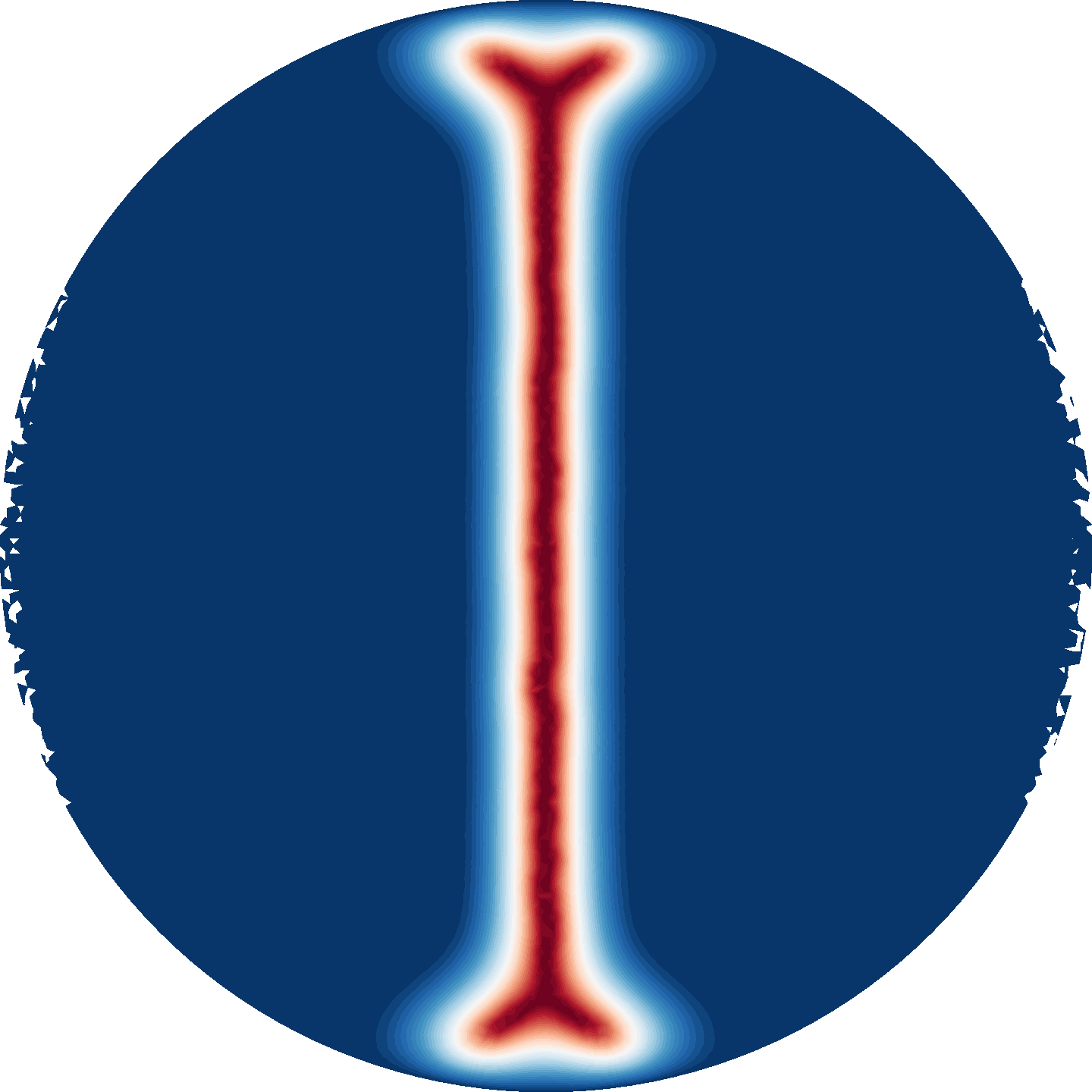}
      \putDiskFig{85}{20}{$\theta = \pi/4$}{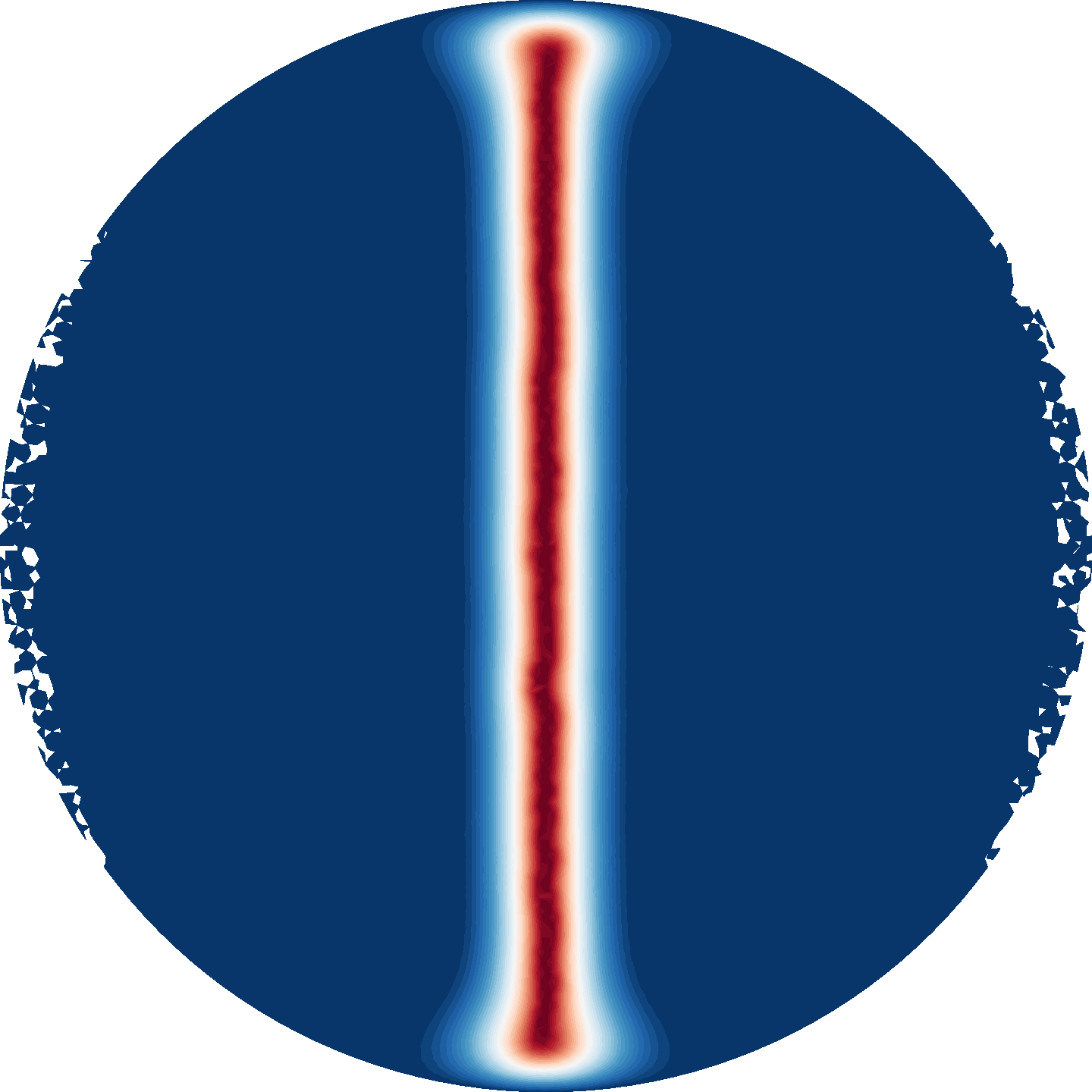}
      \putDiskFig{75}{27}{$\theta = 3\pi/8$}{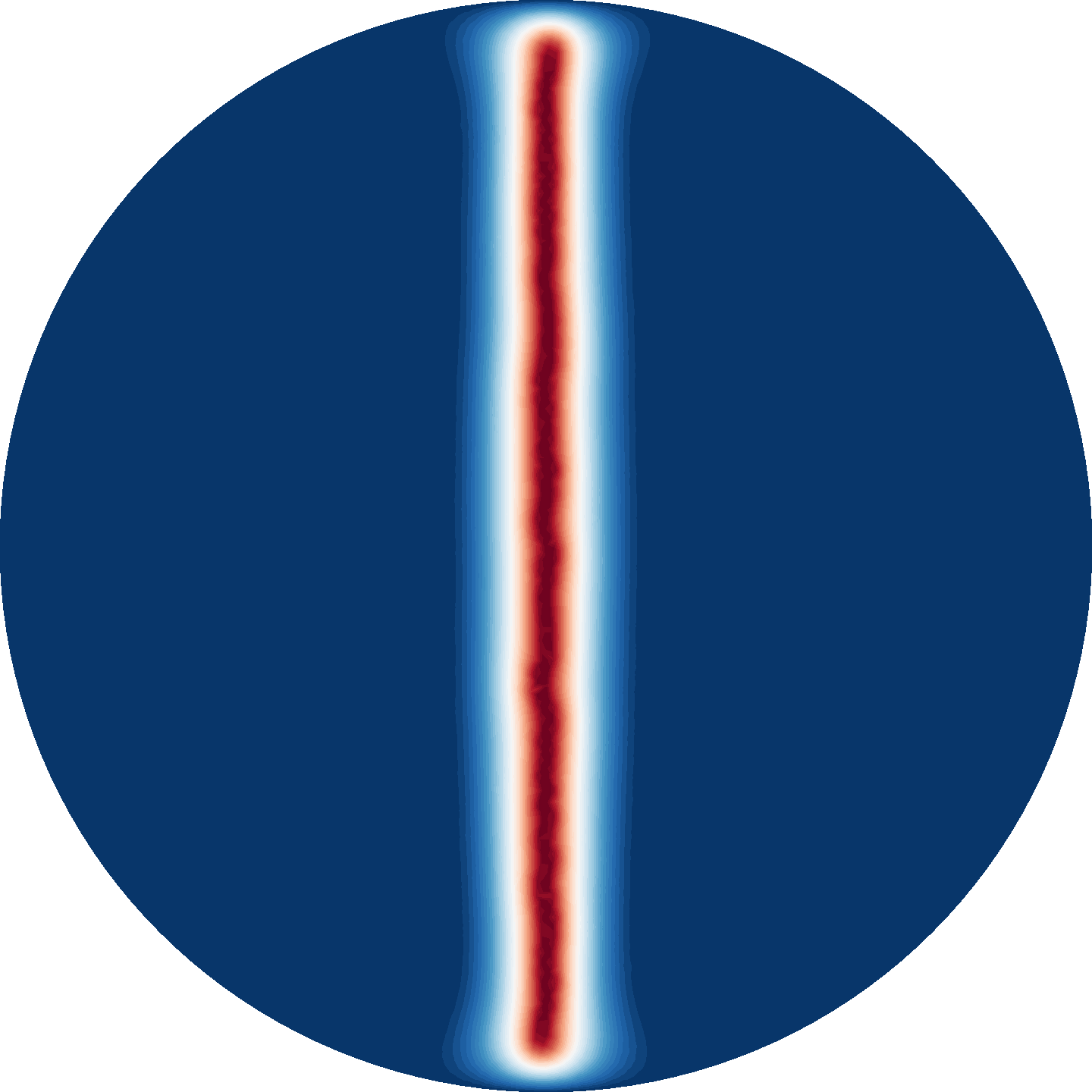}
      \putDiskFig{63}{30}{$\theta = \pi/2$}{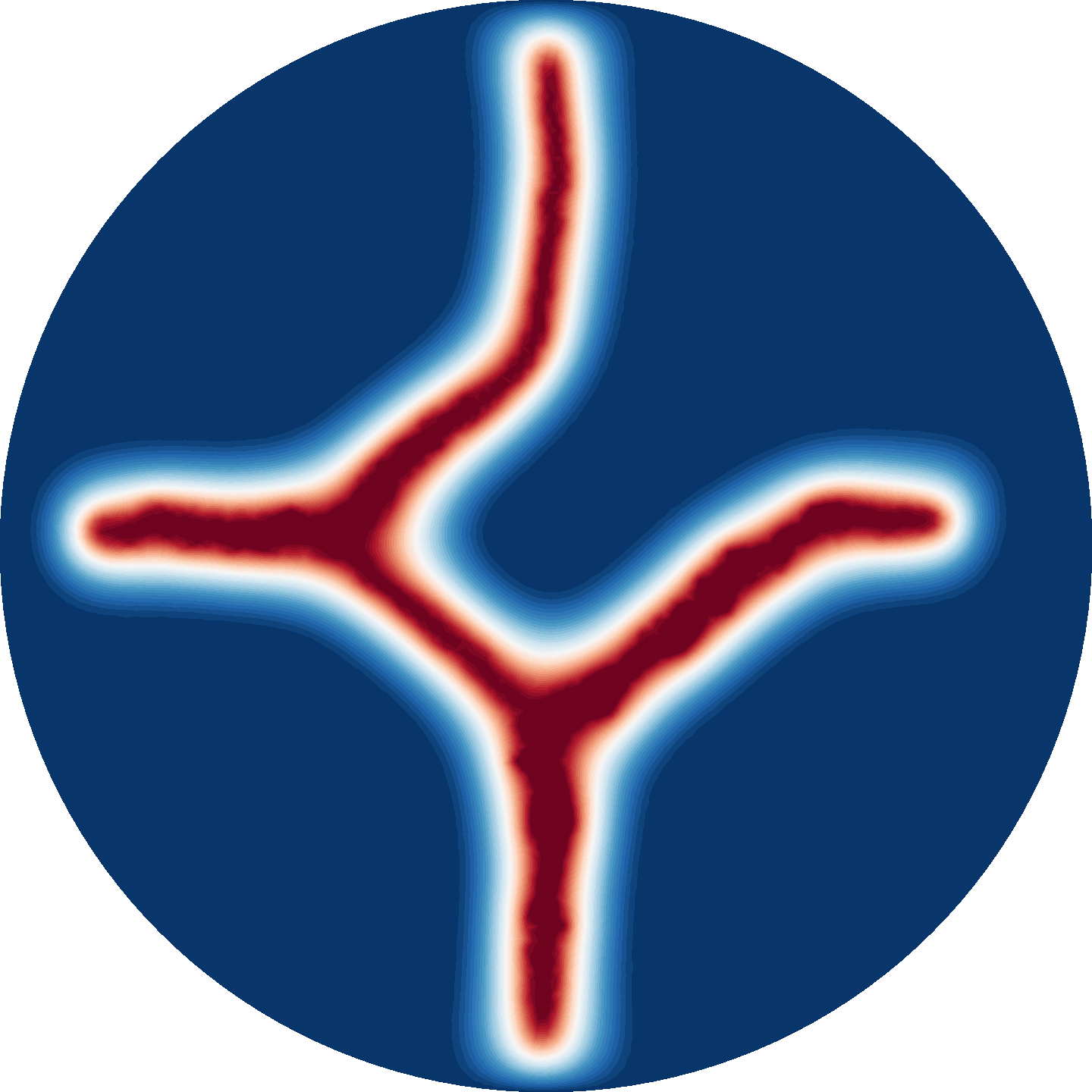}
      \putDiskFig{49}{28}{$\theta = 5\pi/8$}{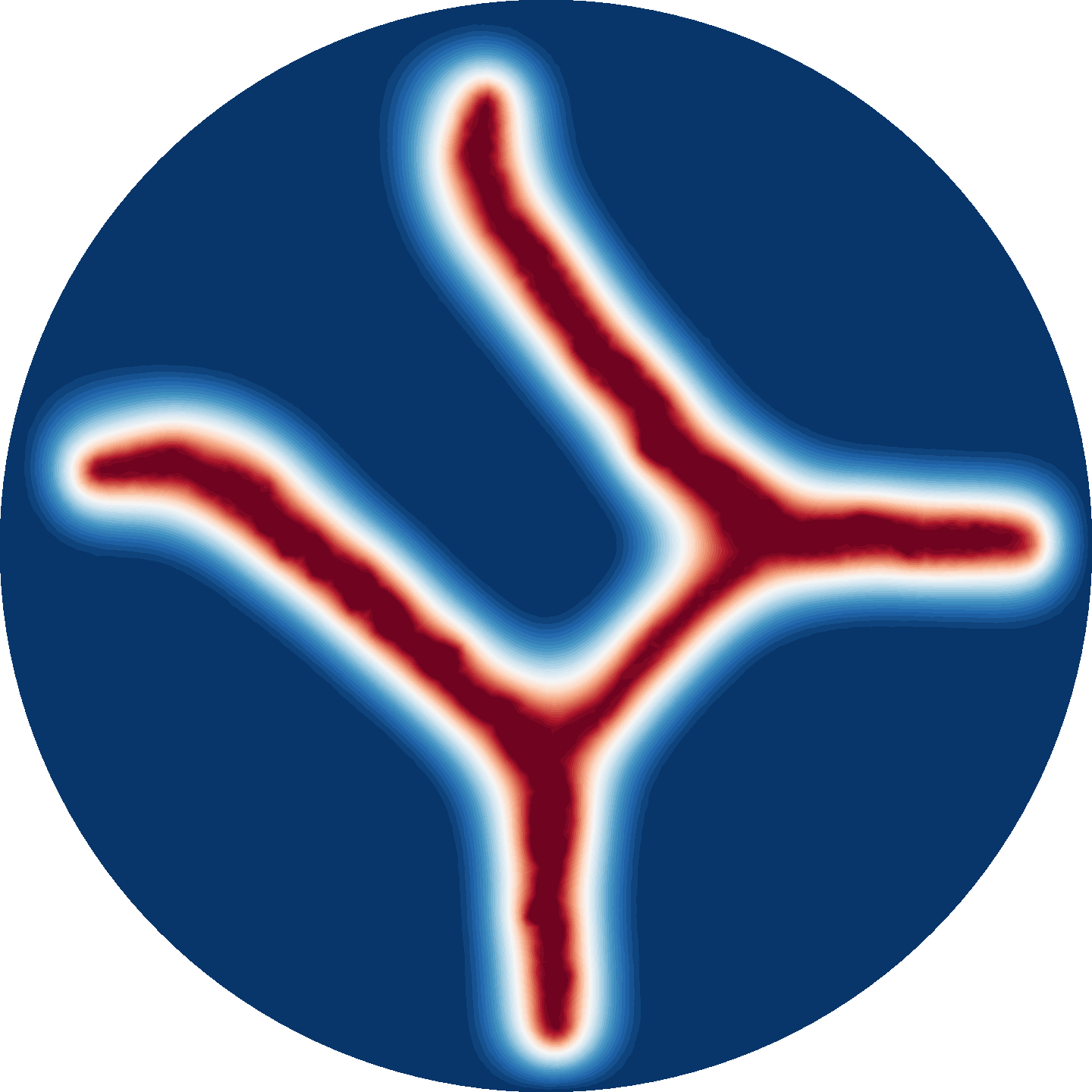}
      \putDiskFig{32}{26}{$\theta = 3\pi/4$}{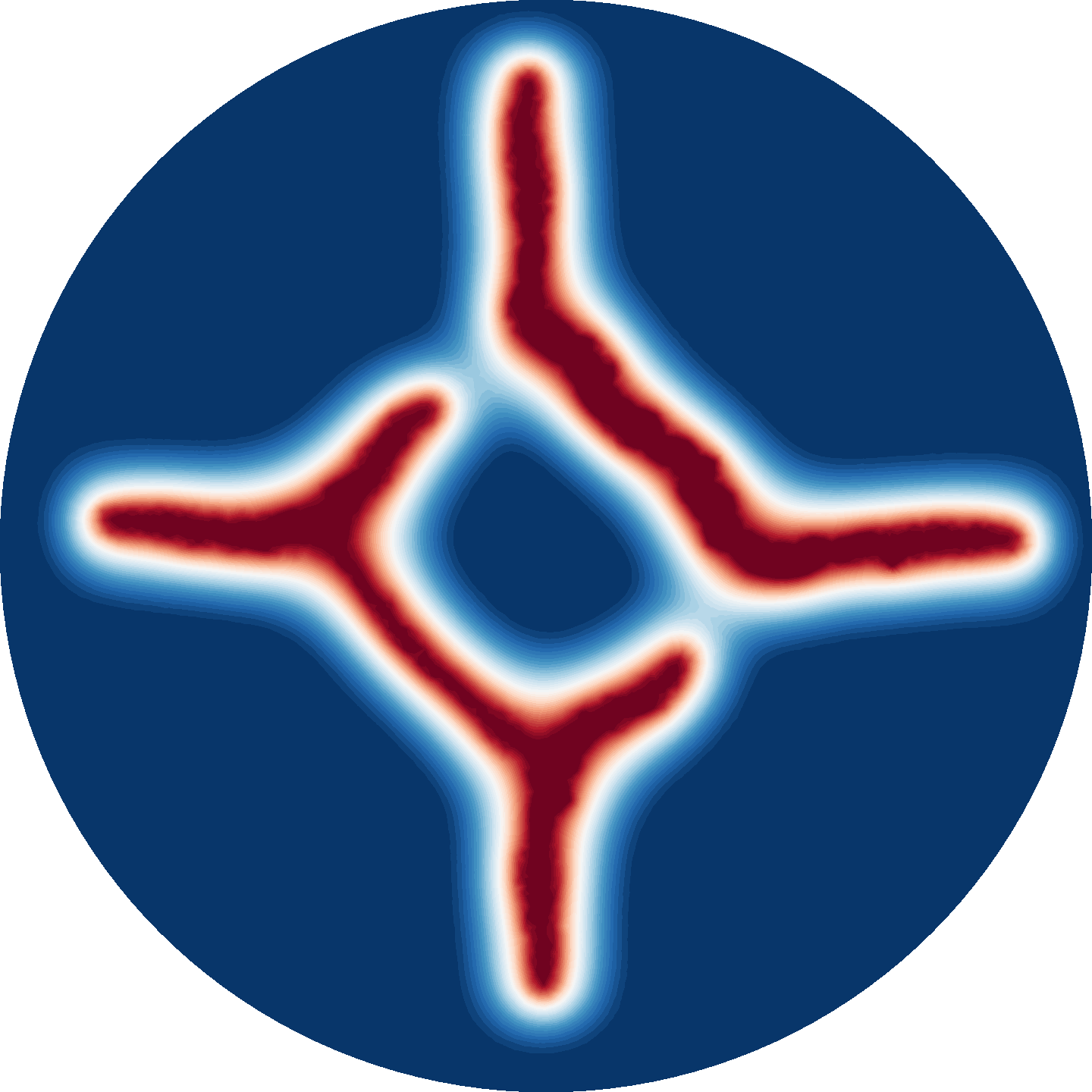}
      \putDiskFig{11}{19}{\hspace{-8mm}$\theta = 7\pi/8$}{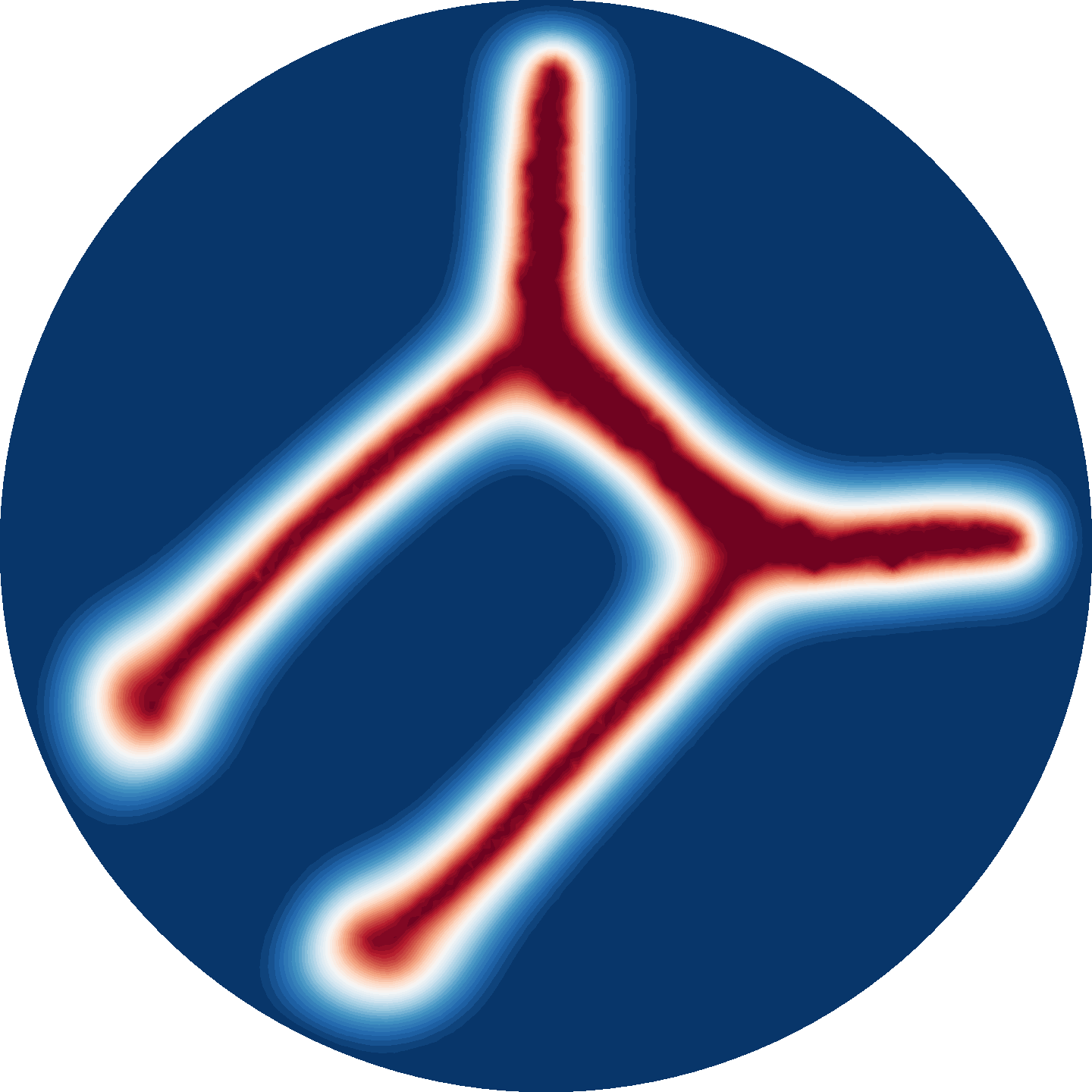}
      \putDiskFig{9}{2.5}{\hspace{-15mm}$\theta = \pi$}{M5-theta_180.00}
    \end{overpic} 
    \\
    \hspace{7mm} $\sepsvtilde$
    \caption{Model \texttt{M5}.}
    \label{} 
  \end{subfigure}
  \vfill
  \caption{Analytical elastic domains  versus numerical (dots) nucleation domains and damage fields after nucleation for the available models.}
  \label{fig_all_nuc} 
\end{figure}

\FloatBarrier
\section{Conclusions and perspectives}
\label{sec_conc} 

In this work we have proposed a novel modeling strategy to endow phase-field fracture models with custom strength criteria. In contrast to prior approaches that either modify the free energy through specific energy splits, directly alter the phase-field evolution equation through non-variational source terms, or introduce additional internal variables such as plastic or nonlinear strains, the key idea here is to introduce a state-dependent dissipation potential that depends on stress/strain invariants, while leaving the free energy responsible for the elastic degradation behavior. 
Developed within the framework of rate-independent systems and generalized standard materials, with the adoption of a new notion of local stability, the proposed energy-based modeling approach clearly distinguishes two material features: stiffness degradation, encoded in the free energy, and strength, encoded in the dissipation potential through a prescribed elastic~domain.
The resulting framework allows elastic degradation and strength criteria to be tuned independently. This feature has been illustrated through representative models, including full and partial elastic degradation together with double-ellipse, Drucker--Prager, and Huber-type strength criteria. The homogeneous response analyses and biaxial disk nucleation tests show that the proposed formulation can reproduce different mixed-mode nucleation behaviors while retaining a clear mechanical interpretation and flexible calibration. 
In this regard, the present model has been shown to overcome the limited flexibility of the volumetric--deviatoric energy split with non-interpenetration discussed in \citet[Tab.~2]{Vicentini2024}, while avoiding the spurious non-zero tangential stiffness retained by several other energy-split formulations in the fully developed crack state.
In addition, analyses of homogeneous responses under non-proportional loading have highlighted the path dependence of the dissipated work, a feature absent from standard phase-field models.

To capture unstable crack propagation events, future developments could employ path-following techniques~\cite{Wambacq2021,Loiseau2026} or time-regularizing strategies, such as inertial terms~\cite{Li2016,Carlsson2019,Correas2024} and/or viscous effects~\cite{Miehe2010,Bilgen2019,Negri2019a}. Moreover, the present framework sets the stage for incorporating more complex material mechanisms into phase-field fracture models in a systematic way.
An additional promising feature of the proposed formulation concerns the stability analysis of homogeneous states. Since the geometry of the elastic domain and the degradation/softening behavior are governed by distinct constitutive ingredients, the present framework may provide a suitable setting for disentangling their respective roles in the onset of instability and localization. In particular, this could help clarify the results obtained in recent works~\cite{Vicentini2024,Zolesi2024}, namely, whether the loss of localization observed for specific strength-domain geometries in existing phase-field formulations is an intrinsic consequence of the domain shape or results from its coupling with the underlying degradation and softening laws.
Finally, a promising research direction that we mention is the extension of the proposed approach to cohesive phase-field fracture models, with the aim of independently controlling the elastic degradation behavior, the strength criterion, and the dissipated fracture energy, while also ensuring that the latter remains independent of the internal length, along the lines of~\citet{Alessi2026a,Alessi2025b,Alessi2026c}.

\section*{Acknowledgements}
The Authors sincerely thank the anonymous Reviewers for their careful reading of the manuscript and for their insightful and constructive comments. Their feedback contributed substantially to improving both the formulation and the presentation of the present work. We also thank Reviewer~3 for bringing the relevant work of \citet{Maggiorelli2025} to our attention.

\bibliographystyle{elsarticle-num-names}
\bibliography{Research-Fracture-2024-Multiaxial_Fracture}

\end{document}